 \documentclass[aps,prc,superscriptaddress,showpacs,floatfix,nofootinbib,twocolumn]{revtex4}

\usepackage{graphicx}   
\usepackage{dcolumn}
\usepackage{bm}

\def\Journal#1#2#3#4{{#1}{\bf #2}, #3 (#4)}

\def\EPJC{{Eur. Phys. J.}~{\bf C}}

\def\NIMA{{Nucl. Instrum. Methods}~{\bf A}}
\def\NPA{{Nucl. Phys.}~{\bf A}}
\def\NPB{{Nucl. Phys.}~{\bf B}}
\def\PLB{{Phys. Lett.}~{\bf B}}
\def\PLC{Phys. Repts.\ }
\def\PRL{Phys. Rev. Lett.\ }
\def\PRD{{Phys. Rev.}~{\bf D}}
\def\PRC{{Phys. Rev.}~{\bf C}}
\def\ZPC{{Z. Phys.}~{\bf C}}
\def\ZPA{{Z. Phys.}~{\bf A}}

\newcommand \bef{\begin{figure}}
\newcommand \eef{\end{figure}}
\newcommand \befw{\begin{figure*}}
\newcommand \eefw{\end{figure*}}
\newcommand \beq{\begin{eqnarray}}
\newcommand \eeq{\end{eqnarray}}
\newcommand \bei{\begin{itemize}}

\newcommand \itb{\item[$\bullet$]}

\newcommand \eei{\end{itemize}}
\newcommand \bet{\begin{table}[hbt]}
\newcommand \eet{\end{table}}

\newcommand \mt{$\langle m_{T}\rangle$ }
\newcommand \Nch{$N_{ch}$ }
\newcommand \Et{$E_{T}$ }
\newcommand \EN{$E_{T}/N_{ch}$ }
\newcommand \dNch{$dN_{ch}/d\eta$ }
\newcommand \dEt{$dE_{T}/d\eta$ }
\newcommand \Np{$N_{p}$ }
\newcommand \Nps{$N_{p}$}

\newcommand \sqn{$\sqrt{s_{_{NN}}}$ }
\newcommand \sqns{$\sqrt{s_{_{NN}}}$}
\newcommand \2{200~GeV}
\newcommand \3{130~GeV}
\newcommand \1{19.6~GeV}
\newcommand \7{17.2~GeV}

\newcommand \etal{ {\it et al.} } 

\begin{document}

\title{Systematic Studies of the Centrality and \sqn Dependence of the \dEt and \dNch in Heavy Ion Collisions at Mid-rapidity}

\newcommand{\abilene}{Abilene Christian University, Abilene, TX 79699, USA}
\newcommand{\acadsin}{Institute of Physics, Academia Sinica, Taipei 11529, Taiwan}
\newcommand{\banaras}{Department of Physics, Banaras Hindu University, Varanasi 221005, India}
\newcommand{\barc}{Bhabha Atomic Research Centre, Bombay 400 085, India}
\newcommand{\bnl}{Brookhaven National Laboratory, Upton, NY 11973-5000, USA}
\newcommand{\caucr}{University of California - Riverside, Riverside, CA 92521, USA}
\newcommand{\ciae}{China Institute of Atomic Energy (CIAE), Beijing, People's Republic of China}
\newcommand{\cns}{Center for Nuclear Study, Graduate School of Science, University of Tokyo, 7-3-1 Hongo, Bunkyo, Tokyo 113-0033, Japan}
\newcommand{\columbia}{Columbia University, New York, NY 10027 and Nevis Laboratories, Irvington, NY 10533, USA}
\newcommand{\dapnia}{Dapnia, CEA Saclay, F-91191, Gif-sur-Yvette, France}
\newcommand{\debrecen}{Debrecen University, H-4010 Debrecen, Egyetem t{\'e}r 1, Hungary}
\newcommand{\fsu}{Florida State University, Tallahassee, FL 32306, USA}
\newcommand{\gsu}{Georgia State University, Atlanta, GA 30303, USA}
\newcommand{\hiroshima}{Hiroshima University, Kagamiyama, Higashi-Hiroshima 739-8526, Japan}
\newcommand{\ihepprot}{Institute for High Energy Physics (IHEP), Protvino, Russia}
\newcommand{\isu}{Iowa State University, Ames, IA 50011, USA}
\newcommand{\jinrdubna}{Joint Institute for Nuclear Research, 141980 Dubna, Moscow Region, Russia}
\newcommand{\kaeri}{KAERI, Cyclotron Application Laboratory, Seoul, South Korea}
\newcommand{\kangnung}{Kangnung National University, Kangnung 210-702, South Korea}
\newcommand{\kek}{KEK, High Energy Accelerator Research Organization, Tsukuba-shi, Ibaraki-ken 305-0801, Japan}
\newcommand{\kfki}{KFKI Research Institute for Particle and Nuclear Physics (RMKI), H-1525 Budapest 114, POBox 49, Hungary}
\newcommand{\korea}{Korea University, Seoul, 136-701, Korea}
\newcommand{\kurchatov}{Russian Research Center ``Kurchatov Institute", Moscow, Russia}
\newcommand{\kyoto}{Kyoto University, Kyoto 606, Japan}
\newcommand{\labllr}{Laboratoire Leprince-Ringuet, Ecole Polytechnique, CNRS-IN2P3, Route de Saclay, F-91128, Palaiseau, France}
\newcommand{\lawllnl}{Lawrence Livermore National Laboratory, Livermore, CA 94550, USA}
\newcommand{\losalamos}{Los Alamos National Laboratory, Los Alamos, NM 87545, USA}
\newcommand{\lpc}{LPC, Universit{\'e} Blaise Pascal, CNRS-IN2P3, Clermont-Fd, 63177 Aubiere Cedex, France}
\newcommand{\lund}{Department of Physics, Lund University, Box 118, SE-221 00 Lund, Sweden}
\newcommand{\muenster}{Institut f\"ur Kernphysik, University of Muenster, D-48149 Muenster, Germany}
\newcommand{\myongji}{Myongji University, Yongin, Kyonggido 449-728, Korea}
\newcommand{\nagasaki}{Nagasaki Institute of Applied Science, Nagasaki-shi, Nagasaki 851-0193, Japan}
\newcommand{\newmex}{University of New Mexico, Albuquerque, NM 87131, USA}
\newcommand{\nmsu}{New Mexico State University, Las Cruces, NM 88003, USA}
\newcommand{\ornl}{Oak Ridge National Laboratory, Oak Ridge, TN 37831, USA}
\newcommand{\orsay}{IPN-Orsay, Universite Paris Sud, CNRS-IN2P3, BP1, F-91406, Orsay, France}
\newcommand{\pnpi}{PNPI, Petersburg Nuclear Physics Institute, Gatchina, Russia}
\newcommand{\riken}{RIKEN (The Institute of Physical and Chemical Research), Wako, Saitama 351-0198, JAPAN}
\newcommand{\rkrbrc}{RIKEN BNL Research Center, Brookhaven National Laboratory, Upton, NY 11973-5000, USA}
\newcommand{\saispbstu}{St. Petersburg State Technical University, St. Petersburg, Russia}
\newcommand{\saopaulo}{Universidade de S{\~a}o Paulo, Instituto de F\'{\i}sica, Caixa Postal 66318, S{\~a}o Paulo CEP05315-970, Brazil}
\newcommand{\seoulnat}{System Electronics Laboratory, Seoul National University, Seoul, South Korea}
\newcommand{\stonybrkc}{Chemistry Department, Stony Brook University, SUNY, Stony Brook, NY 11794-3400, USA}
\newcommand{\stonycrkp}{Department of Physics and Astronomy, Stony Brook University, SUNY, Stony Brook, NY 11794, USA}
\newcommand{\subatech}{SUBATECH (Ecole des Mines de Nantes, CNRS-IN2P3, Universit{\'e} de Nantes) BP 20722 - 44307, Nantes, France}
\newcommand{\tenn}{University of Tennessee, Knoxville, TN 37996, USA}
\newcommand{\titech}{Department of Physics, Tokyo Institute of Technology, Tokyo, 152-8551, Japan}
\newcommand{\tsukuba}{Institute of Physics, University of Tsukuba, Tsukuba, Ibaraki 305, Japan}
\newcommand{\vandy}{Vanderbilt University, Nashville, TN 37235, USA}
\newcommand{\waseda}{Waseda University, Advanced Research Institute for Science and Engineering, 17 Kikui-cho, Shinjuku-ku, Tokyo 162-0044, Japan}
\newcommand{\weizmann}{Weizmann Institute, Rehovot 76100, Israel}
\newcommand{\yonsei}{Yonsei University, IPAP, Seoul 120-749, Korea}
\affiliation{\abilene}
\affiliation{\acadsin}
\affiliation{\banaras}
\affiliation{\barc}
\affiliation{\bnl}
\affiliation{\caucr}
\affiliation{\ciae}
\affiliation{\cns}
\affiliation{\columbia}
\affiliation{\dapnia}
\affiliation{\debrecen}
\affiliation{\fsu}
\affiliation{\gsu}
\affiliation{\hiroshima}
\affiliation{\ihepprot}
\affiliation{\isu}
\affiliation{\jinrdubna}
\affiliation{\kaeri}
\affiliation{\kangnung}
\affiliation{\kek}
\affiliation{\kfki}
\affiliation{\korea}
\affiliation{\kurchatov}
\affiliation{\kyoto}
\affiliation{\labllr}
\affiliation{\lawllnl}
\affiliation{\losalamos}
\affiliation{\lpc}
\affiliation{\lund}
\affiliation{\muenster}
\affiliation{\myongji}
\affiliation{\nagasaki}
\affiliation{\newmex}
\affiliation{\nmsu}
\affiliation{\ornl}
\affiliation{\orsay}
\affiliation{\pnpi}
\affiliation{\riken}
\affiliation{\rkrbrc}
\affiliation{\saispbstu}
\affiliation{\saopaulo}
\affiliation{\seoulnat}
\affiliation{\stonybrkc}
\affiliation{\stonycrkp}
\affiliation{\subatech}
\affiliation{\tenn}
\affiliation{\titech}
\affiliation{\tsukuba}
\affiliation{\vandy}
\affiliation{\waseda}
\affiliation{\weizmann}
\affiliation{\yonsei}
\author{S.S.~Adler}	\affiliation{\bnl}
\author{S.~Afanasiev}	\affiliation{\jinrdubna}
\author{C.~Aidala}	\affiliation{\bnl}
\author{N.N.~Ajitanand}	\affiliation{\stonybrkc}
\author{Y.~Akiba}	\affiliation{\kek} \affiliation{\riken}
\author{J.~Alexander}	\affiliation{\stonybrkc}
\author{R.~Amirikas}	\affiliation{\fsu}
\author{L.~Aphecetche}	\affiliation{\subatech}
\author{S.H.~Aronson}	\affiliation{\bnl}
\author{R.~Averbeck}	\affiliation{\stonycrkp}
% ...
\collaboration{PHENIX Collaboration} \noaffiliation

\date{\today}

\begin{abstract}
The PHENIX experiment at RHIC has measured transverse energy and charged particle multiplicity 
at mid-rapidity in $Au+Au$ collisions at \sqn = 19.6, 130 and 200 GeV as a function of centrality.
The presented results are compared to measurements from other RHIC experiments, and experiments at 
lower energies. The \sqn dependence of \dEt and \dNch per pair of participants is 
consistent with logarithmic scaling for the most central events. The centrality dependence of 
\dEt and \dNch is similar at all measured incident energies. At RHIC energies the ratio of transverse 
energy per charged particle was found independent of centrality and growing slowly with \sqns. 
A survey of comparisons between the data and available theoretical models is also presented.
\end{abstract}

\maketitle

\section{Introduction}
The PHENIX experiment at the Relativistic Heavy Ion Collider (RHIC) at Brookhaven National Laboratory was designed to
measure the properties of matter at extremely high temperatures and densities. Under such conditions, the possibility 
exists to produce states of matter that have not been observed and studied in the laboratory. Perhaps the best known
of these is the {\em quark-gluon plasma} (QGP), matter in which the quarks are not confined within individual baryons
but exist as some form of plasma of individual quarks and gluons. It should be emphasized that the exact properties of
this matter are not known and that the characterization of the deconfined state, if such a state is produced, will 
form an essential part of the RHIC program.

One fundamental element of the study of ultrarelativistic collisions is the characterization of the interaction in terms
of variables such as the energy produced transverse to the beam direction or the number of charged particles. These 
variables are closely related to the collision geometry and are important in understanding global properties of the 
system during the collision.

This paper describes a systematic study of \dEt and \dNch at mid-rapidity by the PHENIX experiment at center-of-mass 
energies \sqns=\1, \3 and \2. The centrality dependence of \dEt and \dNch is characterized by  the number of 
participants, determined with a Glauber model, and is studied as function of the incident energy.
\dEt and \dNch results for all four RHIC experiments are included as part of this study. 
The data taken at 19.6 GeV are particularly interesting because a close comparison to the lower energies 
of the CERN SPS program can be made. Comparisons are also made to previous experiments at the Brookhaven AGS 
and CERN SPS at center-of-mass energies of 4.8 GeV, 8.7 GeV, and 17.2 GeV. Finally, an extensive set of collision models
describing the \Et and \Nch distributions are compared to the existing data.

\section{PHENIX detector\label{sec:dtector}}
PHENIX is one of four experiments located at RHIC~\cite{rhic}. 
The PHENIX detector consists of two central spectrometer arms, 
designated east and west for their location relative to the interaction region,
and two muon spectrometers, similarly called north and south.  
Each central spectrometer arm covers a 
rapidity range of $|\eta|<0.35$ and subtends $90^{0}$ in azimuth. 
The  muon spectrometers both have full azimuthal coverage with a rapidity range of 
$-2.2<\eta<-1.2$ (south) and $1.2<\eta<2.4$ (north).  Additional global 
detectors are used as input to the trigger and for global event characterization 
such as vertex, time of event and centrality determination.
A detailed description of the PHENIX detector can be found in~\cite{phenix}. 
The PHENIX detector subsystems relevant for the physics 
analysis presented in this paper are listed below.

Charged particle multiplicity was measured with two MWPC layers of the Pad Chambers 
(PC)~\cite{pc} called PC1, and PC3. These are located in both central arms at 
the radii of 2.5~$m$ and 5.0~$m$ from the beam axis. The PCs cover the full 
central arm acceptance and have an efficiency for minimum ionizing particles 
greater than 99.5\%. The position resolution of PC1 was measured to be 1.7~$mm$ 
by 3~$mm$ and twice that for PC3. PC1 and PC3 can distinguish between two particle tracks if 
they strike the detector with a separation greater than 4~$cm$ and 8~$cm$, respectively.

For the transverse energy measurements, a PbSc sampling calorimeter 
(EMCal)~\cite{emcal} from the PHENIX central spectrometers was used.
The front face of EMCal is located 5.1~m from the beam axis. 
Scintillation light produced in the PbSc EMCal 
towers is read out through wavelength shifting fibers which 
penetrate the module. The depth of the PbSc calorimeter is $18$ radiation 
lengths ($X_0$) which corresponds to $0.85$ nuclear interaction lengths. 
The PbSc calorimeter has an energy resolution of 
8.1\%/{$\sqrt {E({\rm GeV})}$} $\oplus 2.1$\% for test beam electrons, 
with a measured response proportional to 
the incident electron energy that is within $\pm 2$\% over the range 
$0.3 \leq E_e \leq 40.0$~GeV~\cite{emcal}.

Two identical Beam-Beam Counters (BBC)~\cite{bbc-zdc} consisting of 64 individual Cherenkov 
counters with 2~$cm$ quartz glass radiators each cover the full azimuthal angle 
in the pseudorapidity range $3.0<|\eta|<3.9$. These detectors provide a minimum 
biased (MB) event trigger and timing and are also used for event vertex 
determination. The vertex position resolution for central $Au+Au$ events was
6~$mm$ along the beam axis.

The Zero Degree Calorimeters (ZDC)~\cite{zdc} are hadronic calorimeters located 
on both sides of the PHENIX detector. They cover a rapidity region of $|\eta|>6$ 
and measure the energy of the spectator neutrons with approximately 20\% energy resolution.

The BBC and ZDC were used for the centrality determination.

\section{Data Analysis\label{sec:anaisys}}

The analysis procedures for the \dEt and \dNch measured at \sqns=\3 are described 
in~\cite{phenix_et}~and~\cite{phenix_nch} respectively. In this paper the analysis 
was improved including:
\bei
\itb Inflow and outflow corrections done based on the identified particle data, as opposed to HIJING.
\itb Corrected trigger efficiency of $92.2^{+2.5}_{-3.0}$\% instead of $92.0\pm2\pm1$\%.
\itb Modified definition of the \Et as discussed below.
\eei
The results presented here for \sqns=\3 are consistent with results previously published. 

The same data samples with zero magnetic field were used for both \Et and \Nch measurements 
at each beam energy. The analyzed numbers of events are approximately
$40\times10^3$, $160\times10^3$ and $270\times10^3$ for \sqns=\1, \3 and \2 
respectively.

The main steps of the analysis procedure are discussed below in connection to the systematic 
errors associated with them. Some additional details can be found 
in~\cite{phenix_milov,phenix_bazik,david_thesis,sasha_thesis}. 

\subsection{\Et analysis}
The transverse energy ($E_T$) is defined as:
\begin{equation} 
E_T=\sum_{i} E_i \sin\theta_i. 
\label{eq:ETdef}
\end{equation} 
where $\theta_i$ is the polar angle. The sum is taken over all particles emitted 
into a fixed solid angle in an event. By convention, $E_i$ is taken to be 
$E_i^{tot}-m_{N}$ for baryons, $E_i^{tot}+m_{N}$ for antibaryons and $E_i^{tot}$ 
for all other particles, where $E_i^{tot}$ is the 
total energy of the particle and $m_{N}$ is the nucleon mass 
\footnote{The definition of $E_i$ in our earlier publication 
\cite{phenix_et} is different for antibaryons contribution: $E_i^{tot}$ 
was used instead of $E_i^{tot}+m_{N}$. The current definition increases 
the value of $E_T$ by about 4\%, independent of centrality.}.

The \Et measurement presented in this paper was performed using the PHENIX PbSc EMCal. 
The EMCal absolute energy scale was set using the $\pi^0$ mass peak  
reconstructed from pairs of EMCal clusters. The value was checked 
against a measurement of the minimum ionizing peak for charged particles 
penetrating 
along the tower axis and the energy/momentum ($E/p$) peak of identified 
electrons and positrons. The uncertainty in the absolute energy scale is 
3\% in the \sqn=\1 data and 1.5\% in the \3 and \2 data. 

The EMCal acts as a thin but effective hadronic calorimeter at mid-rapidity 
at a collider \cite{phenix_et}. The mean hadron momenta in the EMCal 
acceptance are approximately 0.4, 0.55 and 0.9 GeV/c for pion, kaons and 
(anti)protons respectively \cite{julia}. Most hadrons stop in the EMCal, 
depositing all their kinetic energy (at $p_T$ less than 0.35 GeV/c for pions, 
0.64 GeV/c for kaons and 0.94 GeV/c for protons). 

The average EMCal response to the different particle species was obtained 
with a GEANT-based \cite{geant} Monte Carlo (MC) simulation of the PHENIX detector 
using the HIJING \cite{hijing_sim} event generator. The HIJING particle 
composition and $p_T$ spectra were tuned to the identified charged particle 
spectra and yields in $Au+Au$ collisions measured by PHENIX \cite{julia,tatsuja} 
at \sqns=\2 and \3. The NA49 results~\cite{na49_1,na49_2,na49_3} were 
used for EMCal response studies for \sqn=\1 data.
The ``deposited'' $E_{T_{EMC}}$ was about 75\% of the total $E_T$ ``striking'' 
the EMCal. This value varied in the $\pm$1.5\% range for different 
centralities and beam energies. 

The uncertainty in the EMCal response to hadrons gave a 3\% error to the total 
$E_T$. This uncertainty was estimated using a comparison between the 
simulated energy deposited by hadrons with different momenta and from the 
test beam data ~\cite{emcal}. An additional error of 1.3\% at \sqn=19.6 and \2 
and 1\% at \sqn=\3 comes from the systematic uncertainties 
in the particle composition and momentum distribution.

$E_T$ was computed for each event (Eq.~\ref{eq:ETdef}) using clusters with 
energy greater than 30 MeV composed of adjacent towers with 
deposited energy of more than 10 MeV\footnote{In \cite{phenix_et} 
thresholds of 20 MeV and 3 MeV were applied for the cluster and 
for the tower, respectively. Energy losses due to thresholds were properly 
accounted for in both analyses.}. 
The energy losses at the EMCal edges and due to energy thresholds, 6\% each, 
were estimated with the absolute uncertainty 1.5\%.

The first main issue for the $E_T$ measurement is the correction for 
losses for particles which originate within the aperture but whose 
decay products miss the EMCal ($\sim$10\%). The second is for the in-flow contribution
($\sim$24\%), which is principally of two types: (1) albedo from the 
magnet poles; (2) particles which originate outside the aperture of the 
calorimeter but whose decay products hit the calorimeter. The in-flow 
component was checked by comparing the MC and the measurements for events 
with a vertex just at and inside a pole face of the axial central-spectrometer 
magnet, for which the calorimeter aperture was partly shadowed. The estimated 
contribution of the in-flow uncertainty to the $E_T$ uncertainty is 3\%~\cite{phenix_et}.

Since $E_T$ measurements are based on the sum of all cluster 
energies in the EMCal, random noise even in a small portion of the total 
number of EMCal towers ($\sim$15,000 in PbSc) may affect the total energy 
in the EMCal, particularly in peripheral collisions. This 
effect was estimated by measuring the total energy in the EMCal in very peripheral 
events with the collision vertex inside the magnet poles. In this case, 
the EMCal is fully shadowed and no energy deposit from beam 
collisions is expected. The estimated contribution was consistent with zero. 
The uncertainty from this effect contributes 3.5\% systematic error to the $E_T$ 
measurement in the most peripheral bin of 45\%-50\% at \sqn=\1, 10\% in the most 
peripheral bin of 65\%-70\% at \sqn=\3 and 6\% to the bin of 65\%-70\% at \sqn=200 GeV.
The contribution to the systematic error for central events was negligible.

\subsection{\Nch analysis}
In the absence of a magnetic field, the particle tracks are straight lines. 
The number of tracks in the event was determined by combining all hits in 
PC3 with all hits 
in PC1. The resulting straight lines were projected onto a plane 
containing the beam line and perpendicular to the symmetry axis of the 
PCs. All tracks intersecting the plane at a radius less 
than 25~$cm$ from the event vertex were accepted. 
95$\pm$1\% of all real tracks in the event point back within this radius. 

The complete set of tracks thus formed contains both real tracks and tracks from
combinatorial background. The 
latter can be determined using a mixed event technique in which each sector in PC1 is
exchanged with its neighbor and the resulting combinatorial background measured.

The average combinatorial background from the mixed event analysis was subtracted from the 
data from the real events. Several corrections were subsequently applied.

A correction of 15.3\%  accounts for nonsensitive mechanical gaps between the PC sectors, 
inactive electronic readout cards, and dead pads in the PC1 and PC3 
detectors. The data were also corrected for the PC efficiency for an isolated hit, 
measured to be 99.5\% using cosmic rays~\cite{pc}. The combined systematic error 
from these corrections was estimated to be 2.5\% for single
east arm and 2.3\% for both east and west arms.

Track losses from the finite double hit resolution of the PCs 
depend on the event multiplicity. Losses occur in both the direct counting 
of tracks and in the combinatorial background subtraction. These two effects 
were studied in great detail using Monte Carlo techniques. To account for 
the track losses in the real event sample, a correction of 15\%, 13\% and 6\%
for the 5\% most central events was applied at \sqns=200, 130, and \1, respectively.

Track loss due to the finite double hit resolution reduces combinatorial background 
in the real events more than in the mixed events. The number of tracks in the mixed events 
must be decreased by 3.6\% to account for this. The uncertainty in the correction related 
to the finite double hit resolution of the PCs
is estimated to be 3.5\% of the number of reconstructed tracks in the most central events 
at \sqns~=~200~GeV. This number 
was deduced from the simulation and cross-checked with an artificial 50\% increase 
of the double hit resolution of PC1 and PC3.

An additional correction is related to the decays of charged particles and feed-down 
from the decays of neutral particles as discussed in~\cite{phenix_nch} where it  
was determined using the HIJING event generator. 
In this paper the measured composition of the produced particles at different 
centralities is used at \sqn=200 and \3~\cite{julia,tatsuja}. The correction related 
to particle decay varies about $\pm$1\% over the full range of measured centralities. 
In mid-central events it is $-1\pm2.9$\% and $+1\pm2.5$\% 
at \sqn=\2 and \3 respectively. At the lowest RHIC energy the correction 
is based on NA49~\cite{na49_1,na49_2,na49_3} measurements at close energy \sqn=\7 
and is about 11$\pm$5.7\% independent of centrality. The difference between 
\sqn = \1 and \3 arises from the decrease of the particle momenta and the width of 
the $\eta$-distribution at lower energy which affects the number of tracks from decays 
of the particles coming from adjacent rapidities. The uncertainty is also larger because the 
correction was based on non-PHENIX data. More details on 
the analysis can be found at~\cite{phenix_nch,david_thesis,sasha_thesis}.

\subsection{Determination of trigger efficiency and \Nps.\label{sec:cent}}
The distribution of the number of participants (\Nps) in $Au+Au$ collisions 
was determined using a Monte Carlo simulation based on the Glauber model. 
The inelastic cross section of $p+p$ collisions used in the Glauber model 
was taken to be 31~mb, 41~mb and 42~mb at \sqn = \1, \3 and \2 
respectively~\cite{PDG} and varied within $\pm$3~mb in order to get the systematic errors.
The nuclear density profile $\rho(r)$ was taken as the 
Woods-Saxon parameterization:
\begin{equation}  
\rho(r) = 1/(1+e^{(r-r_{n})/d}),
\label{eq:WSdensity}
\end{equation} 
where $r_{n}$ is the nucleus radius 
and $d$ is a diffuseness parameter. Based on the measurements of 
electron scattering from $Au$ nuclei~\cite{Hofstadter}, $r_{n}$ was 
set to (6.38$\pm$0.27)~fm and $d$ to (0.54$\pm$0.01)~fm. 

The BBC detectors are located in a region where the number of produced 
particles is proportional to \Np at \sqn=\3 and \2~\cite{phobos6}. 
By comparing measured BBC spectra to simulations, the MB 
trigger efficiency was estimated to be $92.2^{+2.5}_{-3.0}$\% at both \sqn=\2 
and \3, with less than 1\% uncertainty in the difference between these 
two energies.

One can also use the BBC (or ZDC vs. BBC) response to define 
centrality for a given event as a percentile of the total geometrical cross section.
The BBC amplitude distribution and ZDC vs 
BBC signals divided into centrality classes are shown in 
Fig.~\ref{fig:bbc_zdc}.
\befw
\includegraphics[width=0.48\linewidth]{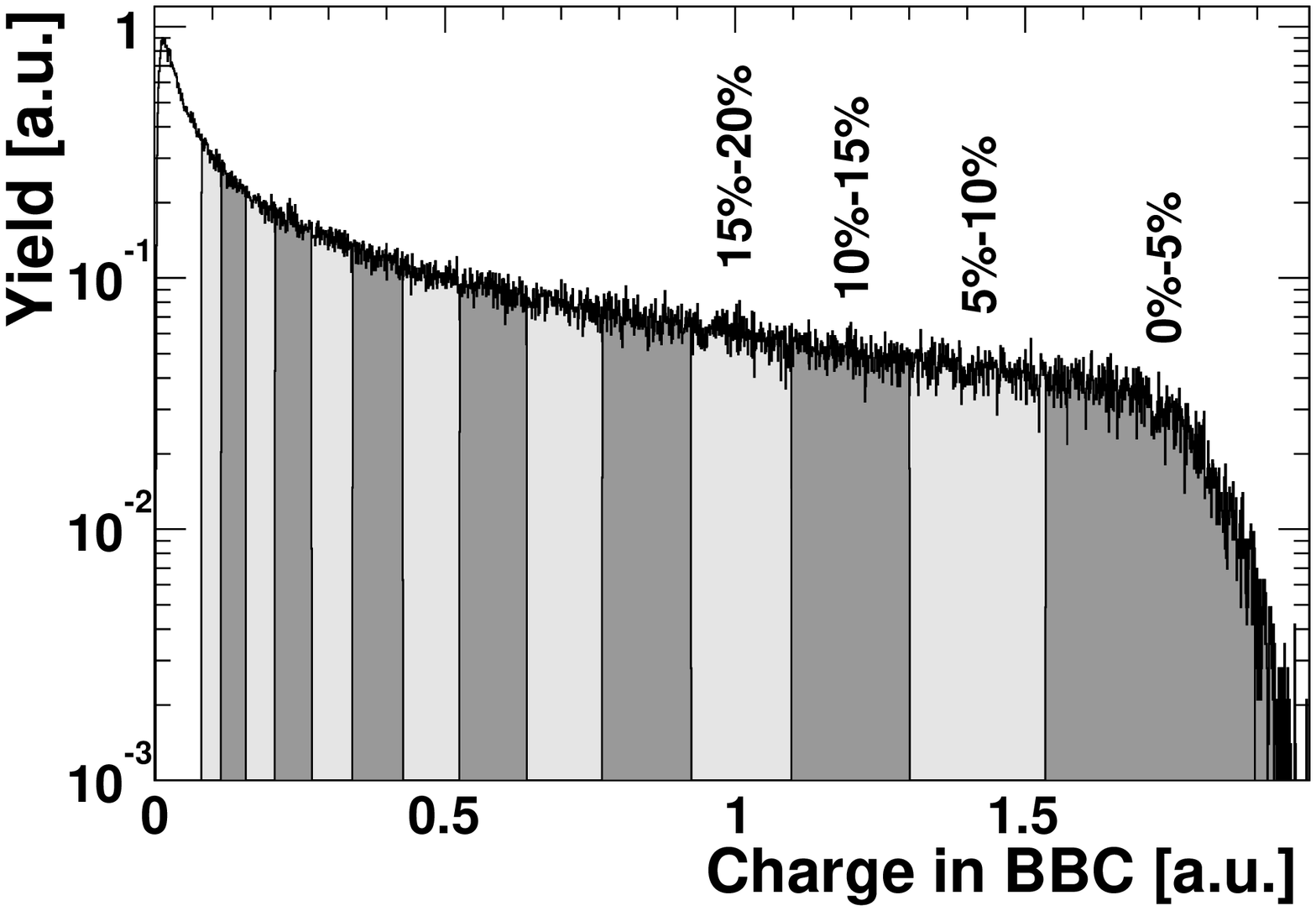}
\includegraphics[width=0.48\linewidth]{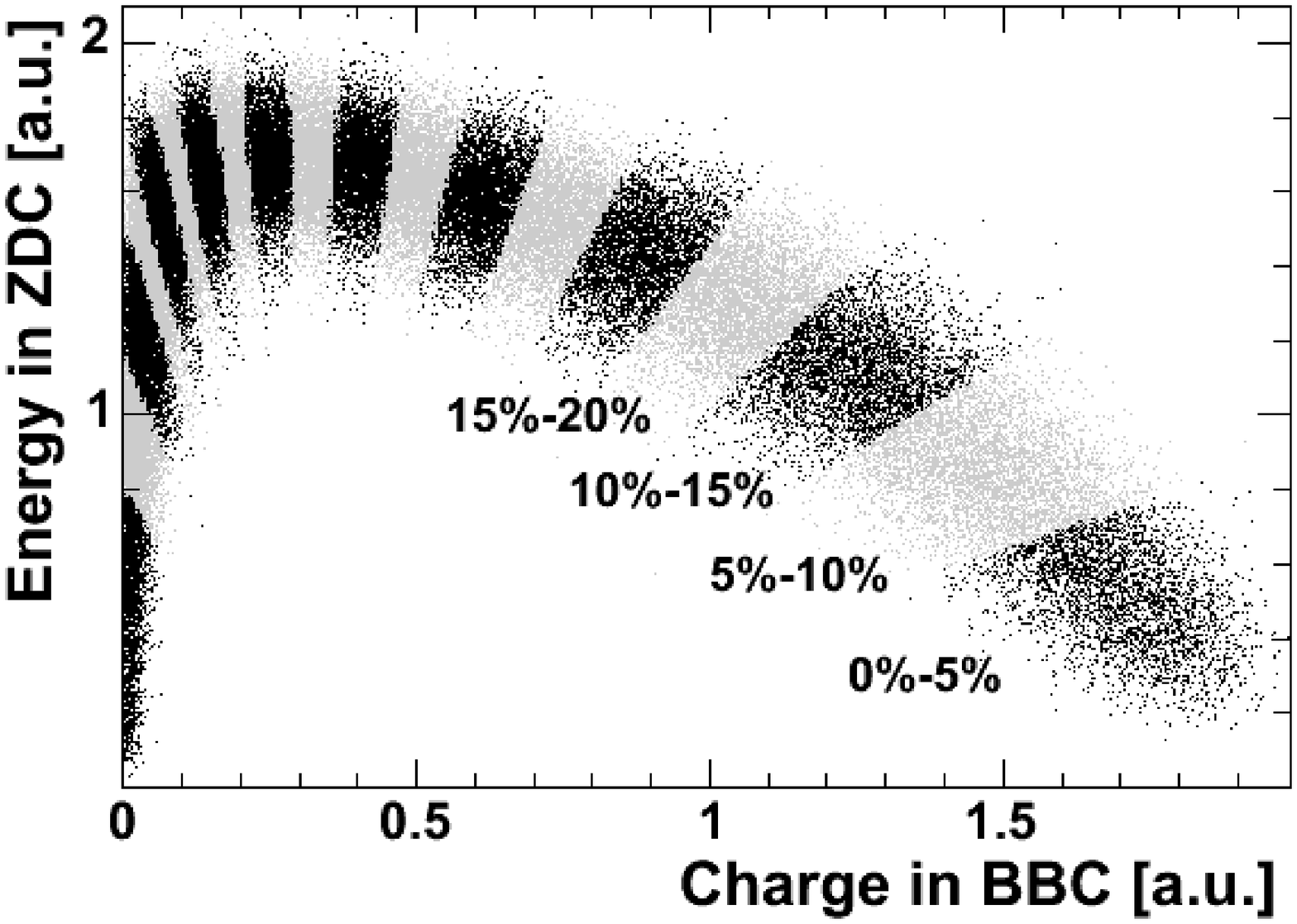}
\caption{Different centrality classes based on the BBC (left) and ZDC vs. BBC 
distributions (right).\label{fig:bbc_zdc}}
\eefw
By matching the detector response simulation to the data, \Np can be 
assigned to each centrality class. The results for \Np vary by less 
than 0.5\% depending on the shape of the cut in the ZDC/BBC space and 
whether the BBC alone was used as a centrality measure. The larger error 
in \Np comes from model uncertainties and can be parameterized as 
$\Delta N_{p}/N_{p}=0.02+3.0/N_{p}$.

At \sqn=\1, the BBC acceptance partially covers the $Au$ nuclei 
fragmentation region where the relation between the particle production and \Np 
is not well known for peripheral events. This makes the MB trigger 
efficiency model dependent. To avoid this problem, an approach based 
on the Glauber model and the Negative Binomial Distribution (NBD) was applied 
to the data from the PHENIX central arm. For the centrality 
associations, the BBC signal can still be used after applying the correction 
described in the text below.

The NBD, written as:
\beq
P(n,\mu,k)=\Gamma(n+k)/(\Gamma(k)n!)\cdot(\mu/k)^{n}/(1+\mu/k)^{n+k}
\label{eq:nbd}
\eeq
represents the number of independent trials $n$ that are required to get a number of 
predetermined successes if the average number of successes per trial is $\mu$. 
The parameter $k$ is related to the variance of the distribution by the equation 
$(\sigma/\mu)^{2}=1/k+1/\mu$. By associating $n$ with the number of particles produced 
in the event such that $n = f(N_{p})$, the NBD describes the distribution of 
hits in a detector~\cite{ags_mjt,mitchell} produced by a given number of \Nps.
In the simplest case when $n \propto N_{p}$, $\langle N_{hit} \rangle =\mu \langle N_{p} \rangle$.
Using probability weights for \Np from the Glauber model, one can construct a 
distribution of the number of hits in a detector. The coefficients $\mu$ and $k$ 
can be obtained by fitting the constructed distribution to the experimentally 
measured distribution. 

The number of hits in the PC1 detector shown in the left panel of Fig.~\ref{fig:nbd_fit} 
\befw
\includegraphics[width=0.48\linewidth]{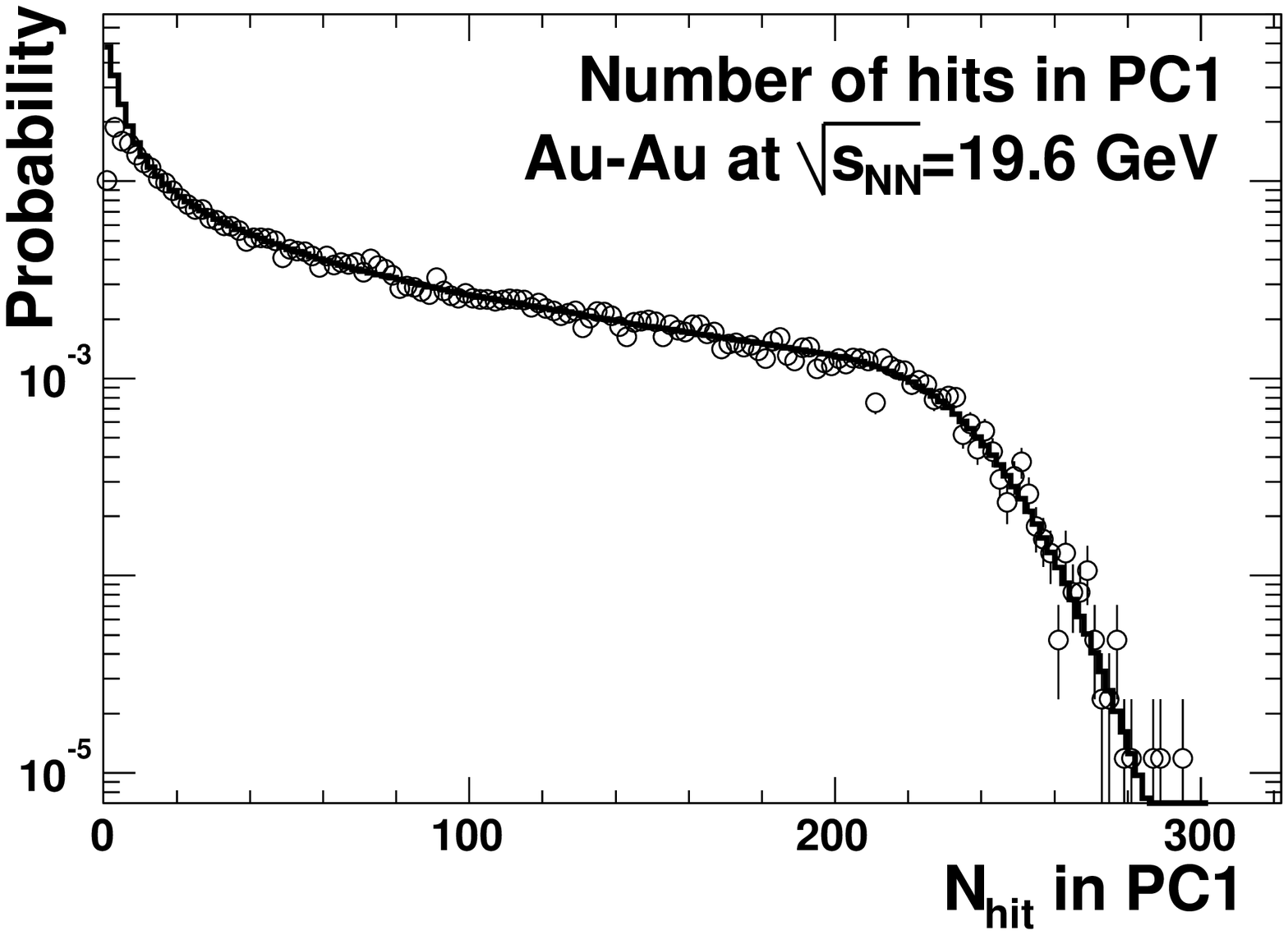}
\includegraphics[width=0.48\linewidth]{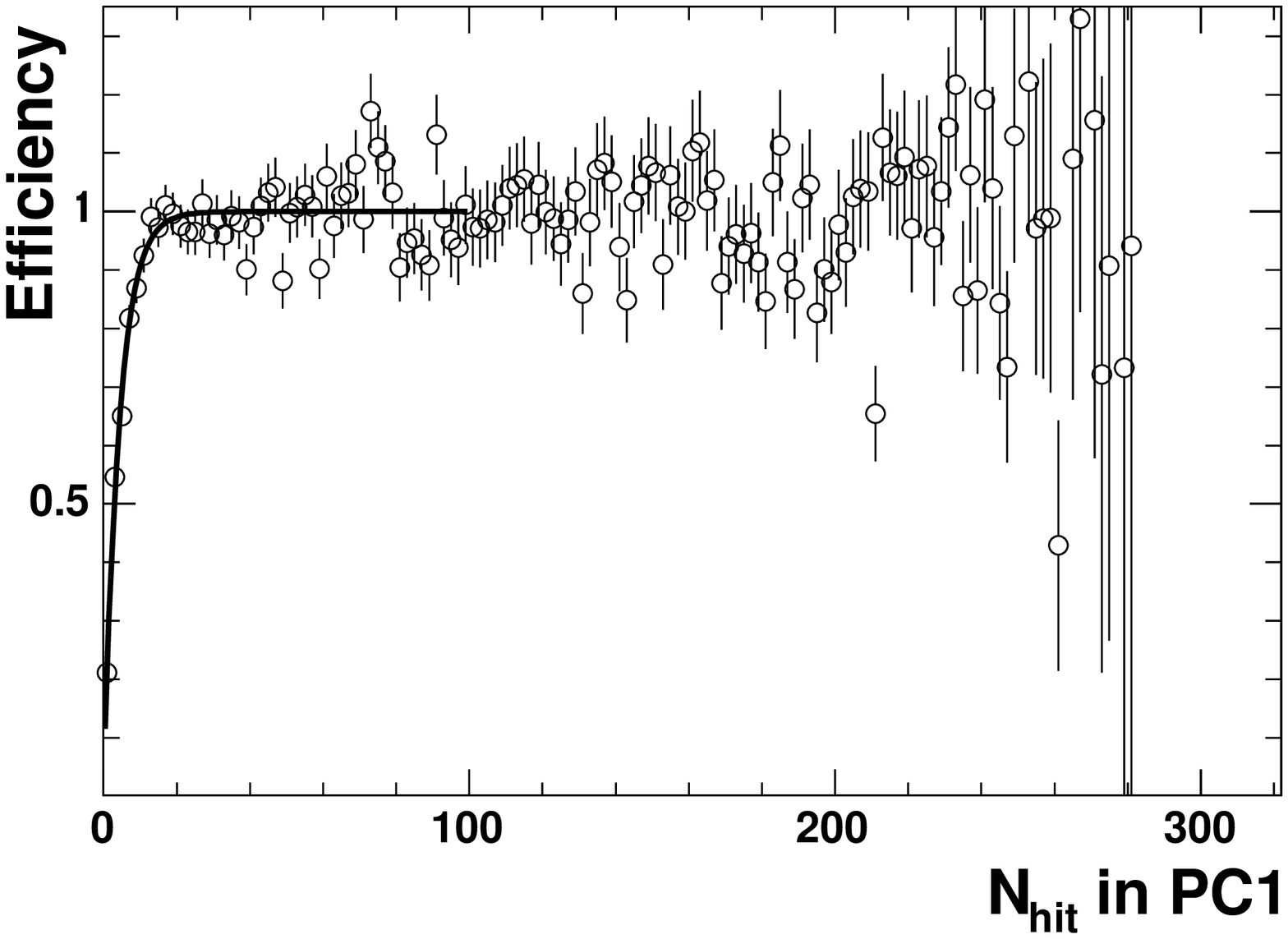}
\caption{Left panel: Glauber/NBD fit (line) to the distribution of the number 
of hits in the PC1 detector at \sqn=\1 (circles). Right panel: MB trigger efficiency 
as a function of the number of hits. The parameterization is to guide the 
eye.\label{fig:nbd_fit}}
\eefw
was used to determine the trigger efficiency. $N_{hit}\propto dN_{ch}/d\eta$ 
can be parameterized as scaling with the number of participants $N_{p}^{\alpha}$, 
where $\alpha$ is between 1.0 and 1.1 as measured by WA98 at the CERN SPS~\cite{wa98}. 
The Glauber/NBD fit to the distribution of the number of hits in PC1 is shown as the 
solid line. The fitting range is constrained above some number of hits, where the trigger 
efficiency is equal to 1. The efficiency as a function of the number of hits 
in the detector can be found by taking the ratio of measured and reconstructed 
distributions. This is shown in the right panel of Fig.~\ref{fig:nbd_fit}.
Intergated over all $N_{hits}$ the MB trigger efficiency was found to be $81.5\pm3$\% at 
\sqn=19.6~GeV. The 1\% uncertainty due to variation of $\alpha$ from 1.0 to 1.1 was included in the systematic error. 
An uncertainty on the difference between \sqn=\1 and \2 was of 1.5\%.

A fraction of events missing in the trigger at all energies belongs to the 
peripheral centrality classes outside the centrality range discussed in this paper.

As a cross check, the same procedure was applied to the BBC response 
at \2. It was found that the MB trigger efficiency in $Au+Au$ and $d+Au$ collisions 
agrees with the procedure based on a full simulation within one standard deviation 
of the systematic error. In $Au+Au$ the \Np in the centrality bins determined using 
Glauber/NBD method agree better than 0.5\% to  
the values used in this paper. In $d+Au$ for a single nucleon-nucleon collision 
the MB trigger efficiency was found to be 57\% consistent 
with $52\pm7$\% measured for PHENIX $p+p$ trigger efficiency at the same 
energy using different method~\cite{phenix_pp}. Finally, the fraction of 
expected $p+Au$ collisions in the $d+Au$ sample agrees with the 
fraction of the events in which the corresponding ZDC detects the spectator 
neutron from the deuteron within better than 1.5\%.

As it was stated above, the BBC detector at \sqn=\1 covers a part of the $Au$ nuclei 
fragmentation region and its response is not linear with \Nps~\cite{phobos6}. Also, the number 
of hits in BBC has a strong vertex dependence mainly due to the fact that the BBC 
samples different parts of the \dNch distribution at different vertices; 
see Fig.~\ref{fig:bbc_north}. The asymmetry of north and south BBC amplitudes in the same 
event was studied to correct for these two effects. Around vertex $z=0$ the asymmetry  
between number of hits in north BBC $N(z)$ and south BBC $S(z)$ is 
$(N(z)-S(z))/(S(z)+N(z)) \propto (d^{2}N_{ch}/d\eta^{2})/(dN_{ch}/d\eta)$ 
reflects the slope of the $\eta$-distribution at BBC rapidity. In order to use the BBC 
signal for \Np determination the observed signals were scaled such that the 
asymmetry between north and south is the same as in the most central events where 
the influence of the fragmentation region is negligible. The vertex dependence 
was also corrected for. The results of the correction are shown in Fig.~\ref{fig:bbc_north}.

\bef
\includegraphics[width=1.0\linewidth]{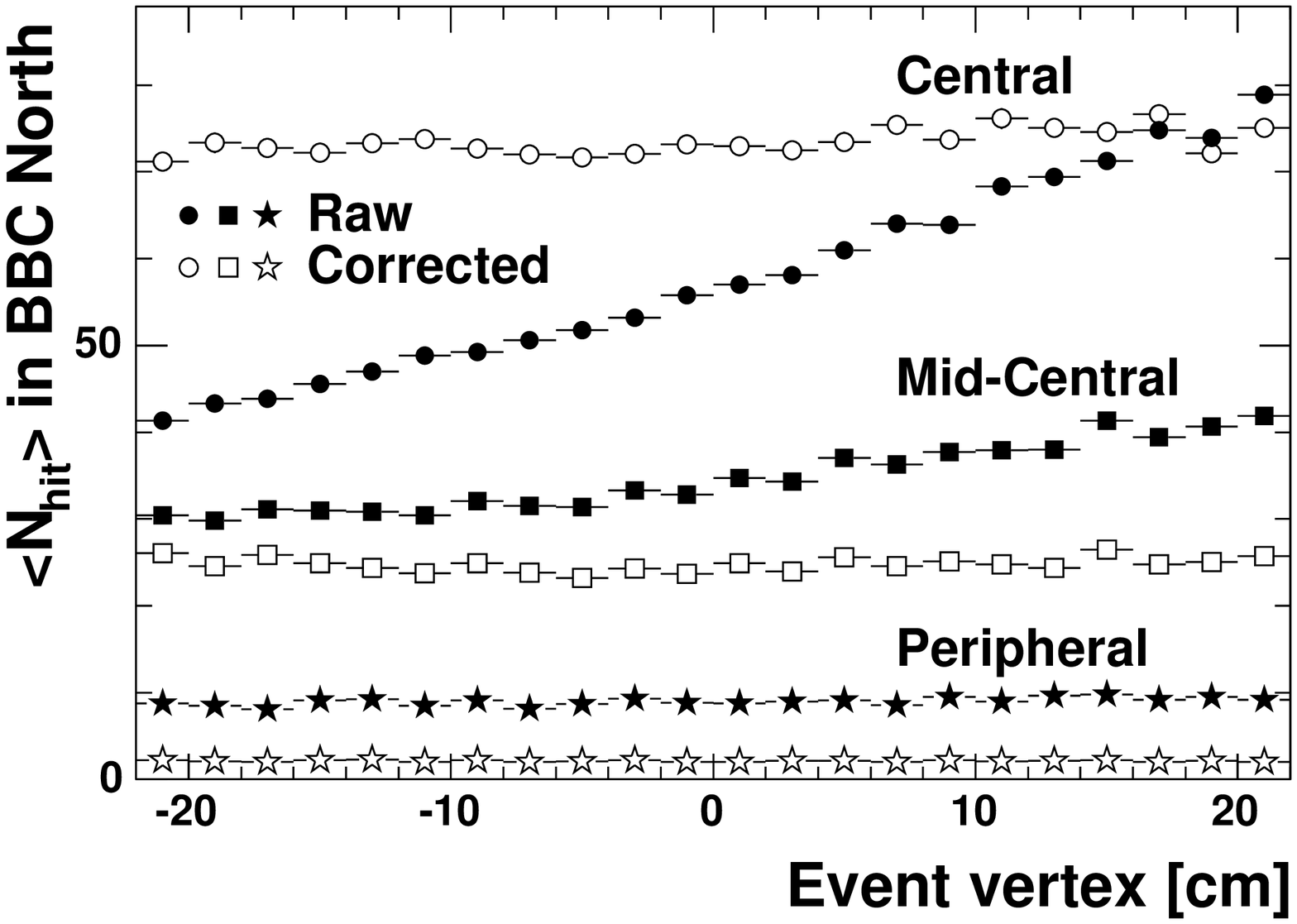}
\caption{\label{fig:bbc_north}
Average number of hits in BBC north vs. event vertex at different 
centralities before the correction (full symbols) and
after correction (open symbols) at \sqns~=~19.6~GeV.}
\eef
The corrected BBC response was used for the centrality determination.
Based on both data and Monte-Carlo simulation, a 
systematic error of 2\% was added to the determination of the 
centrality classes using the BBC correction procedure.

\subsection{Systematic error summary}
Table~\ref{tab:errors} summarizes the systematic errors discussed in this 
section. The ``Energy resp.'' error for 
the $E_T$ measurements combines the uncertainties in absolute energy scale, 
hadronic response and energy losses on the EMCal edges and from energy 
thresholds. The resulting error for each centrality bin is a quadratic sum 
of the errors listed in the Table.

\begin{table}
\caption{Summary of systematic errors given in \%. When the range is given, 
the first number corresponds to the most central bin and the second to the 
most peripheral bin presented in Tables~\ref{tab:results_vs_npart_200}--\ref{tab:results_vs_npart_019}.
\label{tab:errors}}
\begin{ruledtabular}
\begin{tabular}{lcccccc}
\multicolumn{1}{c}{}&\multicolumn{3}{c}{\dEt}  & \multicolumn{3}{c}{\dNch}\\
\sqn [GeV]          &  19.6    & 130    & 200    & 19.6     & 130     & 200     \\
\hline								     	     		
Energy resp.        &  4.7     & 3.8    & 3.9    &          &         &         \\
Bkg. / noise        &  0.5-3.5 & 0.4-10 & 0.2-6  & 1.0      & 1.0     & 1.0     \\
Acceptance          &  2.0     & 2.0    & 2.0    & 2.3      & 2.5     & 2.3     \\
In- \& outflow      &  3.0     & 3.0    & 3.0    & 5.7      & 2.5     & 2.9     \\
Occupancy           &          &        &        & 1.6-0.3  & 3.1-0.1 & 3.5-0.1 \\
\hline										
Centrality          & 2.0      & 0.5    & 0.5    & \multicolumn{3}{c}{same} \\
\Np                 & 2.9-6.7  & 2.8-15.& 2.8-15.& \multicolumn{3}{c}{same} \\
Trigger             & 0.4-8.8  & 0.3-16.& 0.3-16.& \multicolumn{3}{c}{same} \\
\end{tabular}
\end{ruledtabular}
\end{table}

\section{Results \label{sec:result}}
\subsection{PHENIX results}
The distribution of the raw transverse energy, $E_{T_{EMC}}$, into the fiducial 
aperture of two EMCal sectors is shown in the left three panels of 
Fig.~\ref{fig:results_raw} for three RHIC energies \sqn = 19.6, 130 and \2. 
The lower scale represents the fully corrected \Et normalized to one unit of 
pseudorapidity and full azimuthal acceptance. The lower axis in the plot is not labeled beyond \2 
to avoid confusion between the true shape of the \dEt distribution 
and \Et as measured using the limited acceptance of two EMCal sectors.

\befw
\includegraphics[width=0.48\linewidth]{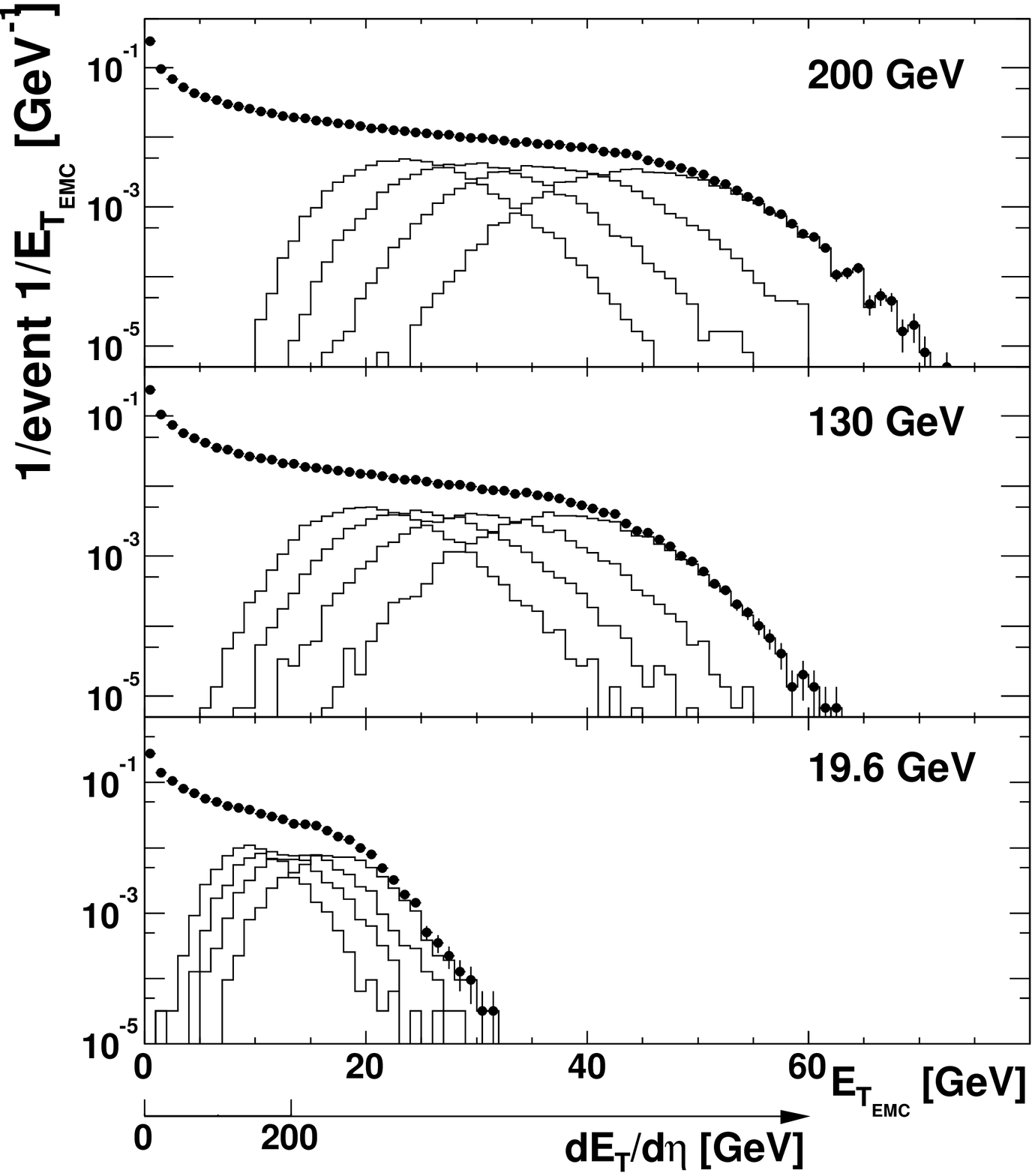}
\includegraphics[width=0.48\linewidth]{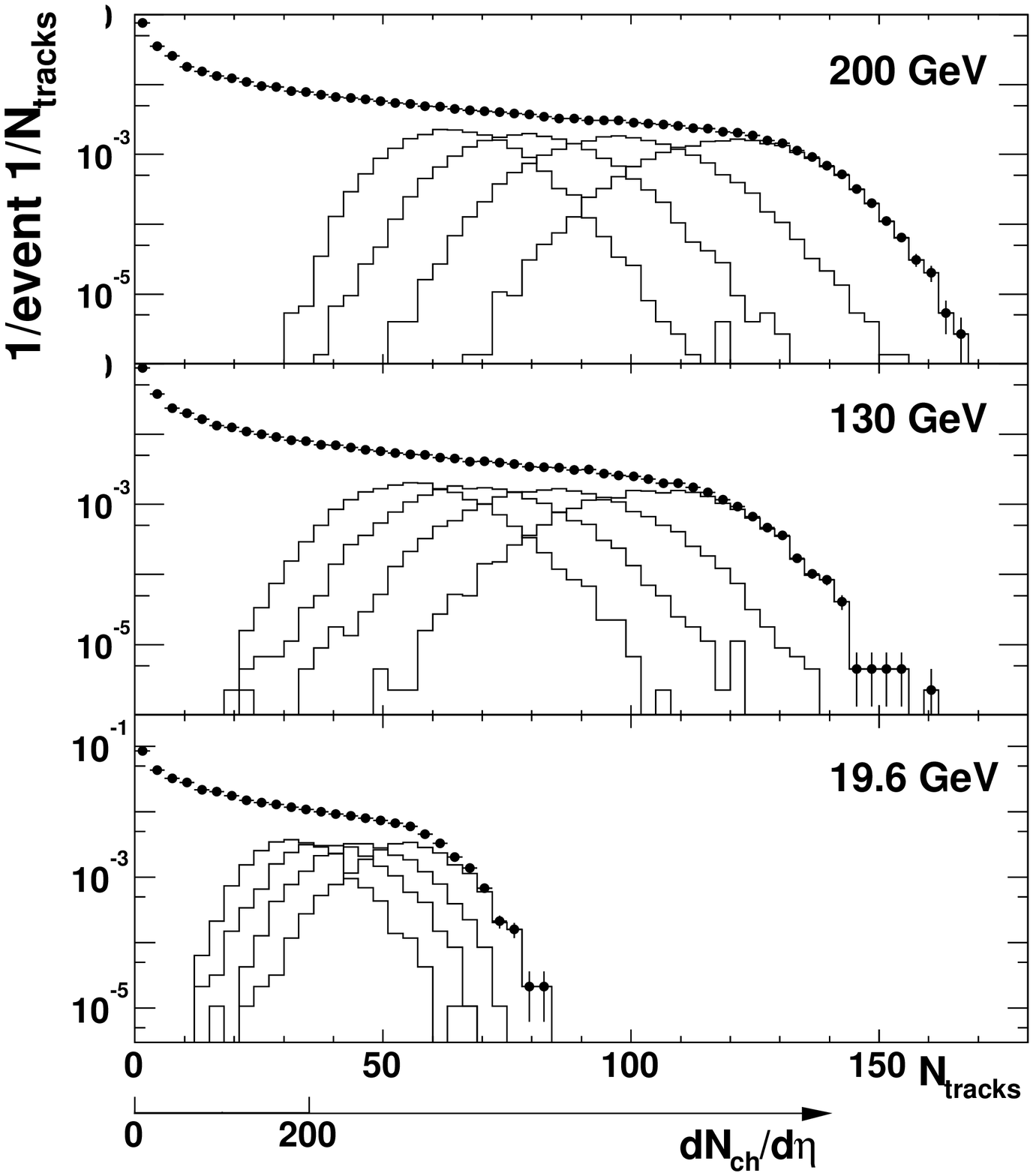}
\caption{The distribution of the raw \Et in two EMCal sectors (left) and the number of 
tracks in the east arm of the PHENIX detector (right) per MB trigger, measured 
at three energies. The lower axis corresponds to mid-rapidity values of 
\dEt and \dNch respectively. Distributions of the four 5\% most central bins 
are also shown in each plot.\label{fig:results_raw}}
\eefw

For the measurements at \sqn = \1 and \2, five EMCal sectors 
(with azimuthal coverage $\Delta \phi = 112^0$) were used, while 
only two sectors ($\Delta \phi = 45^0$) were available during the PHENIX 
run at \sqn = \3. Results obtained with different number of sectors at the same 
energy were consistent within 1.5\%.

The right three panels in Fig.~\ref{fig:results_raw} show the number of tracks 
reconstructed in the east arm of the PHENIX detector after background 
subtraction and all corrections. The lower axis corresponds to measured 
distributions normalized to one unit of pseudorapidity and full azimuthal 
acceptance. For a similar reason as for the \Et measurement, the lower axis is not labeled above \2 
in $dN_{ch}/d\eta$.

For the \Nch measurements at \sqn = \3, only the east arm was used, while for the other 
two energies the measurements were made using both PHENIX central arms. The 
results obtained with two arms at \sqn=\2 and \1 are consistent with each 
other within 1.5\%.

The distributions shown in Fig.~\ref{fig:results_raw} have a characteristic shape 
with a sharp peak that corresponds to the most peripheral events. Missing events caused by the finite 
MB trigger efficiency in peripheral events 
would make this peak even sharper than measured. 
The plateau in all distributions corresponds to mid-central 
events and the fall-off to the most central $Au+Au$ events. The shape of the 
curves in Fig.~\ref{fig:results_raw} in the fall-off region is a product of the 
intrinsic fluctuations of the measured quantities and the limited acceptance of the detector.

The distributions for the four most central bins 0\%-5\% to 15\%-20\% are also shown 
in each panel. The centroids of these distributions were used to calculate the 
centrality dependence of \dEt and $dN_{ch}/d\eta$~\footnote{All plotted and 
quoted numbers correspond to average values in each centrality bin or ratios 
of those averages.}. The statistical uncertainty of all mean values (less than or 
about 1\%) determined by the width of the distributions are small because of 
the large size of the event samples.

The magnitude of \dEt and \dNch at mid-rapidity 
divided by the number of participant pairs as a function of \Np is shown in Fig.~\ref{fig:results_vs_npart} 
and tabulated in Tables~\ref{tab:results_vs_npart_200}--\ref{tab:results_vs_npart_019}. 
The right three panels show the same ratio for \dNch at three RHIC energies.
\befw
\includegraphics[width=0.48\linewidth]{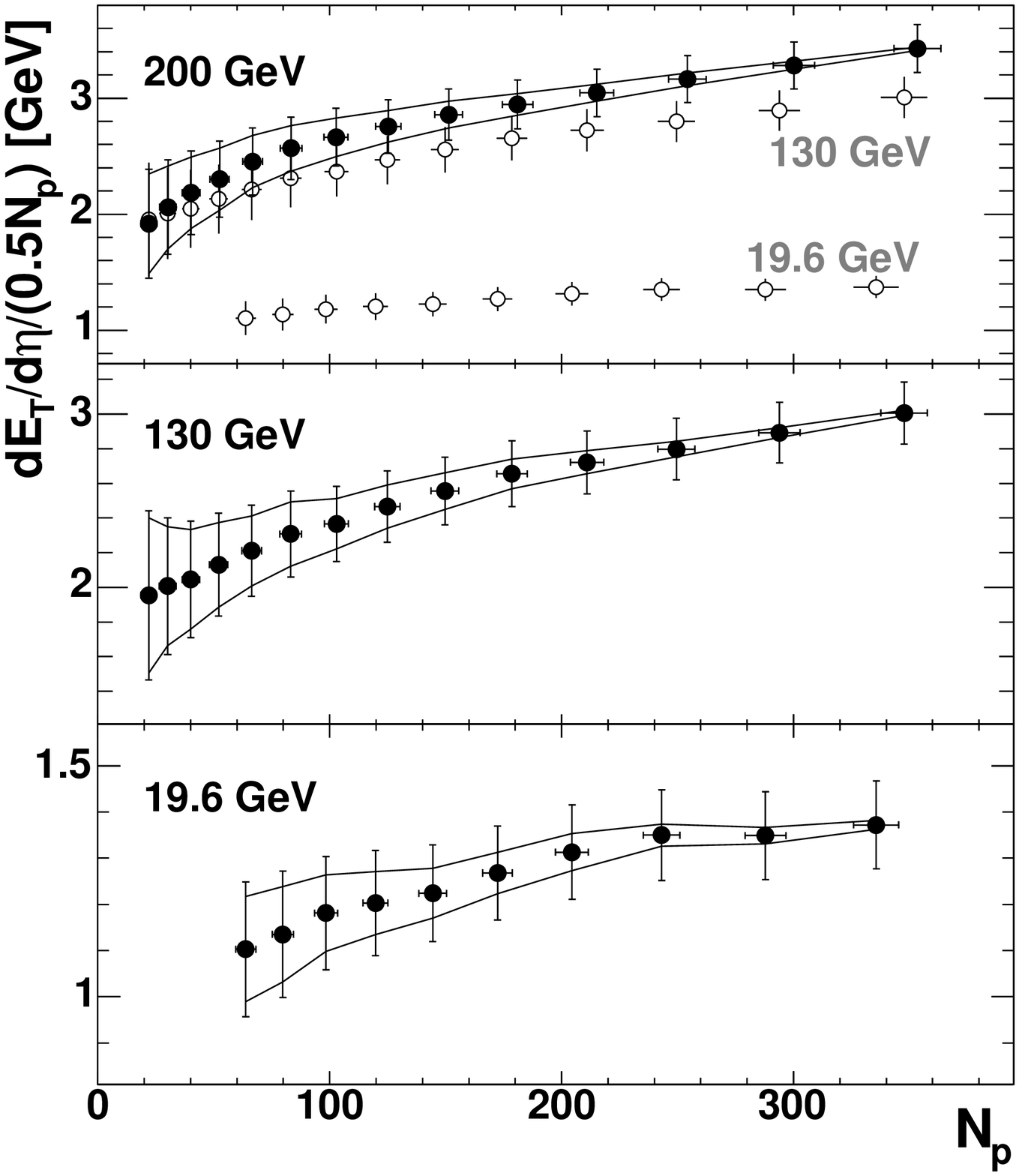}
\includegraphics[width=0.48\linewidth]{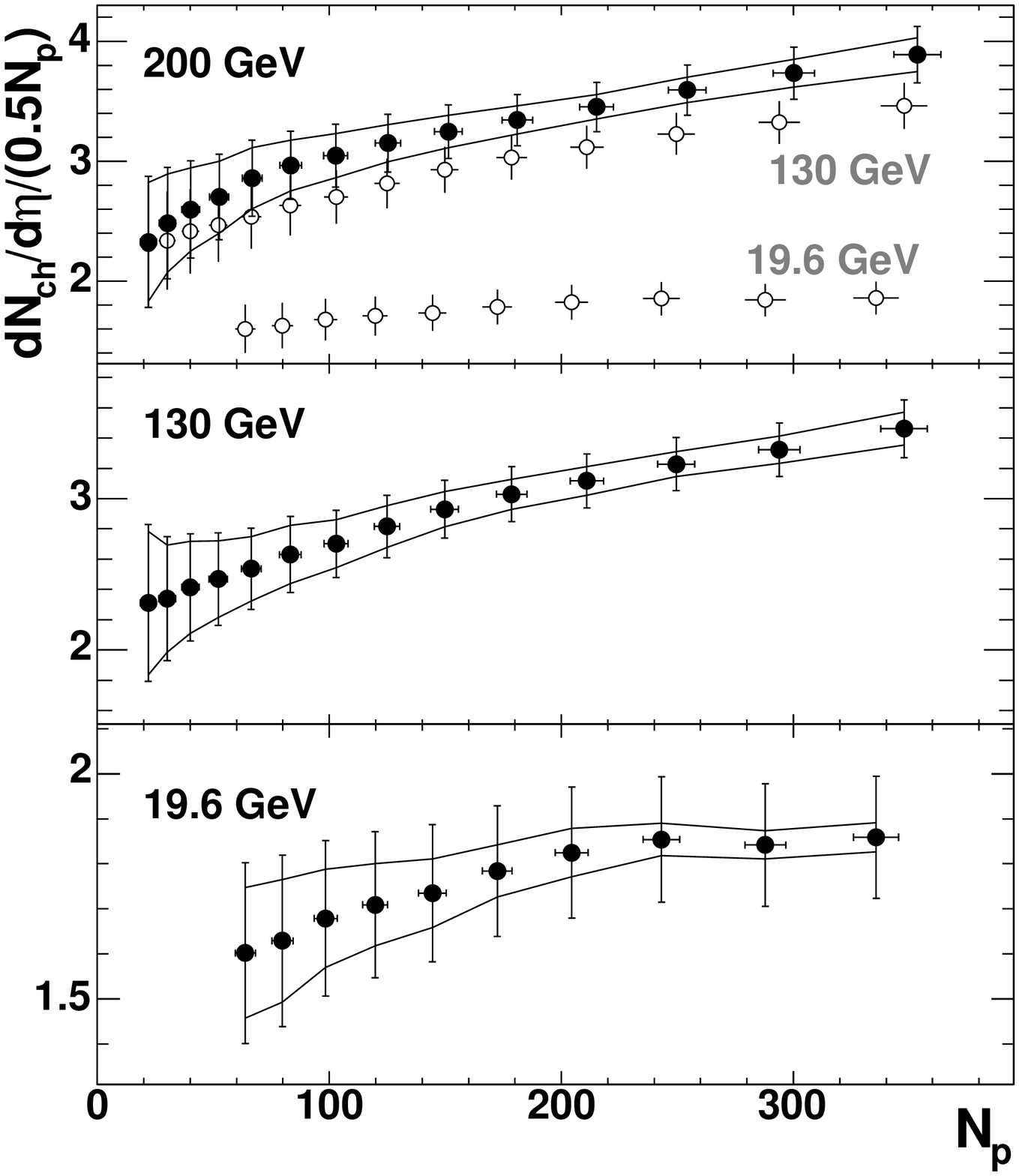}
\caption{\dEt (left) and \dNch (right) divided by the number of participant pairs at 
three RHIC energies. Errors shown with vertical bars are full systematic 
errors. Lines show the part of the systematic error that allows bending or 
inclination of the points. Horizontal errors denote the uncertainty in 
determination of \Nps.\label{fig:results_vs_npart}}
\eefw

The horizontal errors correspond to the uncertainty in \Nps, determined within 
the framework of the Monte Carlo Glauber model. 
The vertical bars show the full systematic errors of the 
measurements\footnote{Here and everywhere errors correspond to one standard 
deviation.} added quadratically to the errors of \Nps. 
The lines denote the corridor in which the points can be inclined or bent. 
The statistical errors are smaller than the size of the markers. The upper panel 
also shows the results of the two lower panels with open markers for 
comparison. 

An important result from Fig.~\ref{fig:results_vs_npart} is an evident 
consistency in the behavior of the centrality curves of \Et shown on the left 
and \Nch shown on the right for all measured energies. Both values 
demonstrate an increase from peripheral (65\%-70\% bin) to the most central events by 50\%-70\%
at RHIC energies \sqns=\3 and \2. For the lowest RHIC energy (\sqns=\1) this 
increase is at the level of systematic uncertainties of the measurement. 
One can note that results from PHOBOS~\cite{phobos_total},
show that the total charged particle multiplicity is proportional to \Np
while the multiplicity at mid-rapidity over \Np increases with \Nps,
indicating that the pseudorapidity distribution gets more narrow for
central events.

The ratios of the \dEt and \dNch per participant pair measured at different 
RHIC energies are shown in Fig.~\ref{fig:ratios_vs_npart} and tabulated in 
Table~\ref{tab:ratios_vs_npart}. In these ratios some common systematic errors cancel.
\befw
\includegraphics[width=0.48\linewidth]{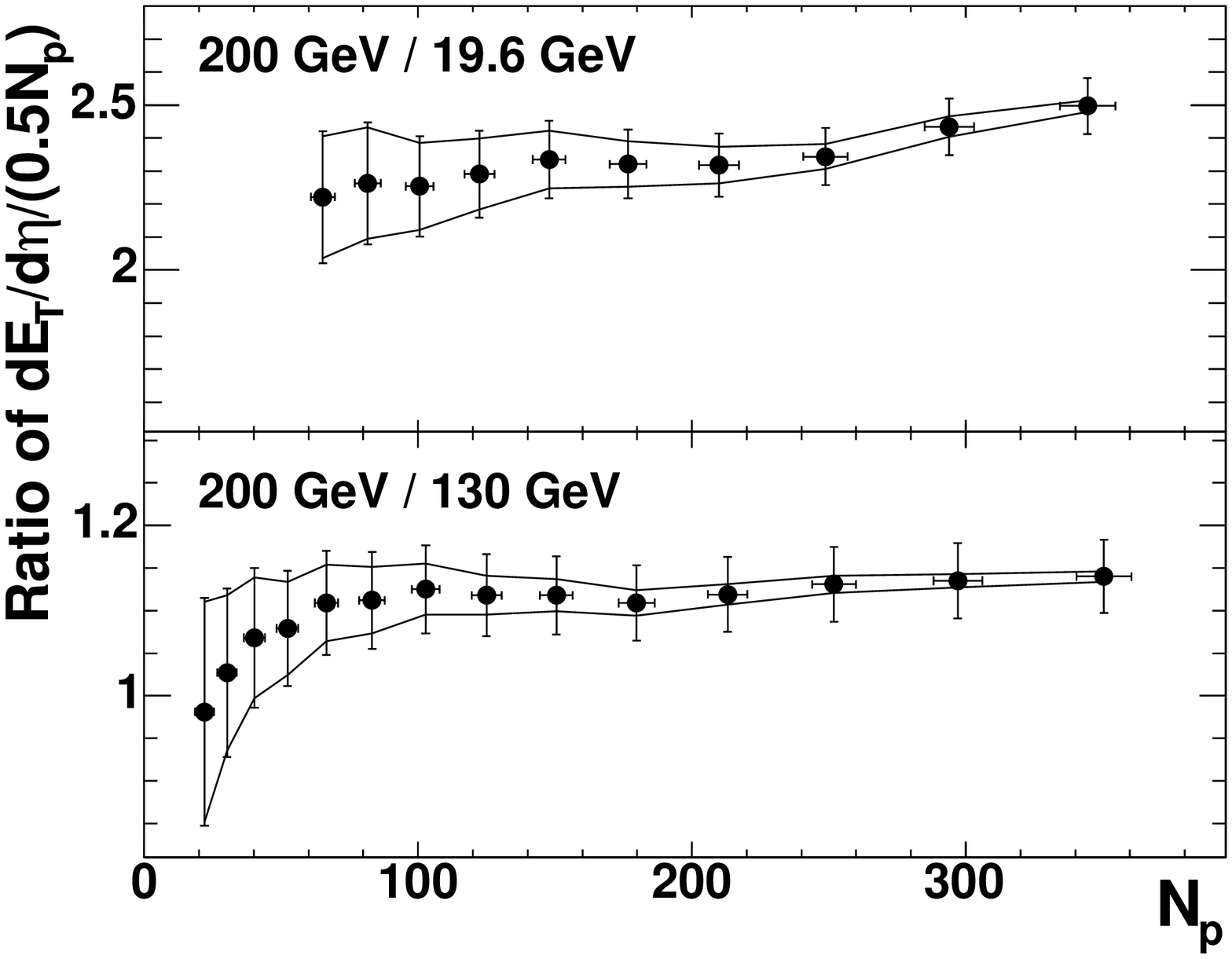}
\includegraphics[width=0.48\linewidth]{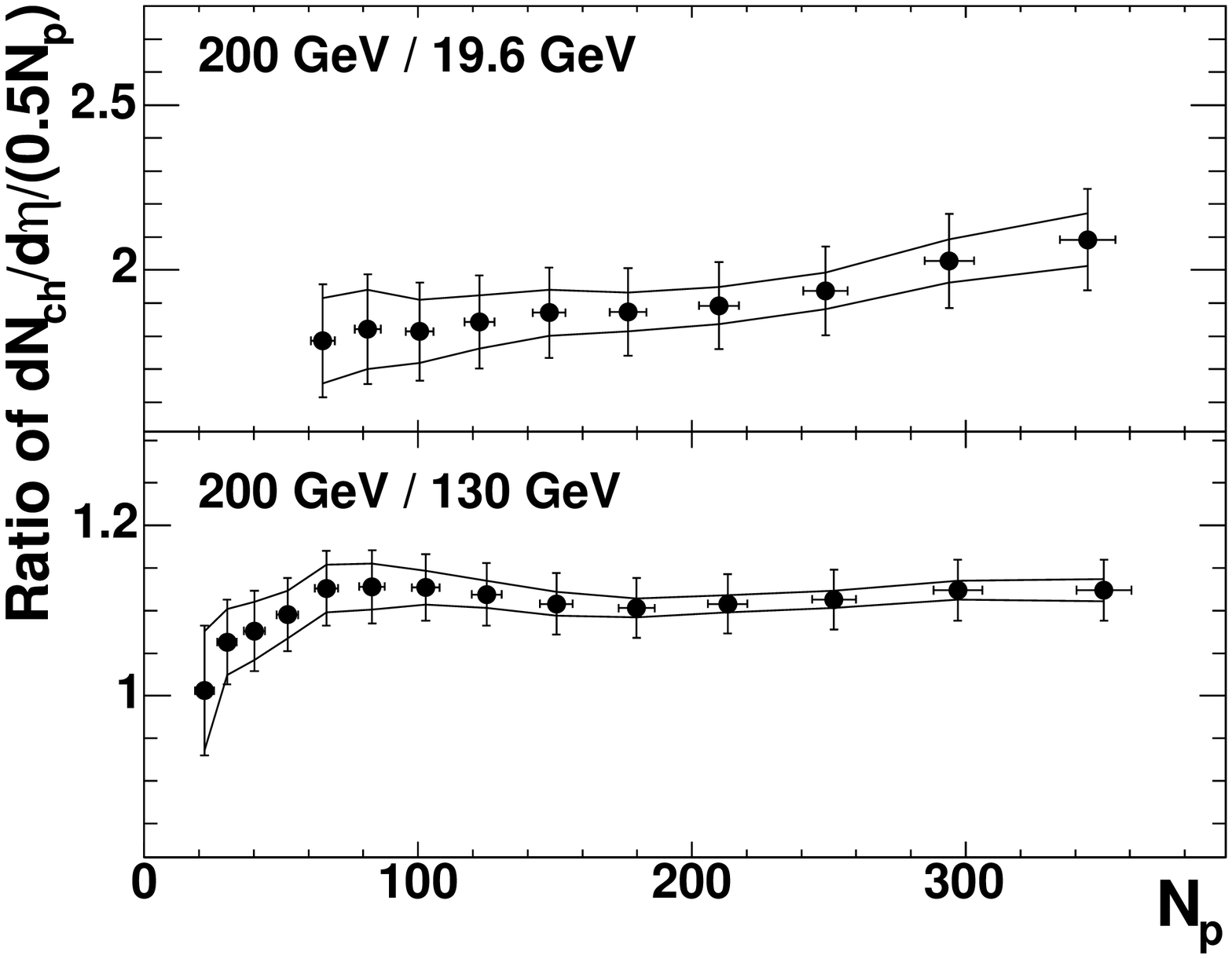}
\caption{Ratios of \dEt (left) and \dNch (right) measured at different RHIC 
energies. The errors shown with vertical bars 
are the full systematic errors. Lines show the part of the systematic error 
that
allows bending or inclination of the points. The horizontal errors denote 
the uncertainty in the determination of \Nps.\label{fig:ratios_vs_npart}}
\eefw

The increase in the \Et production between \1 and \2 (with an average 
factor of 2.3) is larger than for \Nch (with average factor of 1.9). This is consistent
 with an increase in the particle production per participant common to both 
\Et and \Nch and a $\sim$20\% increase in \mt of produced particles contributing 
to the \Et parameter only. See section~\ref{sec:general} and \cite{na49_1,tatsuja}.

The ratio of \2/\1 shows some increase from peripheral to central events; 
however the increase is marginally at the level of the systematic errors of the 
measurement.

The ratio of \2/\3 is flat above \Np $\sim$ 80 and is equal to $1.140\pm0.043$ for 
\Et and $1.126\pm0.036$ for \Nch in the most central bin. A rather sharp 
increase between \Np=22 and 83 in the ratios of both quantities 
is still at the level of systematic uncertainties.

The ratio of the transverse energy and charged particle multiplicity at mid-rapidity as 
a function of centrality is shown in Fig.~\ref{fig:etra_mult_vs_npart} for 
the three energies. The upper plot also shows the results 
displayed in the lower panels for comparison.

\bef
\includegraphics[width=1.0\linewidth]{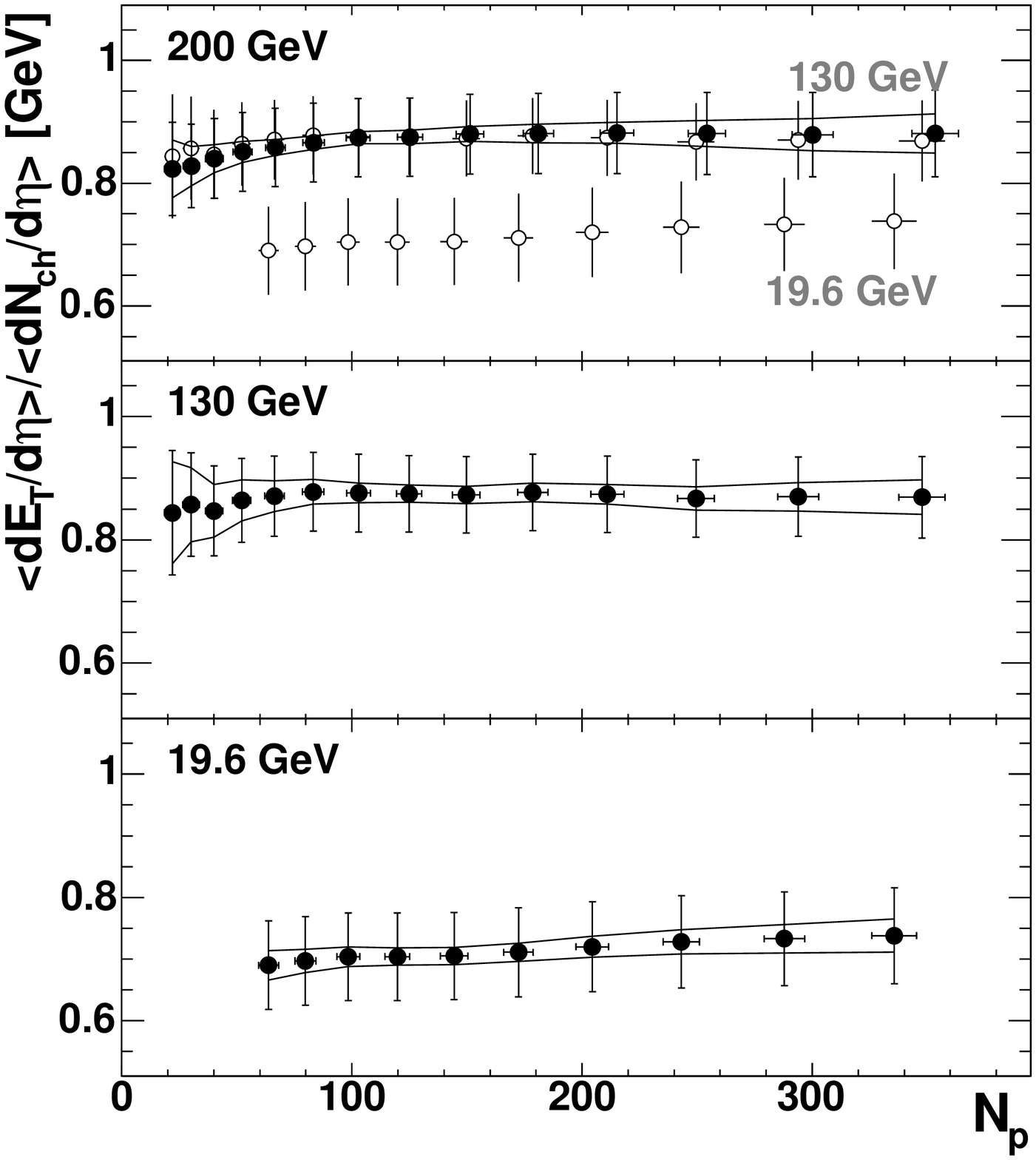}
\caption{\EN vs. \Np at different RHIC energies. The errors shown 
with vertical bars are the full systematic errors. Lines show the part of 
the systematic error that allows bending or inclination of the points. The 
horizontal errors denote the uncertainty in the determination 
of \Nps.\label{fig:etra_mult_vs_npart}}
\eef

The ratio $E_{T}/N_{ch}$\footnote{\EN is used as a shortcut for 
$\langle dE_{T}/d\eta \rangle / \langle dN_{ch}/d\eta \rangle$ at $\eta=0$ in C.M.S..}, sometimes 
called the ``Global Barometric Observable'', triggered considerable discussion~\cite{gulash,raju}.
It is related to the \mt of the produced 
particles and is observed to be almost independent of centrality and 
incident energy of the collisions within the systematic errors of the previous 
measurements. The present paper forges a direct link between 
the highest SPS and lowest RHIC energies, making a more quantitative study of 
\EN possible.

The results presented in Fig.~\ref{fig:etra_mult_vs_npart} and tabulated in 
Tables~\ref{tab:results_vs_npart_200}--\ref{tab:results_vs_npart_019} 
show that the centrality dependence of \EN is weak and lies within the systematic errors 
plotted with lines. There is a clear increase in \EN between 
\sqns=\1 and \2. The \sqn dependence of the results is discussed below.

\subsection{Bjorken Energy Density}
The Bjorken energy density \cite{bjorken} can be calculated using 
\begin{equation} 
\epsilon_{Bj} = \frac{1}{A_{\perp} \tau} \frac{dE_T}{dy}, 
\label{eq:Bj}
\end{equation} 
where $\tau$ is the formation time and $A_{\perp}$ is the nuclei 
transverse overlap area. 

The transverse overlap area of two colliding nuclei was estimated 
using a Monte Carlo Glauber model $A_{\perp} \sim \sigma_x \sigma_y$, 
where $\sigma_x$ and $\sigma_y$ are the widths of $x$ and $y$ position 
distributions of the participating nucleons in the transverse plane. 
The normalization to $\pi R^2$, 
where $R$ is the sum of $r_n$ and $d$ parameters in a Woods-Saxon 
parameterization (Eq.~\ref{eq:WSdensity}), was done for the most central 
collisions at the impact parameter $b=0$.

For the transformation from $dE_T/d\eta|_{\eta=0}$ to $dE_T/dy|_{y=0}$, 
a scale factor of $1.25 \pm 0.05$ was used, see~\ref{sec:general}.

The Bjorken energy density for three RHIC energies is plotted in the left panel of Fig.~\ref{fig:ebj} 
and tabulated in Tables~\ref{tab:results_vs_npart_200}--\ref{tab:results_vs_npart_019}. 
For the 5\% most central collisions, $\epsilon_{Bj} \cdot \tau$ was 
$2.2 \pm 0.2$, $4.7 \pm 0.5$ and $5.4 \pm 0.6$ GeV/($fm^2 \cdot c$) for 
\sqn=19.6, 130 and 200 GeV, respectively. These values increase by 
2\%, 4\% and 5\%, respectively, for the maximal $N_{part}$=394, as obtained 
from extrapolation of PHENIX data points. There is a factor 
of 2.6 increase between the ``SPS''-like energy (\sqn=\1) and the top RHIC energy \sqn=\2.
The comparison of the only published 
$\epsilon_{Bj}$=3.2 GeV/$fm^3$ at SPS for head-on collisions~\cite{na49_4} 
and top RHIC energies, assuming the same $\tau$=1~fm/c, reveals an increase in 
energy density by a factor of only 1.8, which may come from an overestimation 
in the SPS measurement, as shown in the left panel of Fig.~\ref{fig:sqn} and 
discussed in section~\ref{sec:recalc_na49_cent}. 

Another approach is used by STAR in~\cite{star_bj} for the 
estimate of the transverse overlap area of the two nuclei $A_{\perp} \sim N_{p}^{2/3}$ 
in Eq.~\ref{eq:Bj}. This approach accounts only for the common area of colliding 
nucleons, not nuclei. The results are different only in the peripheral bins 
as shown in the right panel of Fig.~\ref{fig:ebj}. For a comparison, the same 
panel shows the result obtained by STAR which agrees with PHENIX result 
within systematic errors, though displaying a smaller increase of the energy 
density with $N_p$.

\befw
\includegraphics[width=0.48\linewidth]{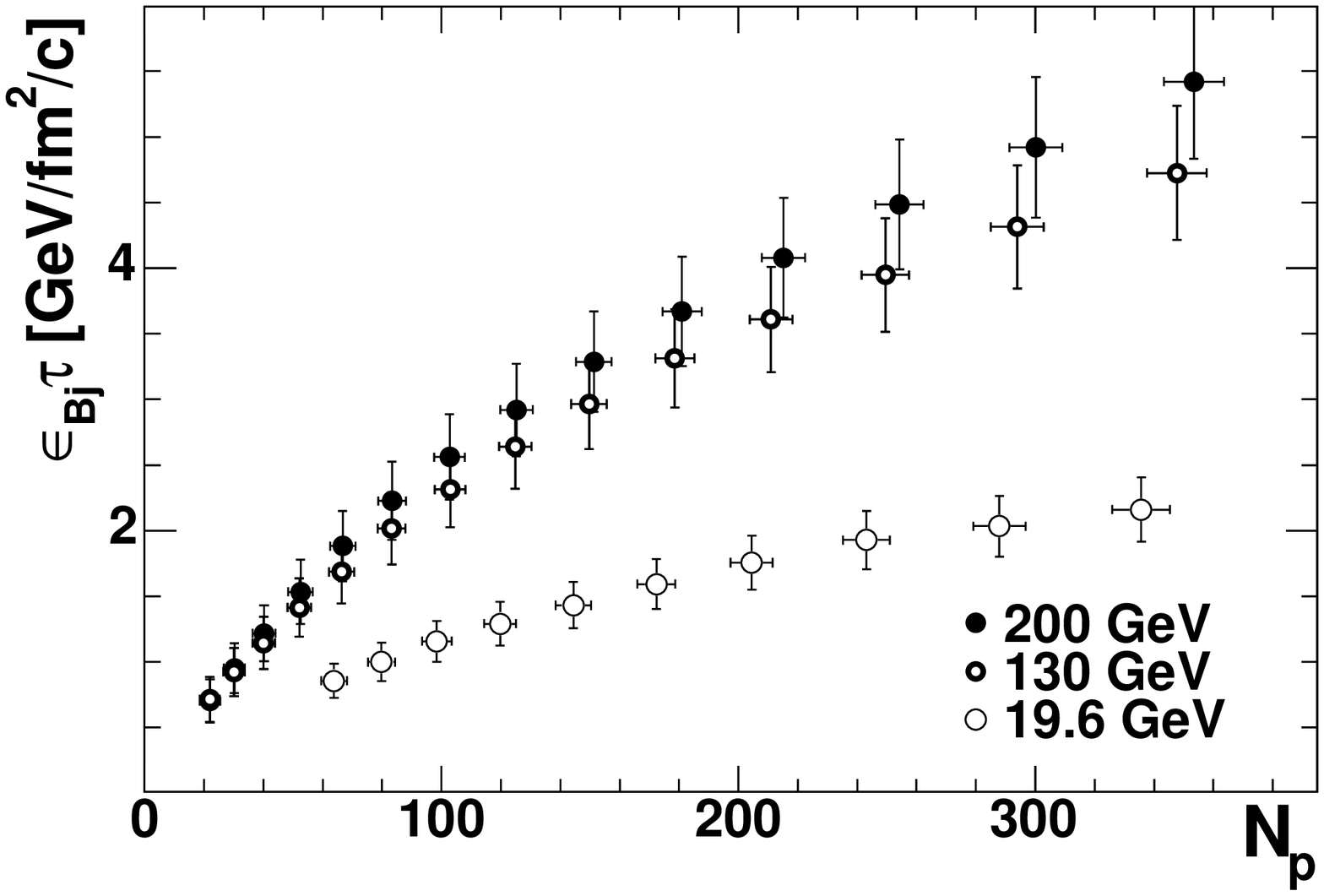}
\includegraphics[width=0.48\linewidth]{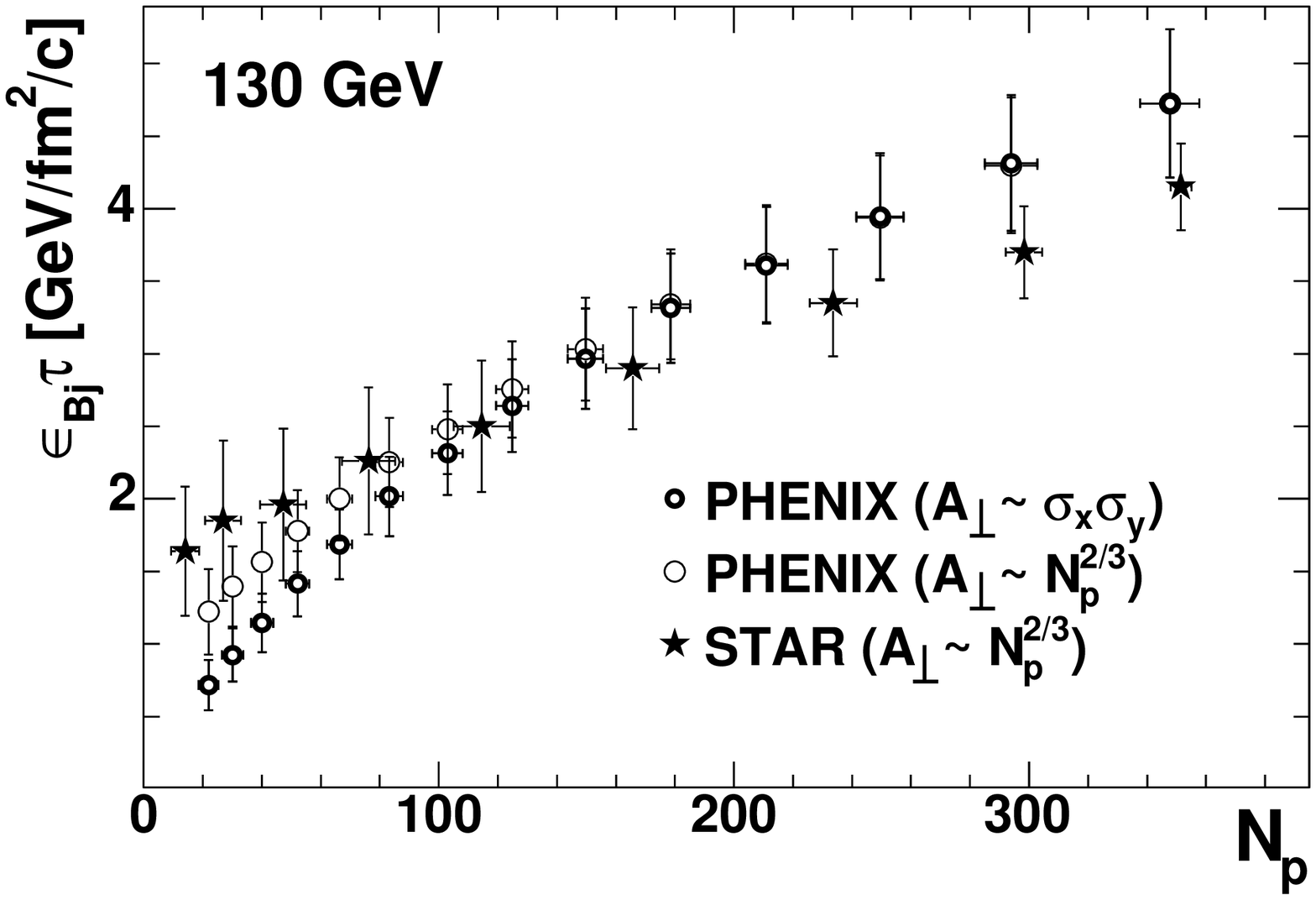}
\caption{$\epsilon_{Bj} \cdot \tau$ deduced from the PHENIX data at three RHIC energies (left) 
and using different estimates of the nuclear transverse overlap at \sqn=\3 (right). 
\label{fig:ebj}}
\eefw

\subsection{Comparison to other measurements}
Comparison to the results of other experiments is complicated by several 
factors. AGS and SPS data were taken in the Laboratory (Lab.) system 
while the RHIC data are in the Center of Mass (C.M.S.) system. Since
 $\eta$ and \Et are not boost invariant quantities, the data should be converted into 
the same coordinate system. Some experiments provide a complete 
set of identified particle spectra from which information about \Et and 
\Nch can be deduced. For other experiments, additional assumptions
are necessary for their published values.
Appendix~A describes how such recalculation was done 
in each particular case.

The PHENIX results for \Nch are compared to the data available from the other 
RHIC experiments. This comparison is shown in the left panels of 
Fig.~\ref{fig:comp_rhic}.
\befw
\includegraphics[width=0.48\linewidth]{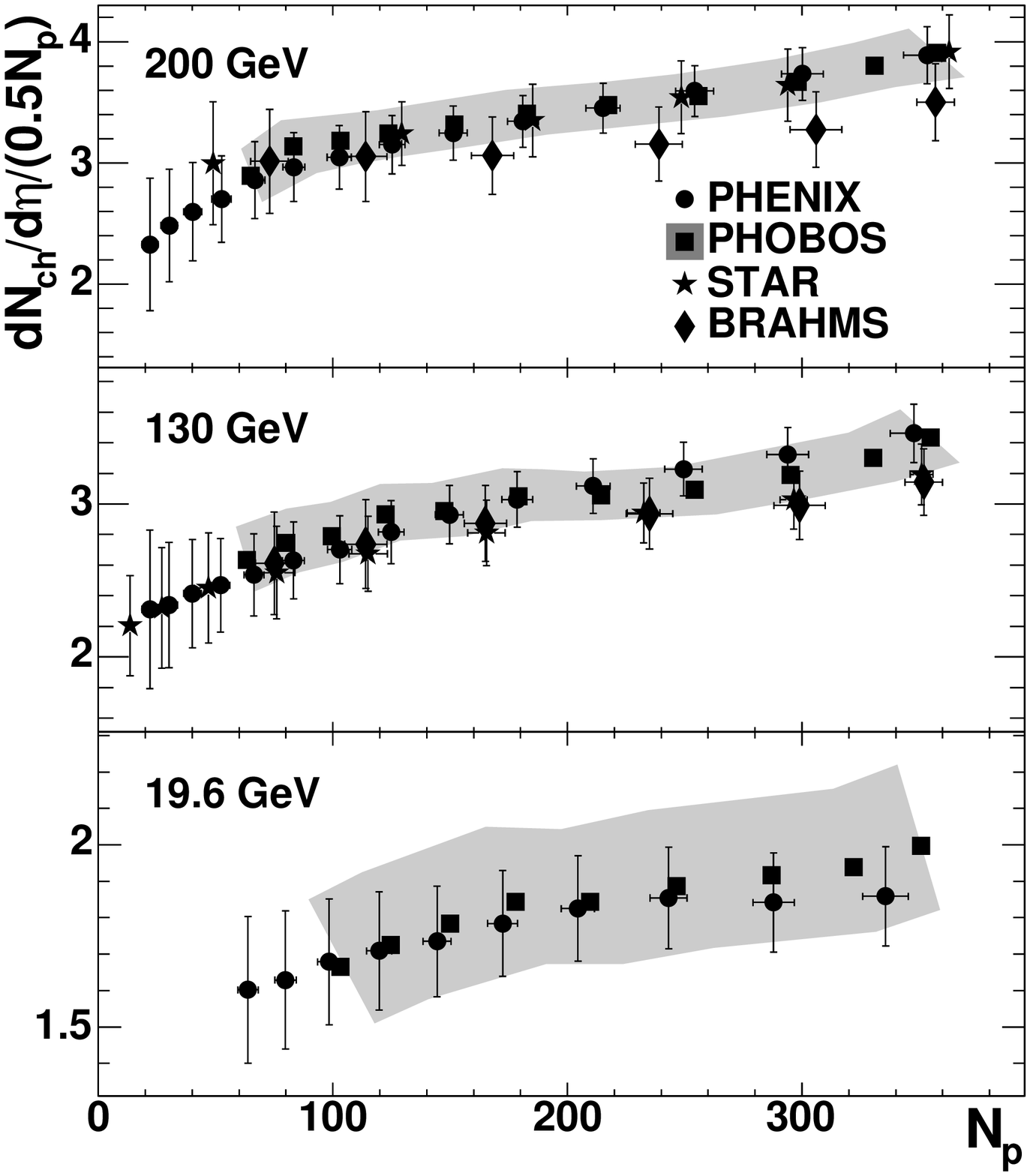}
\includegraphics[width=0.48\linewidth]{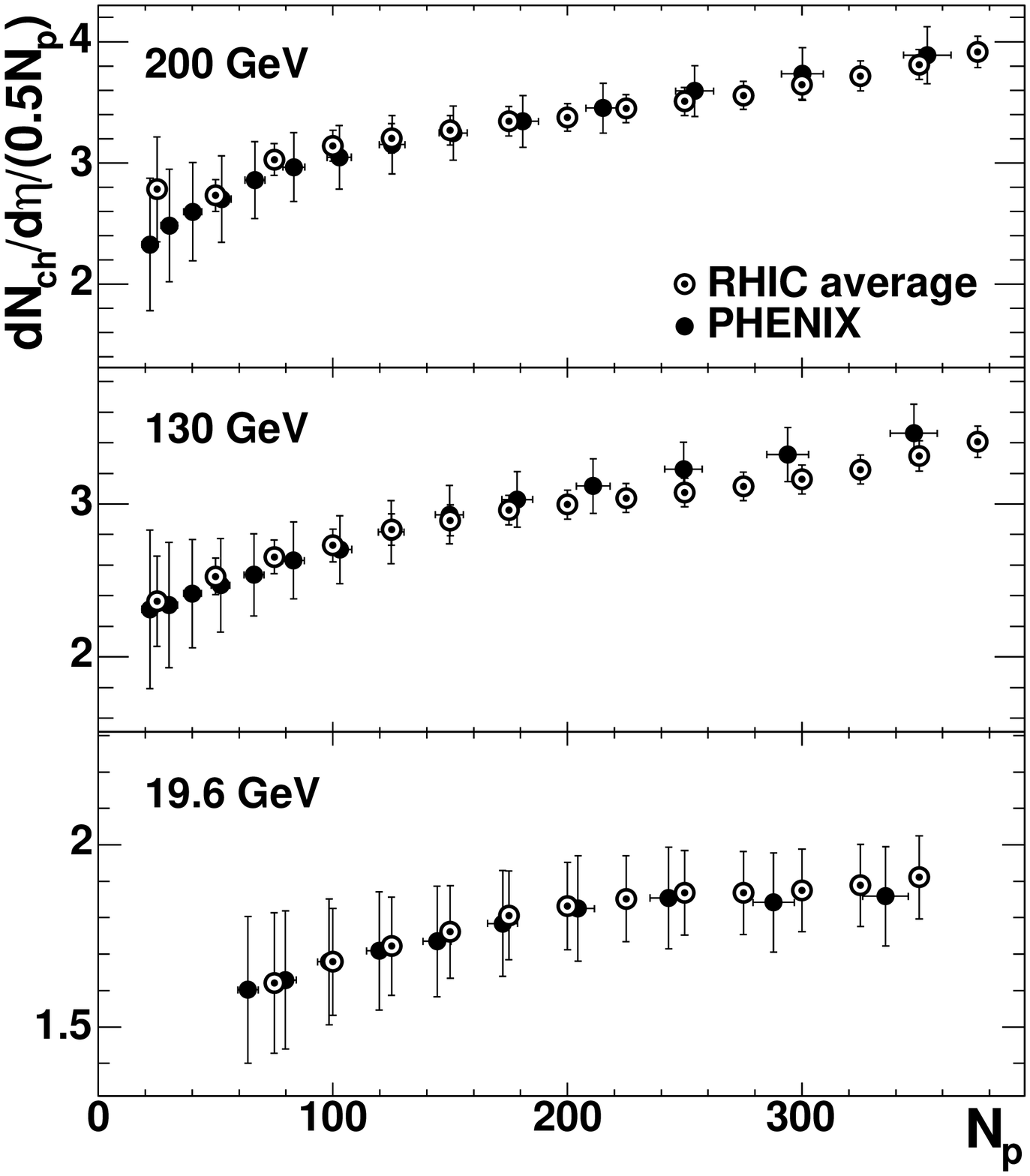}
\caption{Left panel: \dNch per pair of \Np measured by the four RHIC experiments at 
different energies. The shaded area is the PHOBOS systematic error. Right panel: 
RHIC average values (including PHENIX) compared to the PHENIX results.\label{fig:comp_rhic}}
\eefw

There is good agreement between the results of BRAHMS~\cite{brahms1,brahms2}, 
PHENIX, PHOBOS~\cite{phobos1,phobos2,phobos3} and STAR~\cite{star1,star2} using 
\Np based on a Monte-Carlo Glauber model. This agreement is very impressive because 
all four experiments use different apparatuses and techniques to measure the 
charged particle production. The systematic errors of all results are uncorrelated, 
except for those related to the same Glauber model which are small.
That makes it possible to calculate the RHIC average and reduce 
the systematic uncertainty. The averaged results from all four RHIC experiments 
are plotted in the right panel of Fig.~\ref{fig:comp_rhic} and tabulated in 
Table~\ref{tab:averages}. See section~\ref{sec:recalc_average} for the procedure.

Figure~\ref{fig:etra_star} compares $E_T$ results from the PHENIX and STAR~\cite{star_etra}
experiments. The results are consistent for all centralities within systematic errors, though 
STAR \dEt per participant pair has a smaller slope vs. $N_p$ above $\sim$70 participants, and 
\EN shown in the lower panel is consistent for all \Nps.
\bef
\includegraphics[width=1.0\linewidth]{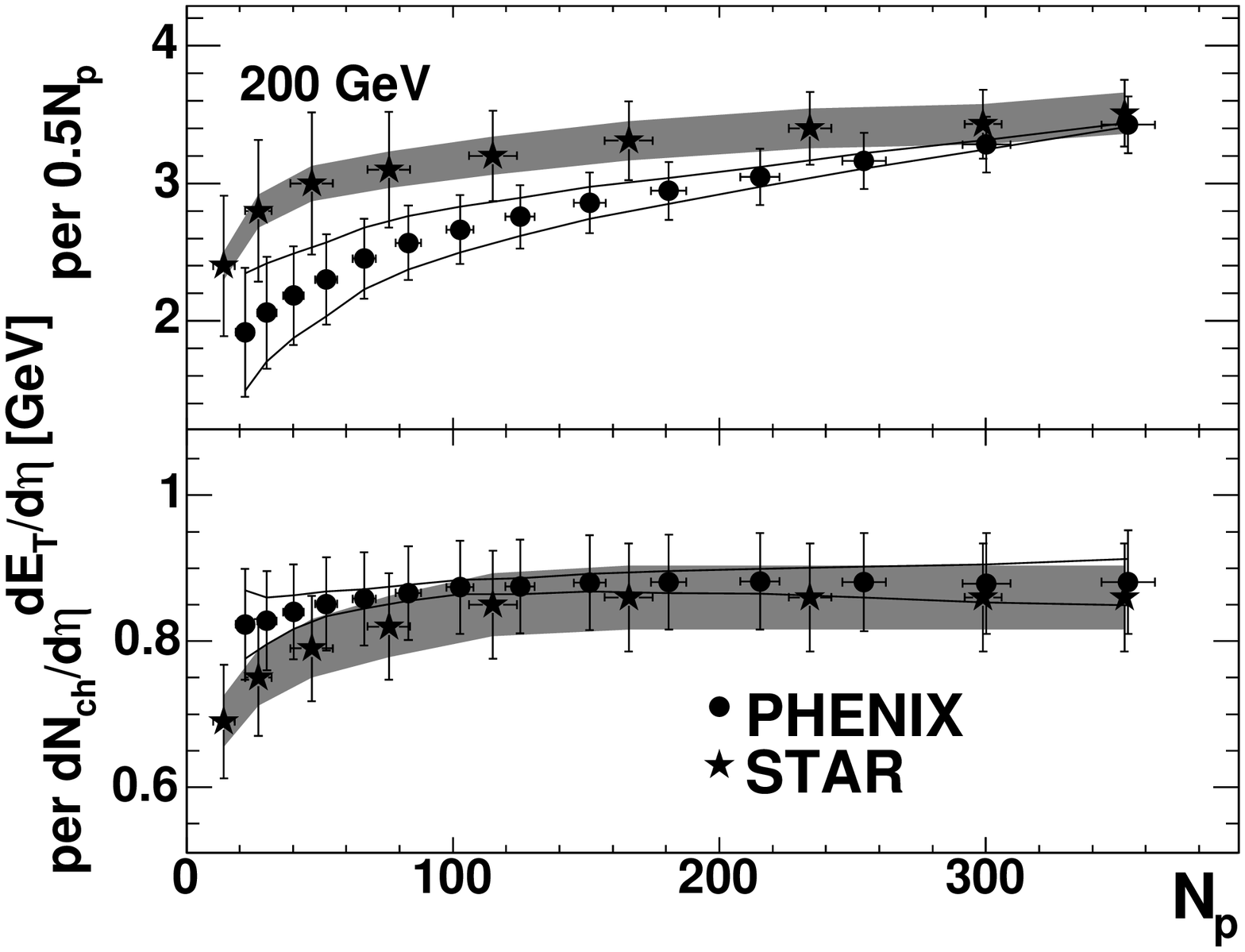}
\caption{\dEt divided by the number of \Np pairs (top) and  \EN (bottom) measured 
by the PHENIX and STAR~\cite{star_etra} experiments at \sqn=200 GeV. PHENIX systematic 
errors are explained in the text. The shaded area is the STAR systematic scaling error.\label{fig:etra_star}}
\eef

The RHIC run at \sqns=\1 allows a connection between 
RHIC and SPS data to be made. The highest SPS energy of 158A~GeV corresponds to 
\sqns=\7 in the C.M.S., making a direct comparison of RHIC and SPS results 
possible. This comparison is shown in Fig.~\ref{fig:comp_sps}. 
See sections~\ref{sec:recalc_na49_cent}~through~\ref{sec:recalc_na50_cent} 
for the details of the data compilation.
\befw
\includegraphics[width=0.48\linewidth]{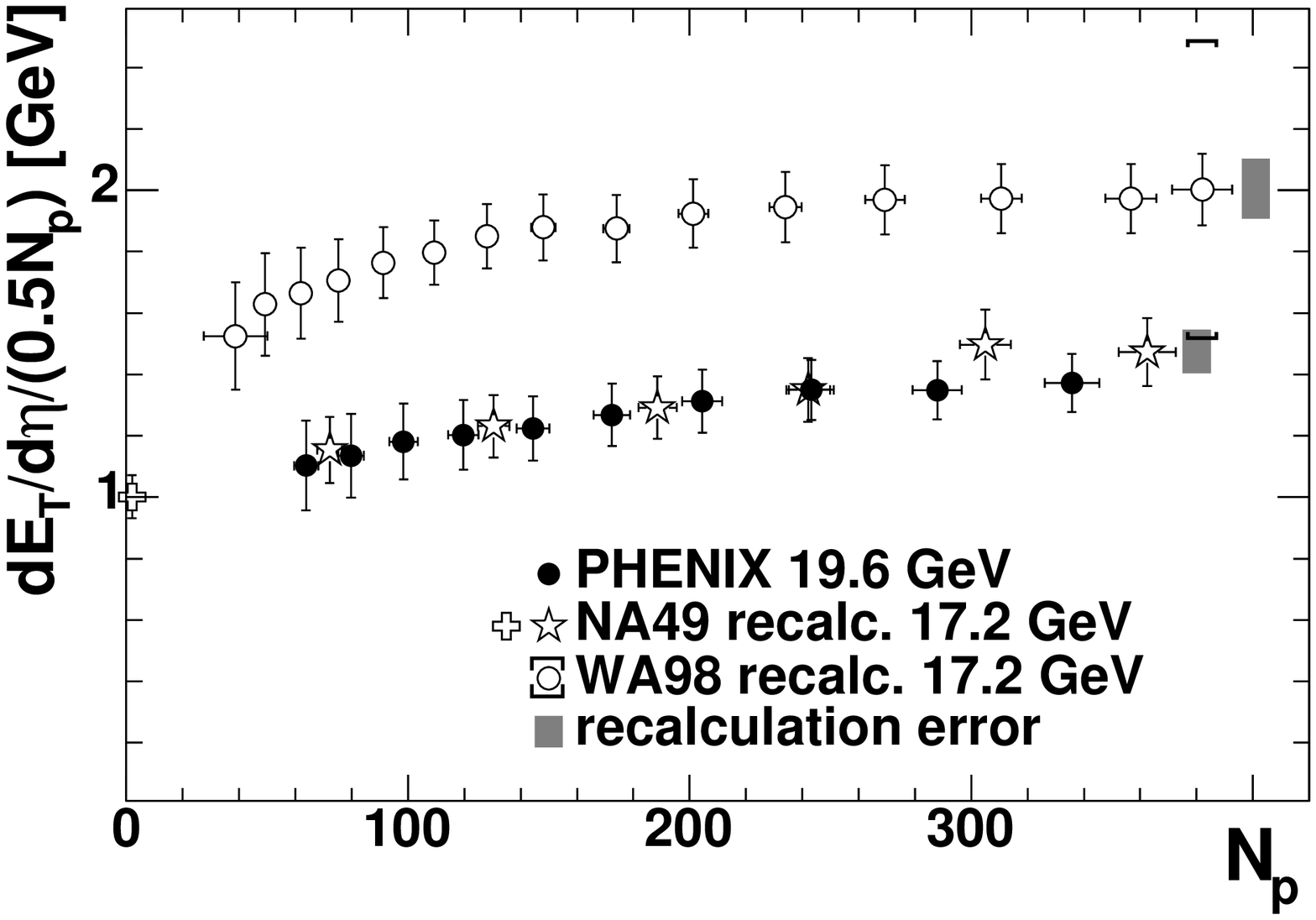}
\includegraphics[width=0.48\linewidth]{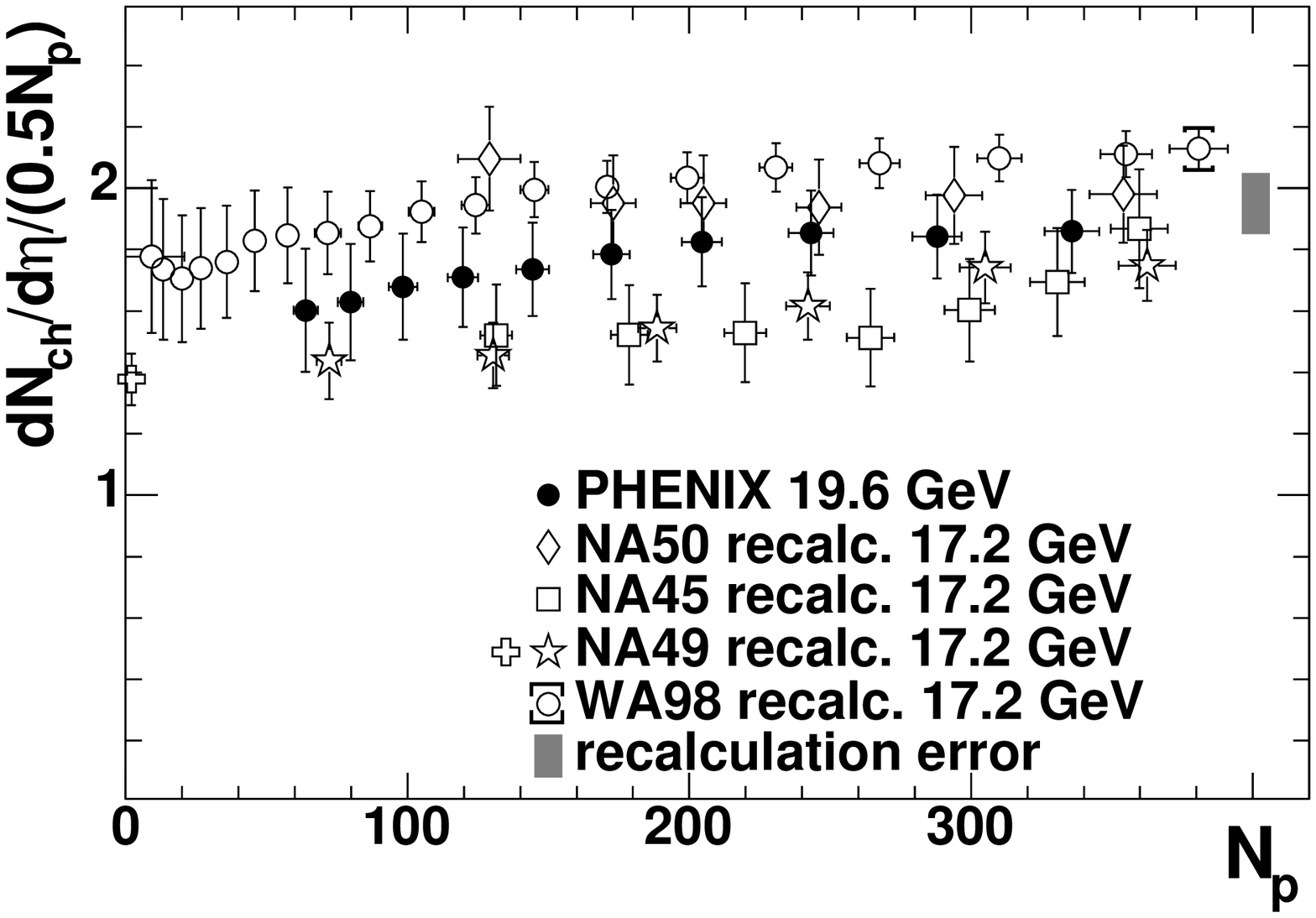}
\caption{\dEt (left) and \dNch (right) divided by the number of \Np pairs 
measured by PHENIX at \sqns=\1 (solid markers) and recalculated from 
the results of the SPS experiments at the highest energy \sqns=\7 (open markers). 
The $p+p$ result of NA49 is marked with an open cross.
\label{fig:comp_sps}}
\eefw

Several comments should be made about this comparison. For both measured 
parameters the PHENIX results and the SPS results agree. The WA98 results 
(see section~\ref{sec:recalc_wa98_cent}) are systematically higher than the results
of other experiments, especially for $dE_{T}/d\eta$. However the WA98 data has 
an additional systematic error common to all points shown for the last bin. 
For \Nch the relative spread of the SPS results is 
larger than for the RHIC results shown in Fig.~\ref{fig:comp_rhic}, 
though overall the \sqns=\7 SPS measurements are consistent with the PHENIX 
result at \sqns=19.6 GeV.

Different SPS and AGS experiments made measurements at lower energies. The 
combined data of AGS, SPS and RHIC provide a complete picture of the 
centrality behavior of \Et and \Nch as a function of the nucleon-nucleon 
energy. The centrality dependence of \dNch at mid-rapidity measured at 
\sqns=4.8, 8.7 and \7 by different experiments is shown in Fig.~\ref{fig:sps_ags}. 
See Table~\ref{tab:averages} for the summary of these results and 
sections~\ref{sec:recalc_na45_cent}~through~\ref{sec:recalc_e802_cent} 
for the details of the data compilation.
\bef
\includegraphics[width=1.0\linewidth]{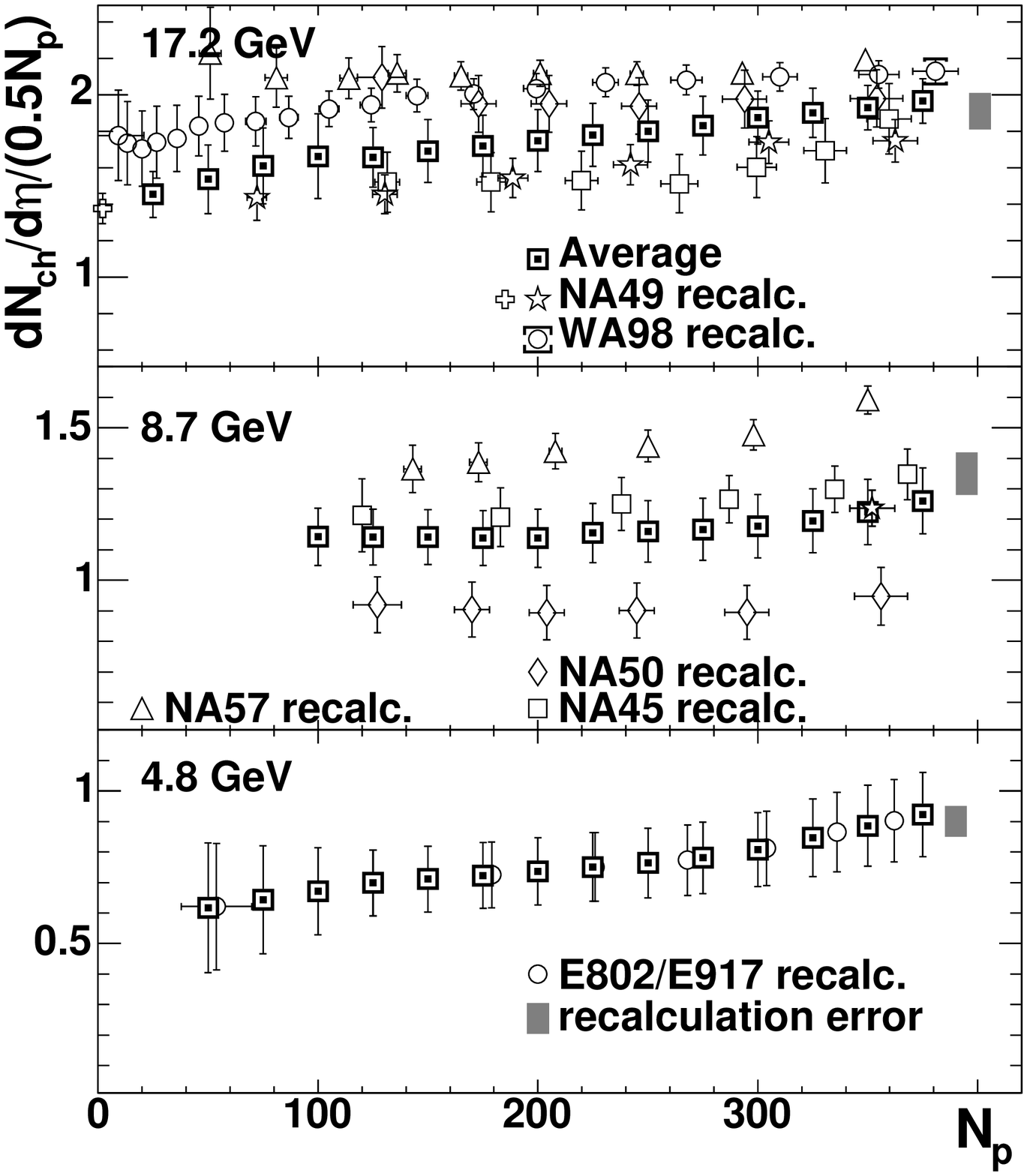}
\caption{\dNch divided by the number of \Np pairs measured by AGS and SPS 
experiments and the average taken at different energies recalculated in the C.M.S.
\label{fig:sps_ags}}
\eef

At the highest SPS energy the averaging procedure is the same as for RHIC 
energies and weighted experimental errors are scaled with the reduced 
$\chi^{2}$-like factor $S$ (described in section~\ref{sec:recalc_average}) 
reaching the value of 1.5 at some points. For the intermediate SPS 
energy \sqns=8.7~GeV, two experiments NA45~\cite{ceres_3} and 
NA50~\cite{na50_1} reported the centrality dependence of \dNch at 
mid-rapidity. The discrepancy in the measurements is close to three times
 the quadratic sum of their systematic error. However the shapes of the two curves are almost 
the same. NA49 has published results (see section~\ref{sec:recalc_na49_cent}) 
which give one point in \dNch at \Np=352. This point favors 
the NA45 result\footnote{The NA57 results at both SPS energies are published 
without systematic errors in~\cite{na57}. They are currently not considered.}. 
The average centrality curve is produced taking into 
account the shape of the centrality curves reported by NA45 and NA50 
and the single NA49 point. See section~\ref{sec:average_eight} for 
the averaging procedure at \sqns=8.7~GeV. The errors are scaled with the 
factor $S$, which reaches a value of 2.5 at some points. The AGS results 
are presented with a curve produced from the combined results of the E802/E917 
experiments (see section~\ref{sec:recalc_e802_cent}). The averaging procedure 
in this case is a simple rebinning of the data.

The average SPS centrality dependence at \sqns=\7 shown in the upper panel 
in Fig.~\ref{fig:sps_ags} and the average curve of the two RHIC 
experiments at \sqns=\1 shown in the lower panel in Fig.~\ref{fig:comp_rhic} 
are very similar. Less than 5\% increase is expected due to the difference in
the incident energy between the highest SPS and the lowest RHIC 
energies (see section~\ref{sec:central_collision} below). 

The average values presented in Figs.~\ref{fig:comp_rhic} 
and \ref{fig:sps_ags} are summarized in Table~\ref{tab:averages}.

\subsection{Dependence on the incident nucleon energy.}
The data compilation made in the previous section allows for
a detailed study of the charged particle production in heavy ion reactions 
at different incident energies of colliding nuclei. Although the data on
transverse energy production is not abundant, a similar comparison can
be made~\cite{phenix_milov,phenix_bazik}.

\subsubsection{Central Collisions}
\label{sec:central_collision}
Figure~\ref{fig:sqn} shows the energy dependence for the \dEt and \dNch 
production per pair of participants in the most central collisions 
measured by different experiments. 
See sections~\ref{sec:recalc_na45_cent}~through~\ref{sec:recalc_fopi} for 
the details of the data compilation.

\befw
\includegraphics[width=0.48\linewidth]{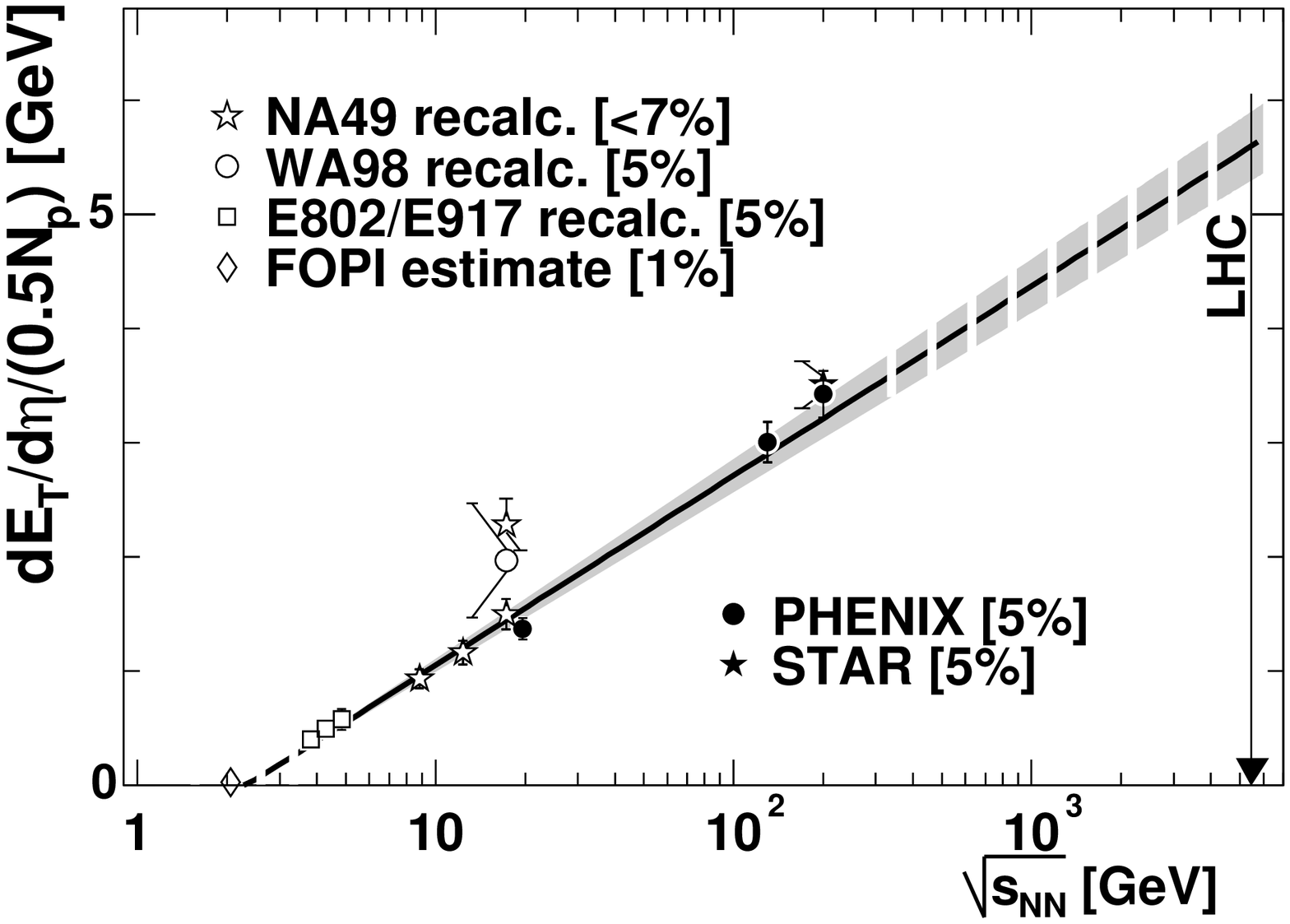}
\includegraphics[width=0.48\linewidth]{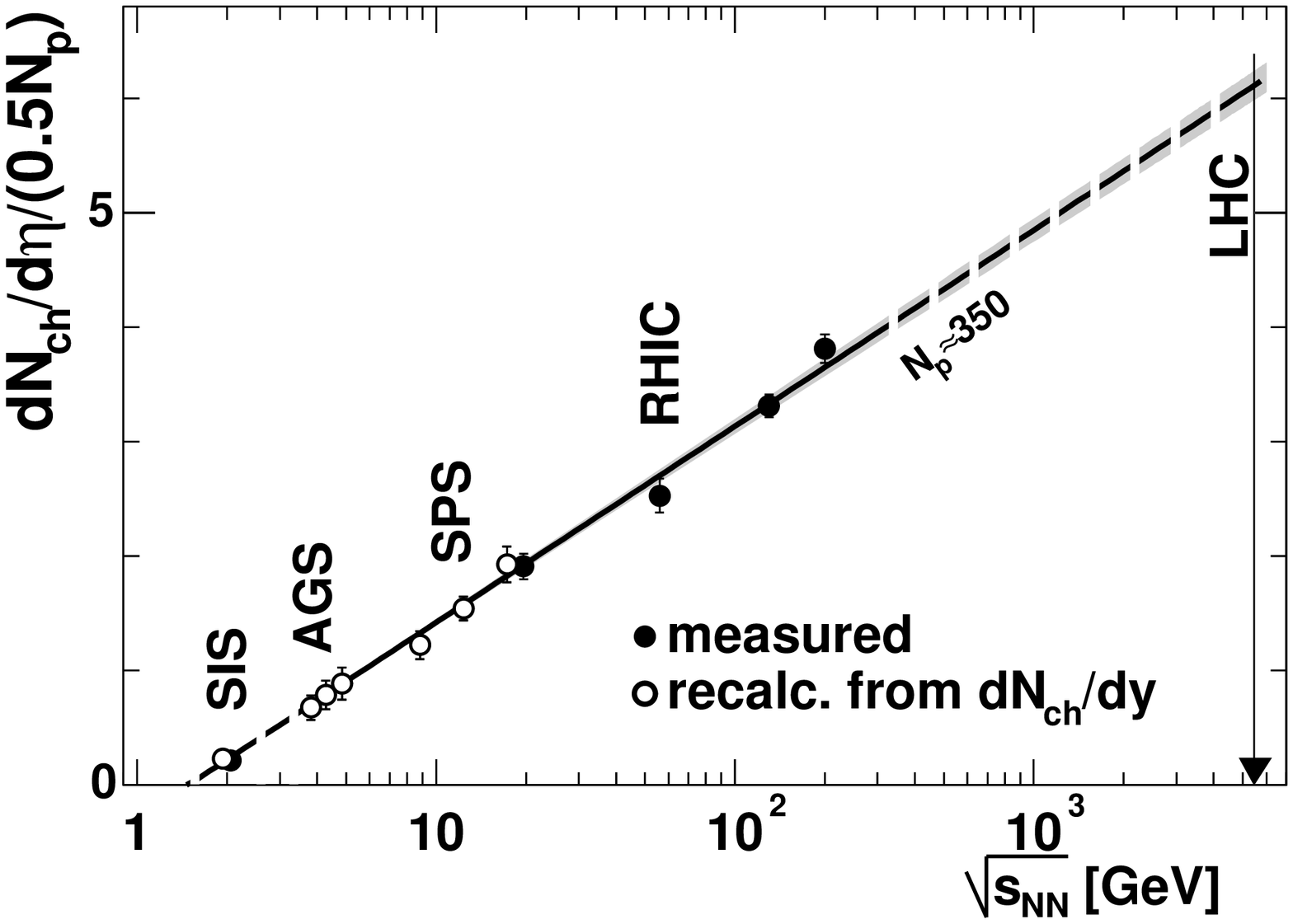}
\caption{\label{fig:sqn}
Left panel: \dEt divided by the number of \Np pairs measured in the most 
central bin (value given in brackets) as a function of incident nucleon energy. The line is a logarithmic fit. 
The band corresponds to a 1$\sigma$ statistical deviation of the fit parameters. 
Right panel: the same for $dN_{ch}/d\eta$.  The values of \Nch are the average values 
corresponding to \Nps~=~350. The single point at \sqn=~56~GeV is explained 
in section~\ref{sec:recalc_phobos}.
}
\eefw
The results shown in Fig.~\ref{fig:sqn} are consistent with logarithmic 
scaling as described in~\cite{phenix_milov,sasha_thesis,david_thesis}. 
Use of the logarithmic function is phenomenological and is suggested 
by the trend of the data in the range of available measurements.
The agreement of the fits with the data in both panels is very good, especially in 
the right panel where the averaged values are used for \Np=350. The single point 
of NA49~\cite{na49_4} is excluded from the \Et fit (see 
section~\ref{sec:recalc_na49_cent}). The results 
of the fit $dX/d\eta=(0.5N_{p}\cdot A) ln(\sqrt{s_{NN}}/\sqrt{s_{NN}^{0}})$ are:\\
for \Et  $\sqrt{s_{NN}^{0}} = 2.35\pm0.2$~GeV  and $A = 0.73\pm0.03$~GeV \\
for \Nch $\sqrt{s_{NN}^{0}} = 1.48\pm0.02$~GeV and $A = 0.74\pm0.01$.\\

The parameter $\sqrt{s_{NN}^{0}}$=2.35~GeV obtained from the \Et fit is 
slightly above, although within 3$\sigma$ from the minimum possible 
value of \sqns=$2\times a.m.u.$=1.86~GeV. The measurement closest to it 
at \sqns=2.05~GeV done by the FOPI experiment allows
to estimate the amount of \dEt produced to be 5.0~GeV in the most central collisions 
corresponding to \Np=359. Section~\ref{sec:recalc_fopi} gives details of the estimate. 
This does not disagree with the extrapolation of the fit but 
does indicate that the logarithmic parameterization requires higher order 
terms to describe how the \Et production starts at very low \sqns.

The right panel of Fig.~\ref{fig:sqn} shows the logarithmic fit to the \Nch data. 
It agrees well with all \dNch results plotted for \Nps=350. Unlike that for $E_{T}$, 
the fit parameter $\sqrt{s_{NN}^{0}}$ for \Nch is 1.48$\pm$0.02~GeV which is 
lower than the minimum allowed \sqns. This suggests that above $2 \times a.m.u.$ 
the \Nch production as a function of \sqn should undergo threshold-like behavior, 
unlike the \Et production which must approach zero smoothly due to energy 
conservation. 

The FOPI measurement at \sqns=1.94~GeV and 2.05~GeV agrees with the 
extrapolation of the fit at energy very close to $2 \times a.m.u.$. It is an 
interesting result that colliding nuclei with kinetic energies of
0.037~GeV and 0.095~GeV per nucleon in the C.M.S. follow the same particle 
production trend as seen at AGS, SPS and RHIC energies.

A fit to the charged particle multiplicity shows a factor of 2.2 increase in \dNch 
per participant in the most central events from the highest energy at the AGS (\sqn=4.8~GeV) to the 
highest energy at the SPS (\sqn=\7) and a factor of 2.0 from the highest SPS energy to the highest RHIC energy (\sqn=\2).
Assuming the same behavior extends to the LHC highest energy \sqns=5500~GeV one would expect  
\dNch = $(6.1 \pm 0.13)\cdot (0.5 N_{p})$ 
and the increase in particle production from the highest RHIC energy to be 
$\sim$60\% for the most central events. 
With the greater energy, the rapidity width should increase by $\sim$60\%
i.e. the total charged particle multiplicity at LHC would increase by 
a factor of $\sim$2.6 from the top RHIC energy.

The ratio of \EN for the most central bin as a function of \sqn is 
shown in Fig.~\ref{fig:et_nc_sqn}.
\bef
\includegraphics[width=1.0\linewidth]{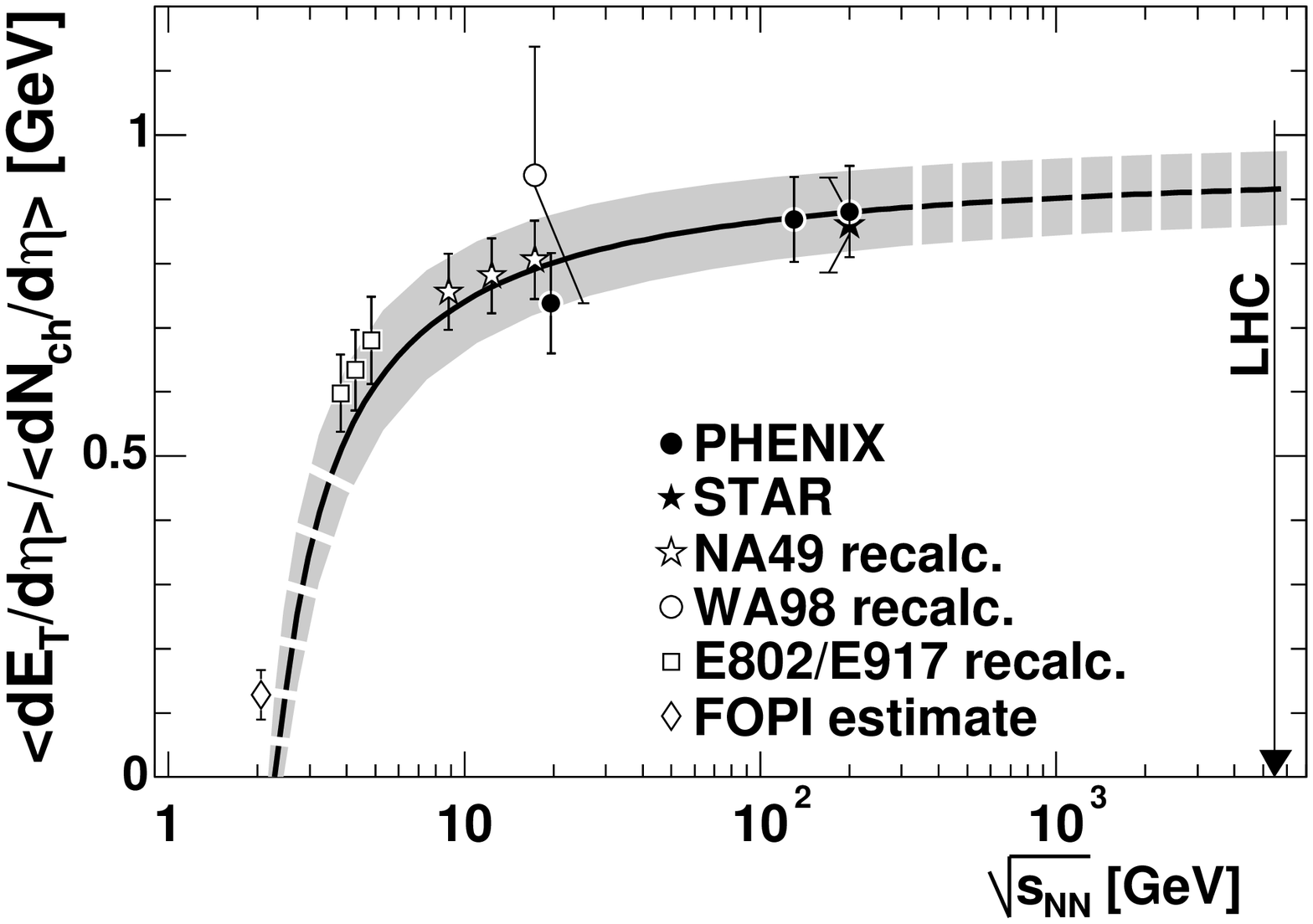}
\caption{Ratio of \Et over \Nch for the most central events as a function 
of \sqn recalculated into C.M.S.. The line is the ratio of two fits shown 
in Fig.~\ref{fig:sqn}. The band corresponds to one standard deviation of 
the combined error.\label{fig:et_nc_sqn}}
\eef
Note that the line shown in the figure is not the fit to the data 
points. Rather, it is calculated from the fits shown in Fig.~\ref{fig:sqn}. 
The calculation agrees well with the data.

There are two regions in the plot which can be clearly separated. The
region from the lowest allowed \sqn to SPS energy is characterized
 by a steep increase of the \EN ratio with \sqns. In this region the increase 
in the incident energy causes an increase in the \mt of the produced particles. 
The second region starts 
from the SPS energies and continues above. In this region, the \EN ratio is very weakly 
dependent on \sqns. The incident energy is converted into particle 
production at mid-rapidity rather than into increasing the particle $\langle m_{T} \rangle$. 

The shape of the \EN curve in the first region is governed by the 
difference in the $\sqrt{s_{NN}^{0}}$ parameter between \Et and $N_{ch}$. In the 
second region it is dominated by the ratio of the $A$ parameters in the fits. 
This ratio is close to 1~GeV. Extrapolating to LHC energies one gets
a \EN value of (0.92$\pm$0.06)~GeV.

\subsubsection{Centrality shape}
\label{sec:centrality_shape}
Another interesting question is how the shapes of the centrality curves 
of \Et and \Nch change with \sqns. 

One approach previously used in a number of papers is to describe the shape of the 
centrality dependence as a sum of ``soft'' and ``hard'' contributions such that the 
``soft'' component is proportional to \Np and the ``hard'' component to the the number of 
binary collisions $N_{c}$: $A \times N_{p} + B \times N_{c}$. A disadvantage of this 
approach is that the contributions called ``soft'' and ``hard'' do not necessarily 
correspond to the physical processes associated with these notations. 
Another approach is to assume that the production of \Et or \Nch is proportional to 
$N_{p}^{\alpha}$, although the parameter $\alpha$ does not have any physical meaning.

The results of $B/A$ and $\alpha$ obtained from the fits 
to the data at different \sqn are summarized in Table~\ref{tab:alphas}. Although the numbers 
tend to increase with beam energy, the values presented in Table~\ref{tab:alphas} 
are consistent with each other within the systematic errors.

\begin{table}
\caption{$B/A$ ratio and parameter $\alpha$ from the fit to the data. Errors are 
calculated assuming a change of the slope of centrality curves within the limits of the bending 
errors for PHENIX and full errors for the averaged data (Table~\ref{tab:averages}).\label{tab:alphas}}
\begin{ruledtabular}
\begin{tabular}{cccc}
\sqn &        \dEt         &           \dNch       &      \dNch         \\
 GeV &        PHENIX       &          PHENIX       &      Average       \\
 \multicolumn{4}{c}{$B/A$}\\
\hline
200  & $0.49^{+.69}_{-.22}$ & $0.41^{+.57}_{-.21}$ &$0.28^{+.18}_{-.15}$\\
130  & $0.41^{+.52}_{-.23}$ & $0.41^{+.45}_{-.23}$ &$0.26^{+.18}_{-.11}$\\
19.6 & $0.37^{+.48}_{-.22}$ & $0.21^{+.30}_{-.15}$ &$0.23^{+.73}_{-.23}$\\
17.2 &			    &                      &$0.31^{+.46}_{-.24}$\\
8.7  &			    &                      &$0.12^{+.64}_{-.20}$\\
\multicolumn{4}{c}{parameter $\alpha$}\\
\hline
200  & 1.20$\pm$0.07 & 1.18$\pm$0.08 & 1.16$\pm$0.06\\
130  & 1.14$\pm$0.08 & 1.17$\pm$0.08 & 1.14$\pm$0.05\\
19.6 & 1.13$\pm$0.07 & 1.09$\pm$0.06 & 1.10$\pm$0.11\\
17.2 &               &               & 1.11$\pm$0.08\\
8.7  &               &               & 1.06$\pm$0.13\\
4.8  &               &               & 1.20$\pm$0.24\\
\end{tabular}
\end{ruledtabular}
\end{table}

The availability of higher quality data would make it possible to derive a more conclusive 
statement about the shape of the curves plotted in Figs.~\ref{fig:comp_rhic}~and~\ref{fig:sps_ags}. 
With the present set of data usually limited to \Np above 50, 
a large part of the centrality curve is missing or smeared by systematic errors.
To avoid this, one can compare $Au+Au$ collisions to $p+p$ (\Nps=2) at the same energy.

\bef
\includegraphics[width=1.0\linewidth]{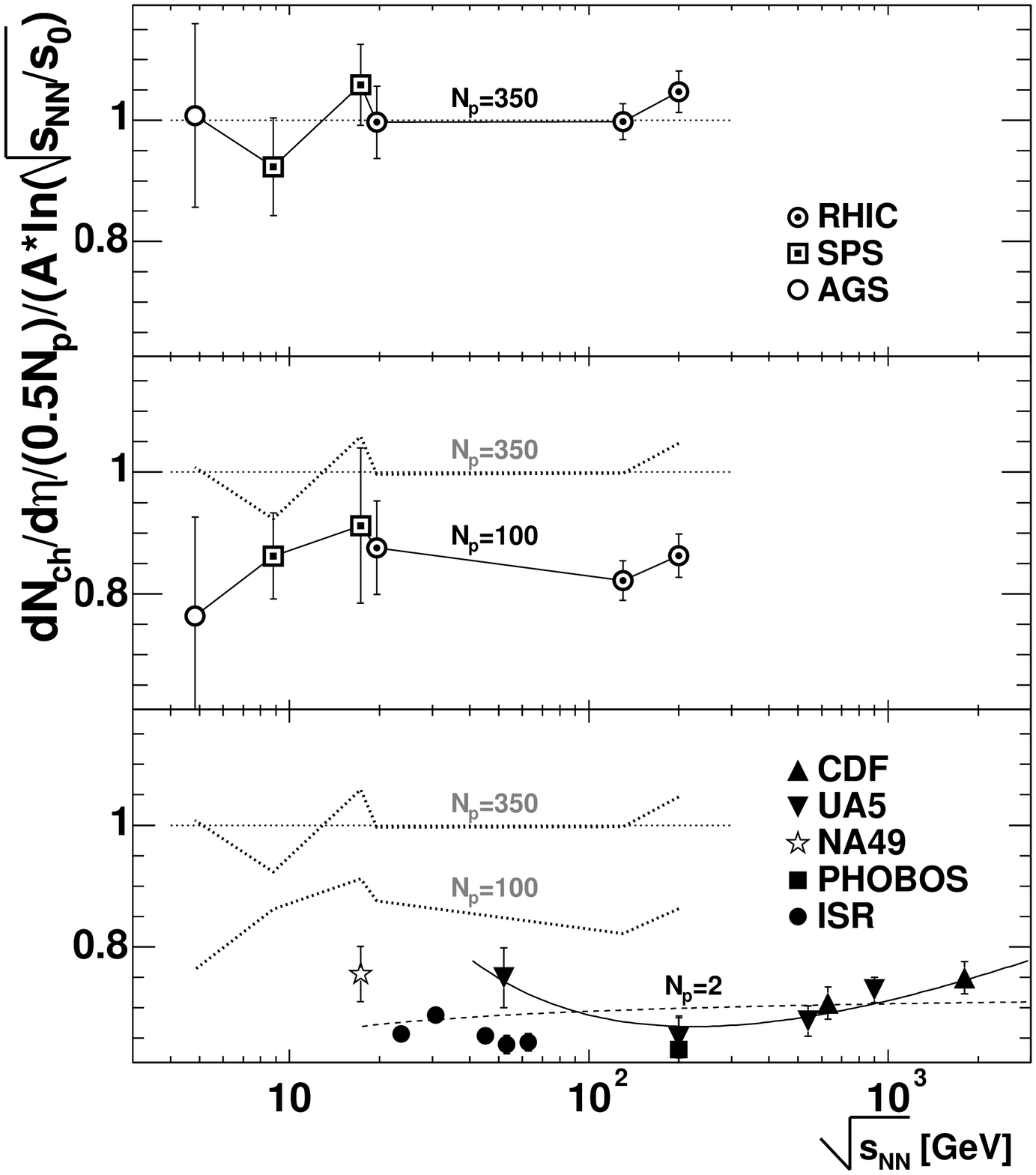}
\caption{The three panels show $dN_{ch}/d\eta/(.5N_{p})$ divided by the logarithmic parameterization from 
Fig.~\ref{fig:sqn}. The panels correspond to \Nps~=~350,~100~and~2 ($p+p$) from top to bottom. 
$Au+Au$ points are connected with lines also shown in lower panels for comparison. 
The $Au+Au$ data is tabulated in Table~\ref{tab:averages}. $p+p$ data and parameterizations 
$dN/d\eta=2.5-0.25ln(s_{NN})+0.023ln(s_{NN})^{2}$ (solid line) and 
$dN/d\eta=0.27ln(s_{NN})-0.32$ (dashed line) are taken from~\cite{ua5,isr}. 
\label{fig:centr}}
\eef

Figure~\ref{fig:centr} shows $dN_{ch}/d\eta/(0.5N_{p})$  
divided by the parameterization plotted in the right panel of Fig.~\ref{fig:sqn}.
The top panel shows the most central events with \Nps~=~350. All points are consistent 
with 1 demonstrating an agreement of the fit to the data. The points are connected with 
a line for visibility. The middle panel shows results for mid-central events with \Nps~=~100 
connected with a solid line. The dotted line is the same line as in the top panel for \Nps~=~350.
The points for \Nps~=~100 are lower than \Nps~=~350 by a factor of $0.8-0.9$, over the plotted 
range of incident energies. The lower panel shows $p+p$ data corresponding to \Nps~=~2 measured 
by several experiments. Dotted lines are the same as appear in the upper two panels for \Nps~=~350~and~100 
and the $p+p$ parameterizations are taken from~\cite{ua5,isr}. In the range of RHIC energies these 
points are lower by a factor of $0.65-0.75$ than the most central events.

These results indicate that the centrality curves normalized to the most central collisions 
have a similar shape for all RHIC energies within the errors of available measurements.

\subsection{Comparison to models}
A variety of models attempting to describe the behavior of \Et and \Nch as a function 
of centrality at different \sqn are available. 
An updated set of model results were collected from several theoretical groups 
to make a comparison as comprehensive as possible.

Figures~\ref{fig:theo1} 
through~\ref{fig:theo_ratio1} show the comparison between the existing theoretical 
models\footnote{Models 
are presented as the best fit by the polynomial of the lowest degree which 
is closer than 1\% to any theoretical point provided by the authors of the models. 
The polynomial is plotted in the range where points are provided.}
and the data for 19.6, 130 and 200 GeV. Brief descriptions of the models and 
their main characteristics are given below.

\befw
\includegraphics[width=1.0\linewidth]{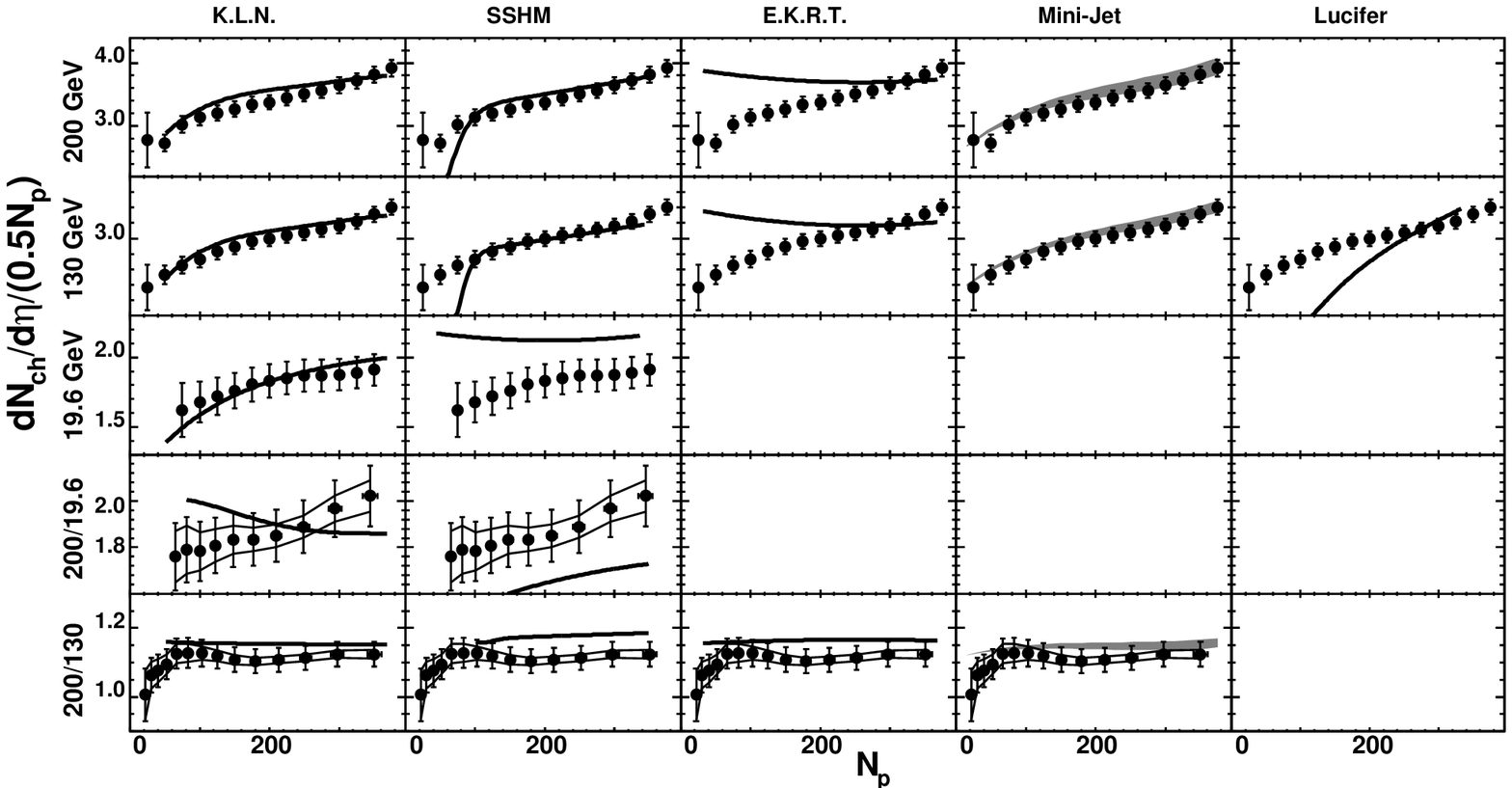}
\caption{\dNch per pair of participants compared to theoretical models. 
{\it KLN}~\cite{kln}, {\it SSHM}~\cite{accardi}, {\it EKRT}~\cite{ekrt}, 
{\it Minijet}~\cite{minijet} and {\it LUCIFER}~\cite{kahana}. The band shows 
the range of prediction for the Minijet model.\label{fig:theo1}}
\eefw

One of the more commonly used Monte Carlo event generators 
is {\it HIJING}~\cite{hijing_sim, new_hijing}. This model, like several others,
uses pQCD for initial minijet production, and the Lund string 
model~\cite{lund_string} for jet fragmentation and hadronization. 
{\it HIJING} also includes jet quenching and 
nuclear shadowing. 
This type of model typically has two components, a soft part  
proportional to \Np and a hard part proportional to $N_{coll}$, which
partly motivated the discussion in section~\ref{sec:centrality_shape}.  
There are also so-called 
saturation models which also rely on pQCD and predict that at some fixed scale
the gluon and quark phase-space density saturates, thus limiting the number of
produced quarks and gluons. An example of this type of model is 
{\it EKRT}~\cite{ekrt}, which is referred to as a final state saturation model.
In this paper, comparisons are also made to another parton 
saturation type model, {\it KLN}~\cite{kln}, an
initial state saturation model, and also to models related to {\it HIJING}, 
namely 
{\it Minijet}~\cite{minijet} and {\it AMPT}~\cite{ampt}. {\it AMPT} is a 
multiphase transport model, and extends {\it HIJING} by including explicit 
interactions between initial minijet partons and also final state hadronic interactions. 
{\it Minijet} follows the same two-component model
as {\it HIJING} but also incorporates an energy dependent cut-off scale,
similar to the saturation models.

The other models are listed briefly below. 
{\it SSHM} and {\it SFM} did not have a designated short identifier, so
 they were named somewhat arbitrarily here, based on the physics the models 
incorporate.
{\it SSHM (Saturation for Semi-Hard Minijet)}~\cite{accardi} is also a 
two-component model: pQCD-based for semi-hard partonic interactions, while 
for the soft particle production it uses the wounded nucleon model.
{\it DSM}~\cite{dsm}, the Dual String Model, is basically the Dual 
Parton Model~\cite{dpm}, with the inclusion of strings. 
{\it SFM (String Fusion Model)}~\cite{perez}, is a string model which 
includes hard collisions, collectivity in the initial state (string fusion), 
and rescattering of the produced secondaries. 
Finally, there are the hadronic models, {\it LUCIFER}~\cite{kahana}, a cascade model, with input fixed
 from lower energy data, and 
{\it LEXUS}~\cite{lexus}, a Linear EXtrapolation of Ultrarelativistic 
nucleon-nucleon Scattering data to nucleus-nucleus collisions.

The available model results range from predicting (or postdicting) \dNch 
at one energy to predicting both
\dNch and \dEt at 19.6, 130 and 200 GeV. The models have varying success in 
reproducing the data. 

In Fig.~\ref{fig:theo1}, it can be seen that {\it KLN} 
is among the most successful at 
describing the \dNch centrality dependence for all three energies. 
However, at \sqn=\1 the theoretical curve is steeper than the data. 
This results in a reversed 
centrality dependence relative to the data for the \2 to \1 ratio.
{\it SSHM} describes the 130 and 200 GeV data well, for centralities 
above \Np$\sim$100, which is the approximate limit of applicability for 
this and other 
saturation models. For the less central events, the model values are lower 
than the data. At 19.6 GeV, the model values are 
significantly higher than the data. 
The saturation model
{\it EKRT} describes the central points at both energies but overshoots the 
more peripheral data points and thus does not reproduce the general centrality 
dependence of the data. 
For the non-saturation models included in this figure, {\it Minijet} reproduces both the 
overall scale, as well as the centrality and energy dependence of 
the data rather well, while the cascade model 
{\it LUCIFER} describes the central 
points at 130 GeV well, but undershoots the less central values at this energy.

Most of the models included in Fig.~\ref{fig:theo2}, provided values 
for all three energies: 19.6, 130 and 200 GeV. 
{\it SFM} is in reasonable agreement with the 130 and 200 GeV data, but gives
much larger values than the data at 19.6 GeV.
{\it AMPT} is in overall good agreement with the data for the two higher energies,
except for the increasing trend in \dNch at the most peripheral events, which
is not seen in the data. At the lower energy the \Nch centrality behavior is 
underestimated. 
{\it LEXUS} rather severely 
overshoots the data for all energies, indicating
that nucleus-nucleus effects are not accounted for.
The {\it HIJING} models (version 1.37 and a new version with implemented baryon junctions, {\it HIJING B-$\bar{B}$}) 
only provide points at 130 and 200 GeV and are in 
reasonable agreement with the data at those energies, but generally give 
somewhat lower values. The curves shown include quenching and shadowing implemented in {\it HIJING}.
{\it DSM} describes 19.6 GeV reasonably well for all centralities, and the more
central bins for 130 and 200 GeV, but overpredicts the values for 
semi-central and peripheral events.

\befw
\includegraphics[width=1.0\linewidth]{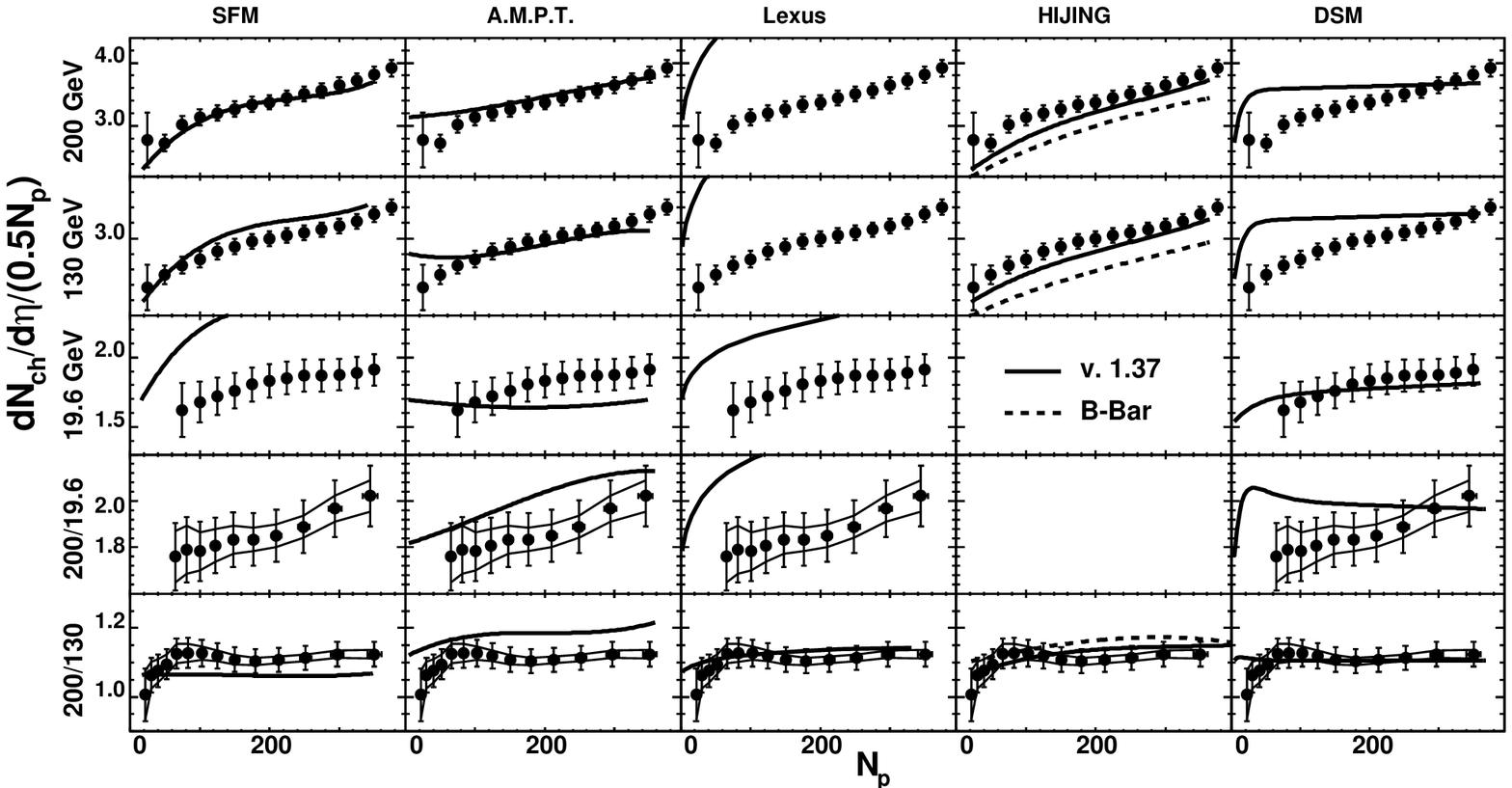}
\caption{Theoretical models compared to \dNch per pair of participants. {\it SFM}~\cite{perez}, {\it AMPT}~\cite{ampt}, {\it LEXUS}~\cite{lexus},{\it HIJING}~\cite{hijing_sim, new_hijing} and {\it DSM}~\cite{dsm}. \label{fig:theo2}}
\eefw

Figure~\ref{fig:theo_ratio1} shows the results for the models that provide 
data for both \dNch and $dE_{T}/d\eta$. For $dE_{T}/d\eta$, {\it LEXUS} and {\it SFM} consistently overshoot the 
data for all energies. In the ratio \EN, {\it LEXUS} gives values that are 
too low except at the lowest energy 19.6 GeV. That might indicate that the 
hadronization mechanism allows too little energy per particle. 
The {\it SFM} gives values that are too large, except for the most peripheral bin, 
which suggests that the particles are assigned transverse masses that are too large.
The {\it HIJING} versions and the related {\it AMPT} model are in reasonable 
agreement with the data for both \dEt and $E_{T}/N_{ch}$\footnote{Note that the 
{\it HIJING} versions available at the time the data were collected and used
for predictions were in worse agreement with the data~\cite{phenix_bazik}. This was
before energy loss and minijet separation/cut-off scale parameters were updated.}.

Also shown in Fig.~\ref{fig:theo_ratio1} are the ratios of results at 
\2 to \1, and 
\2 to \3, for $dE_{T}/d\eta$. These results, especially the comparison of the \2 to \1 data, is intended 
to make a 
more precise check of the \sqn dependence of the models. 
{\it SFM} fails to describe the \1 data and thus can not describe
the energy dependence
probed by these ratios, unlike {\it LEXUS} which however does not agree
well with the individual data curves for 19.6, 130 and 200 GeV.
{\it AMPT} and the {\it HIJING} versions reproduce the values of the ratios well,
as expected since they are in reasonable agreement with the individual curves.
{\it AMPT} and {\it Hijing} are also successful in describing the \EN ratio, as illustrated
in the lower panels of Fig.~\ref{fig:theo_ratio1}.

\befw
\includegraphics[width=0.84\linewidth]{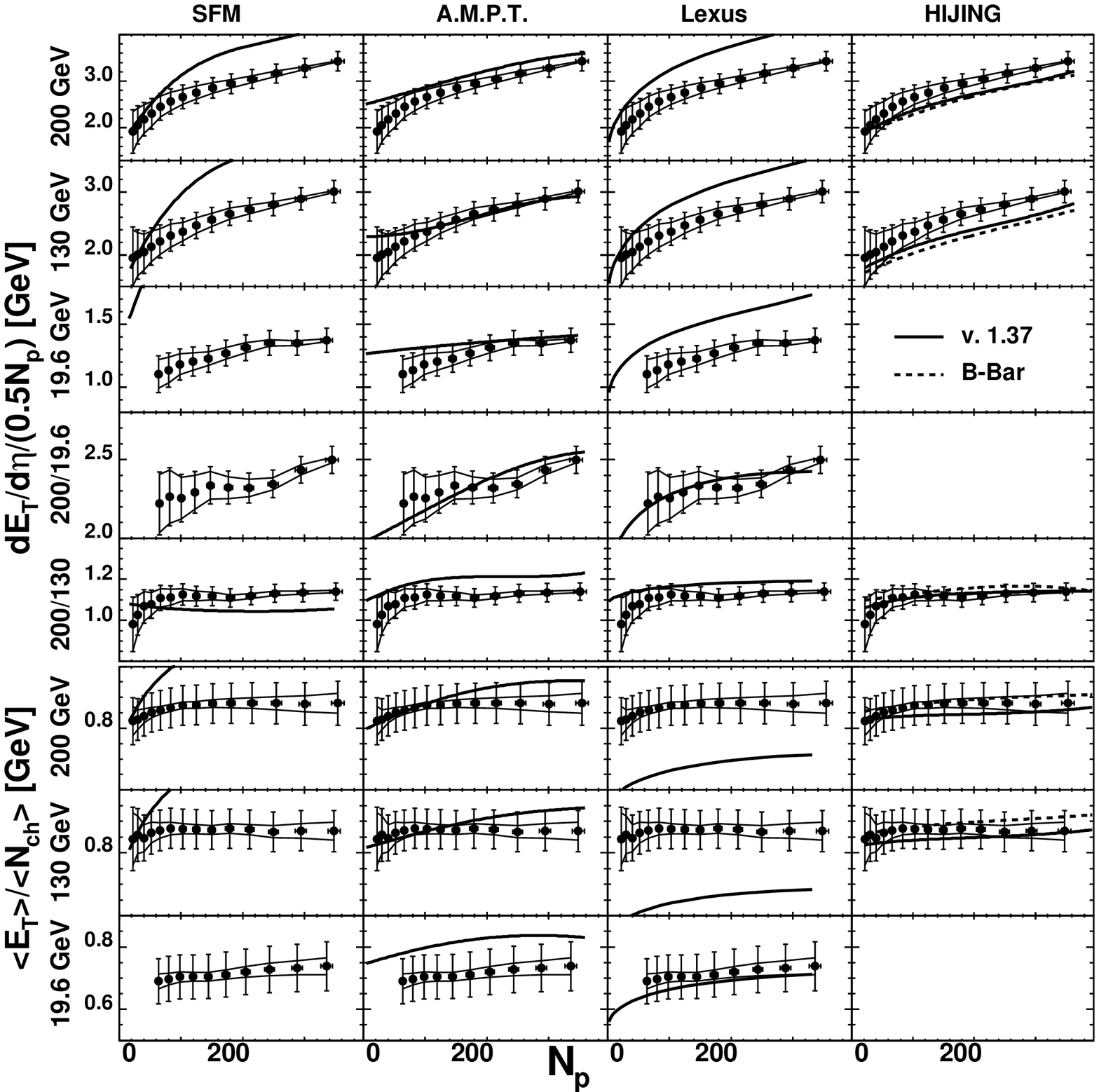}
\caption{Theoretical models compared to \dEt per pair of participants (upper panels) 
and per produced charged particle (lower panels). {\it SFM}~\cite{perez}, {\it AMPT}~\cite{ampt}, 
{\it LEXUS}~\cite{lexus} and {\it HIJING}~\cite{hijing_sim, new_hijing}.\label{fig:theo_ratio1}}
\eefw

To summarize, most models reproduce at least some of the data fairly well,
but most also fail in describing all the data. 
Since the model results typically are given without systematic errors, it is 
not entirely straightforward to quantify the level of agreement or 
disagreement with the data. Qualitatively, the models that are most successful 
in describing both \dEt and \dNch in terms of the overall trends, both 
regarding centrality dependence and energy dependence, are {\it AMPT}, 
and the {\it HIJING} versions. {\it KLN} and {\it Minijet} unfortunately do not give 
information on \dEt but are successful in describing the \dNch results.
The \dNch results thus can either be described by the initial state saturation scenario ({\it KLN}) or
by the mini-jet models that need an energy-dependent mini-jet   
cut-off scale as described in~\cite{new_hijing, minijet} to reproduce the data.

\section{Summary\label{sec:summary}}

This paper presents a systematic study of the energy and centrality
dependence of the charged particle multiplicity and transverse energy 
at mid-rapidity at \sqn = 19.6, 130 and 200 GeV. 

The yields, divided by the number of participant nucleons, show a consistent
centrality dependence (increase from peripheral to central) between \dEt and 
\dNch for all energies. Furthermore, the increase in the ratio \EN from \1 to \2 
is consistent with a 20\% increase in \mt with increasing \sqns. 
The ratio \EN shows only a weak centrality dependence at RHIC energies.

For the \sqn dependence, comparisons were made not only among RHIC results
but also including data from lower energy fixed-target experiments at SPS, 
AGS and SIS. A phenomenological fit, scaling logarithmically with \sqns, 
describes both \dEt and \dNch for the most central collisions well for all energies.

Using the fit results, one can separate two regions with different 
particle production mechanisms. The region below SPS energy is characterized by a 
steep increase in \EN$\sim$\mt with \sqns, whereas for the energies above 
SPS \EN is found to be weakly dependent on \sqns.

Within the systematic errors of the measurements 
the shape of the centrality curves of $dN_{ch}/d\eta/(0.5N_{p})$ vs. \Np 
were found to be the same in the range of RHIC energies and to scale with 
$\ln(\sqrt{s_{NN}})$. The same must be true for \Et because \EN has a very 
weak centrality dependence.

Based on the \dEt measurements, the Bjorken energy density 
estimates were performed and $\epsilon_{Bj} \cdot \tau$ was determined to be
$5.4 \pm 0.6$ GeV/($fm^2 \cdot c$) at \sqn = \2 for the most central bin. 
This is in excess of what is
believed to be sufficient for a phase transition to the new state of matter. 
The energy density increases by about a factor of 2.6 from the top SPS energy 
to the top RHIC energy.

Finally, a comparison between the RHIC \dNch and \dEt data and a collection 
of models was performed. A few models, notably {\it HIJING} and {\it AMPT},
reproduce both \dEt and \dNch rather well for several energies.


\begin{acknowledgments}
We thank the staff of the Collider-Accelerator and Physics
Departments at Brookhaven National Laboratory and the staff
of the other PHENIX participating institutions for their
vital contributions.  
We are grateful for information provided by the model authors. 
In particular, we thank A.~Accardi, S.~Barshay, S.~Jeon, S.~Kahana, 
D.~Kharzeev, Z.~Lin, N.~Armesto~Perez, R.~Ugoccioni, V.T.~Pop, 
and X.N.~Wang for helpful correspondence.
We acknowledge support from the
Department of Energy, Office of Science, Nuclear Physics
Division, the National Science Foundation, Abilene Christian
University Research Council, Research Foundation of SUNY, and
Dean of the College of Arts and Sciences, Vanderbilt
University (U.S.A), Ministry of Education, Culture, Sports,
Science, and Technology and the Japan Society for the
Promotion of Science (Japan), Conselho Nacional de
Desenvolvimento Cient\'{\i}fico e Tecnol{\'o}gico and Funda\c
c{\~a}o de Amparo {\`a} Pesquisa do Estado de S{\~a}o Paulo
(Brazil), Natural Science Foundation of China (People's
Republic of China), Centre National de la Recherche
Scientifique, Commissariat {\`a} l'{\'E}nergie Atomique,
Institut National de Physique Nucl{\'e}aire et de Physique
des Particules, and Institut National de Physique
Nucl{\'e}aire et de Physique des Particules, (France),
Bundesministerium f\"ur Bildung und Forschung, Deutscher
Akademischer Austausch Dienst, and Alexander von Humboldt
Stiftung (Germany), Hungarian National Science Fund, OTKA
(Hungary), Department of Atomic Energy and Department of
Science and Technology (India), Israel Science Foundation
(Israel), Korea Research Foundation and Center for High
Energy Physics (Korea), Russian Ministry of Industry, Science
and Tekhnologies, Russian Academy of Science, Russian
Ministry of Atomic Energy (Russia), VR and the Wallenberg
Foundation (Sweden), the U.S. Civilian Research and
Development Foundation for the Independent States of the
Former Soviet Union, the US-Hungarian NSF-OTKA-MTA, the
US-Israel Binational Science Foundation, and the 5th European
Union TMR Marie-Curie Programme.

\end{acknowledgments}

\appendix
\section{Recalculation of the non-PHENIX experimental data \label{sec:recalc}}

Comparisons of \dEt and \dNch between different experiments can be made only if the 
results are presented in the same coordinate system since these values are
not boost invariants. In some cases a full set of identified particles measured by 
one experiment can be recalculated into \Et and $N_{ch}$.
Each case that involves handling non-PHENIX published data 
is separately explained in this Appendix.

\subsection{General\label{sec:general}}
Figure~\ref{fig:cms_lab} shows simulated rapidity distributions for \Et and 
\Nch in the C.M.S. and Lab. frames. Plots presented here are for illustration purposes 
only. The invariant distributions which do not change their shape under 
transition from Lab. to C.M.S. are $dm_{T}/dy$ and $dN_{ch}/dy$, while all others do.
\befw
\includegraphics[width=0.48\linewidth]{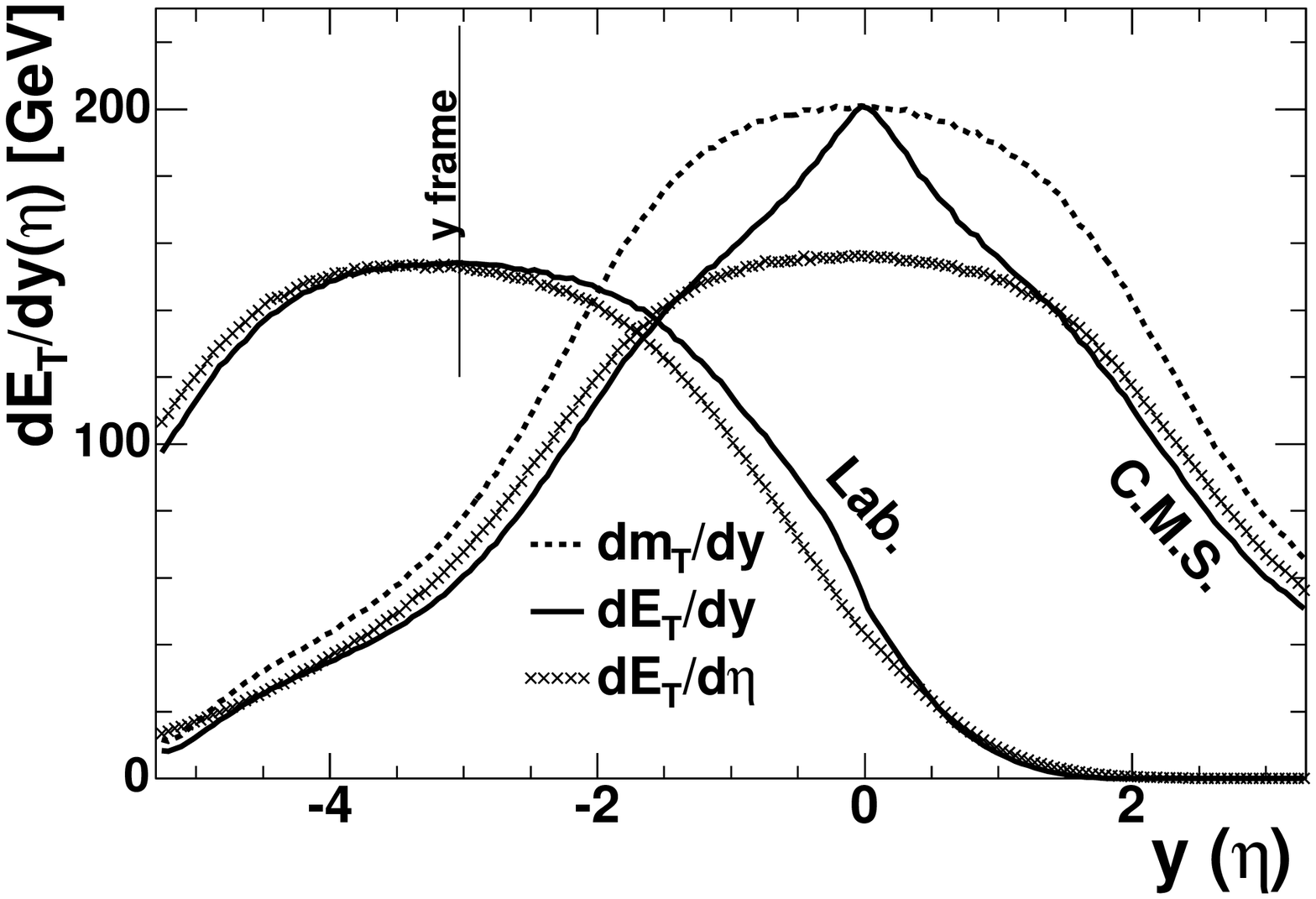}
\includegraphics[width=0.48\linewidth]{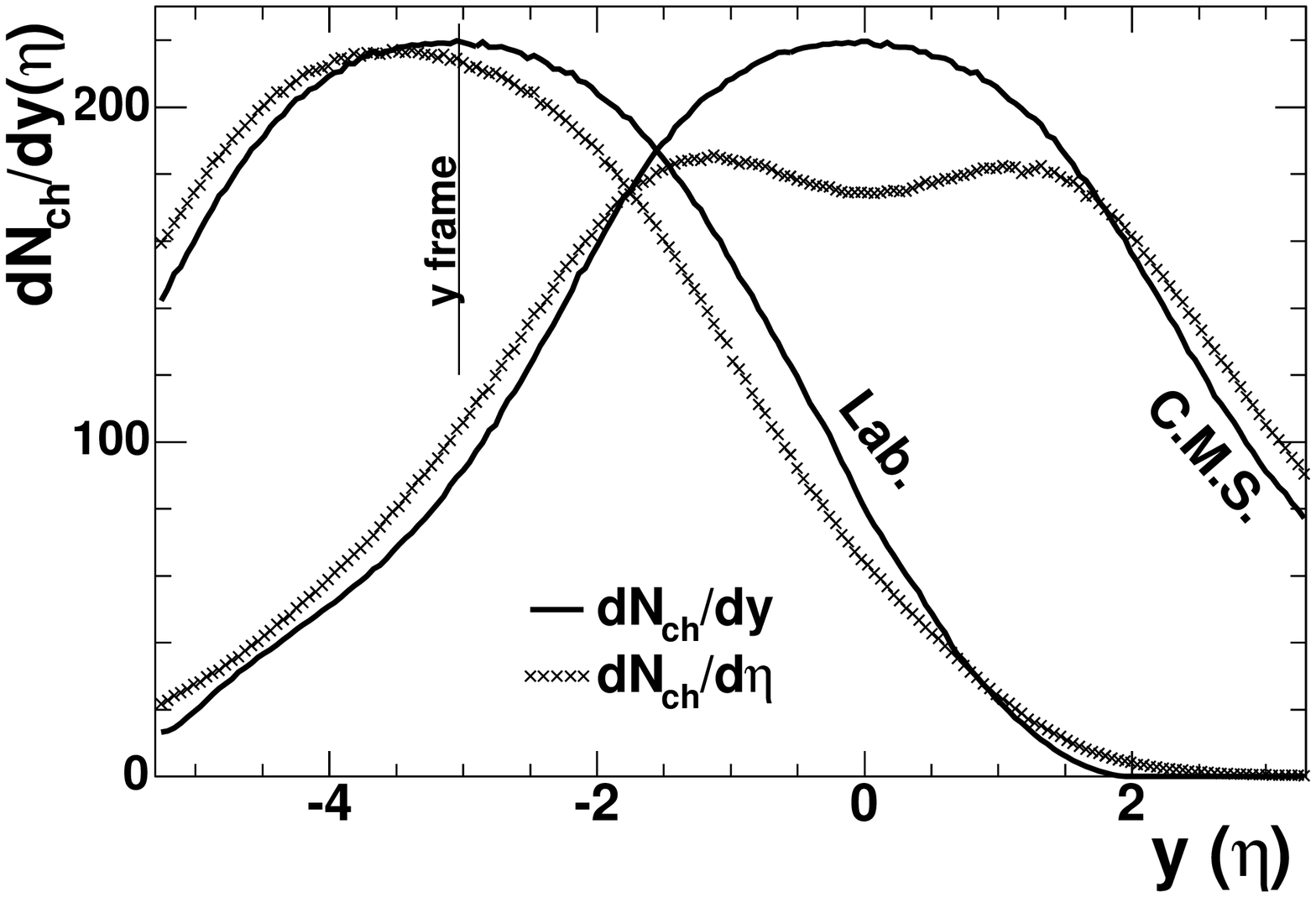}
\caption{Simulated \Et (left) and \Nch (right) distributions in rapidity and 
pseudo-rapidity units in C.M.S. and Lab. systems.\label{fig:cms_lab}}
\eefw

In the C.M.S. system, the transition from ``$\eta$'' to ``$y$'' at 
mid-rapidity requires a scaling factor between 1.2 and 1.3. An accurate 
determination of this coefficient from the published data of other experiments 
is not always possible, therefore for SPS and AGS energies a coefficient of 
1.25 was used. Because of the definition of \Et used in this paper, $dE_{T}/dy \approx dm_{T}/dy$, 
 around mid-rapidity, where $m_{T}$ is a quadratic sum of the particle mass and 
transverse momentum: $m_{T} = \sqrt{m^{2}+p_{T}^{2}}$.

In the Lab. system $dN_{ch}/dy\approx dN_{ch}/d\eta$ and $dE_{T}/dy\approx dE_{T}/d\eta$ 
at maximum rapidity. A 1.04 conversion factor was assigned to 
the transition from ``$\eta$'' to ``$y$'' in the Lab. system.

An error of 5\% was assigned to any converted value. This error also absorbs 
uncertainties on various assumptions used in the calculations. For example, the contribution of 
neutral particles to the total \Et is assumed to be:
\beq
E_{T}^{\pi^{0}} = (E_{T}^{\pi^{+}}+ E_{T}^{\pi^{-}})/2. \nonumber \\
E_{T}^{K^{0}}   =  E_{T}^{K^{+}}  + E_{T}^{K^{-}}       \nonumber \\
E_{T}^{n} + E_{T}^{\bar{n}} = E_{T}^{p}+E_{T}^{\bar{p}}
\eeq

\subsection{Averaging procedure\label{sec:recalc_average}}
Average values were calculated for \Np=25, 50,... 375. The centrality bin 
corresponding to a given \Np can be different in different experiments. 
\dNch per participant and the associated error were deduced by a weighted average 
interpolation from the two nearest values of each experiment. 
The closest value was required to be within a proximity of 
25 participants from the \Np value. 
The error bars are multiplied by the $S$-factor, where 
$S=\sqrt{\chi^{2}/n.d.f.}$ if $\chi^{2}/n.d.f.>1$ or $S=1$ otherwise. See the Particle Data Group reference~\cite{PDG} for details.

\subsection{NA49\label{sec:recalc_na49_cent}}
The identified particle yields in the most central events at mid-rapidity at \sqns=\7 are shown in Fig.~6 in reference~\cite{na49_1}.

\begin{table}
\caption{Results on identified particle yields in the most central events 
published by NA49 at mid-rapidity at \sqns=\7 taken from Fig.~6 in reference~\cite{na49_1}.\label{tab:na49_1}}
\begin{ruledtabular}
\begin{tabular}{lcccccc}
Particle  & $\pi^{+}$ & $\pi^{-}$ & $K^{+}$ & $K^{-}$ & $p$  & \={p}\\
$dN/dy$   &    167    &  165      &   32    &   15    &  33  &    5 \\
error     &    10     &  10       &    4    &    5    & 1.5  &  0.5 \\
\end{tabular}
\end{ruledtabular}
\end{table}

The total yields per participant and number of participants are taken from Fig.~10 
in reference~\cite{na49_1}.
\begin{table}
\caption{Total yields of identified particles per participant and mean momentum at mid-rapidity in different centrality bins 
published by NA49 (P) at \sqns=\7, from Figs.~8~and~10 in reference~\cite{na49_1}. Recalculated values (R) are plotted in 
Fig.~\ref{fig:comp_sps},~\ref{fig:sps_ags}.\label{tab:na49_2}}
\begin{ruledtabular}
\begin{tabular}{clccccccc}
P & \Np                                      & 362  &  305 &  242 &  189 &  130 &  72  &    2 \\
P & error                                    &   10 &   15 &   15 &   15 &   10 &   5  &      \\
P & $\langle dN^{\pi}/dy/N_{p}\rangle$       & 1.65 & 1.64 & 1.55 & 1.48 & 1.40 & 1.42 & 1.42 \\
P & error                                    & 0.08 & 0.08 & 0.08 & 0.08 & 0.08 & 0.08 & 0.08 \\
P & $\langle p^{\pi^{+}}_{T}\rangle$ [GeV/c] & 0.29 & 0.31 & 0.30 & 0.30 & 0.30 & 0.29 & 0.28 \\
P & $\langle p^{\pi^{-}}_{T}\rangle$ [GeV/c] & 0.27 & 0.31 & 0.29 & 0.30 & 0.30 & 0.30 & 0.28 \\
P & $ dN^{K^{+}}/dy/N_{p}$                   & 0.27 & 0.27 & 0.22 & 0.19 & 0.16 & 0.14 & 0.10 \\ 
P & error                                    & 0.02 & 0.02 & 0.02 & 0.02 & 0.02 & 0.02 & 0.02 \\
P & $\langle p^{K^{+}}_{T}\rangle$ [GeV/c]   & 0.57 & 0.54 & 0.50 & 0.49 & 0.53 & 0.55 & 0.45 \\
P & $ dN^{K^{-}}/dy/N_{p}$                   & 0.15 & 0.15 & 0.12 & 0.10 & 0.09 & 0.07 & 0.06 \\
P & error                                    & 0.01 & 0.01 & 0.01 & 0.01 & 0.01 & 0.01 & 0.01 \\
P & $\langle p^{K^{-}}_{T}\rangle$ [GeV/c]   & 0.57 & 0.55 & 0.53 & 0.51 & 0.55 & 0.51 & 0.42 \\
P & $\langle p^{p}_{T}\rangle$ [GeV/c]       & 0.77 & 0.74 & 0.75 & 0.73 & 0.70 & 0.65 & 0.54 \\
P & $ dN^{\bar{p}}/dy/N_{p}$                 & 0.03 & 0.03 & 0.03 & 0.03 & 0.03 & 0.02 & 0.02 \\
P & error                                    & .005 & .005 & .005 & .005 & .005 & .005 & .005 \\
P & $\langle p^{\bar{p}}_{T}\rangle$ [GeV/c] & 0.87 & 0.84 & 0.80 & 0.75 & 0.78 & 0.70 & 0.48 \\
R & $dE_{T}/d\eta/(.5N_{p})$ [GeV]           & 1.47 & 1.50 & 1.35 & 1.29 & 1.23 & 1.15 & 1.00 \\
R & error [GeV]                              & 0.11 & 0.11 & 0.10 & 0.10 & 0.10 & 0.11 & 0.07 \\
R & $dN_{ch}/d\eta/(.5N_{p})$                & 1.75 & 1.74 & 1.62 & 1.54 & 1.46 & 1.44 & 1.38 \\
R & error                                    & 0.12 & 0.12 & 0.11 & 0.11 & 0.11 & 0.12 & 0.08 \\
\end{tabular}
\end{ruledtabular}
\end{table}
Using shapes of the $dN/dy$ distributions shown in Fig.~7 in reference~\cite{na49_1} for different 
centrality bins quantities tabulated in~\ref{tab:na49_1}~and~\ref{tab:na49_2} can be converted into \dEt and \dNch 
per participant pair at mid-rapidity. Systematic errors on particle yields are given in Table~1 in the same 
reference. The systematic error on this value is not mentioned in the paper, therefore they
were taken from~\cite{na49_3}. The results used in this paper are also given in Table~\ref{tab:na49_2}.

For the same and lower \sqns, the identified particle yields and \mt were reconstructed using 
formula (1) and Fig.~1 in reference~\cite{na49_2} and also Table II and formula (1) and 
(2) in reference~\cite{na49_3}. The data obtained from the Tables and the fits are
summarized in Table~\ref{tab:na49_4}. Using $dN/dy$ and \mt the values of \dEt and \dNch 
were recalculated in the C.M.S. frame. 
The accuracy of the procedure was verified by the consistency 
of results presented in Table~\ref{tab:na49_4} and Table~\ref{tab:na49_5}.
\begin{table}
\caption{Temperatures of the identified particles published by NA49 at different \sqns, as 
extracted from~\cite{na49_2,na49_3}. The yields are results of the fits of the 
parameterizations given in these publications.
\label{tab:na49_4}}
\begin{ruledtabular}
\begin{tabular}{lccccccccc}
\multicolumn{10}{c}{\sqns=\7}\\	      	      	    	   	   	   	     	       	 	     	 
 & $\pi^{+}$&$\pi^{-}$&$K^{+}$&$K^{-}$&$p$&$\bar{p}$& $\Lambda$&$\bar{\Lambda}$&$d$ \\
$T$ [GeV]  & .180 & .180 &  .232 & .226 & .127 & .122 &  .127 & .122 & .127  \\
error [GeV]& 0.01 & .010 &  .007 & .011 & .004 & .002 &  .004 & .002 & .004  \\
$dN/dy$    & 170. & 175. &  29.6 & 16.8 &  23. &  1.4 &   16. &  3.5 & 0.32  \\
error      &   9. &   9. &   1.5 &  0.8 &  7.4 & 0.23 &   6.1 & 0.67 & 0.23  \\
\hline
\multicolumn{10}{c}{\sqns=12.4GeV}\\				    			     
 & $\pi^{+}$&$\pi^{-}$&$K^{+}$&$K^{-}$&$p$&$\bar{p}$& $\Lambda$&$\bar{\Lambda}$&$d$ \\
$T$ [GeV]  & .179 & .179 &  .230 & .217 & .133 & .120 &  .133 & .120 & .133  \\
error [GeV]& 0.01 & .010 &  .008 & .007 & .003 & .001 &  .003 & .001 & .003  \\
$dN/dy$    & 132. & 140. &  24.6 & 11.7 &  29. &  0.7 &  17.5 &  0.8 & 0.85  \\
error      &   7. &   7. &   1.2 & 0.6  &  6.2 & 0.06 &   4.4 & 0.08 & 0.28  \\      
\hline
\multicolumn{10}{c}{\sqns=8.7 GeV}\\				    			     
 & $\pi^{+}$&$\pi^{-}$&$K^{+}$&$K^{-}$&$p$&$\bar{p}$& $\Lambda$&$\bar{\Lambda}$&$d$ \\
$T$ [GeV]  & .169 & .169 &  .232 & .226 & .130 & .137 &  .130 & .137 & .130  \\
error [GeV]& 0.01 & 0.01 &  .007 & .007 & .002 & .004 &  .002 & .004 & .002  \\
$dN/dy$    & 96.6 & 106. &  20.1 &  7.6 &  40. & 0.28 &  17.2 & 0.28 & 1.25  \\
error      &  6.  &   6. &   1.0 &  0.4 &  5.8 & 0.08 &   2.9 & 0.08 & 0.37  \\   
\end{tabular}
\end{ruledtabular}
\end{table}

\begin{table}
\caption{Recalculated NA49 results, as plotted in Figs.~\ref{fig:sqn},~and~\ref{fig:et_nc_sqn}.\label{tab:na49_5}}
\begin{ruledtabular}
\begin{tabular}{lcccc}
\sqn [GeV]                     &    17.2       &     12.4      &      8.7      \\
\Np                            & 363.$\pm$10   & 352.$\pm$10   & 352.$\pm$10   \\
$dE_{T}/d\eta/(.5N_{p})$ [GeV] & 1.50$\pm$0.11 & 1.16$\pm$0.09 & 0.94$\pm$0.07 \\
$dN_{ch}/d\eta/(.5N_{p})$      & 1.86$\pm$0.08 & 1.54$\pm$0.07 & 1.24$\pm$0.06 \\
\EN   [GeV]                    & 0.81$\pm$0.06 & 0.78$\pm$0.06 & 0.76$\pm$0.06 \\
\end{tabular}
\end{ruledtabular}
\end{table}
The single \Et point in Fig.~\ref{fig:sqn} is taken from~\cite{na49_4} as 405 GeV 
and scaled up by 10\%, then divided by pairs of \Nps~=~390 as explained in the text. 
This point does not agree with the value of \Et deduced from \cite{na49_1,na49_2,na49_3}.

\subsection{WA98\label{sec:recalc_wa98_cent}}
The centrality dependence of the \Et, \Nch and \EN were read from the plots in 
Figs.~7~and~14 in reference~\cite{wa98} and converted to the C.M.S. frame. 
Results are summarized in Table~\ref{tab:wa98}.

\squeezetable
\begin{table*}
\caption{Published (P) WA98 results at \sqns=\7 taken from Figs.7~and~14 in reference~\cite{wa98} 
and recalculated (R) results plotted in Figs.~\ref{fig:comp_sps},~\ref{fig:sps_ags},~\ref{fig:sqn},~and~\ref{fig:et_nc_sqn}. 
Additional systematic errors are shown in the plots.\label{tab:wa98}}
\begin{ruledtabular}
\begin{tabular}{clccccccccccccccccccc}
P & \multicolumn{5}{l}{\Np}                             &  382 &  357 &  311 &  269 &  234 &  201 &  174 &  148 &  128 &  109 &   91 &   75 &   62 &   49 &   39\\
P & \multicolumn{5}{l}{error}                           &   11 &    9 &    7 &    7 &    6 &    5 &    5 &    4 &    4 &    4 &    4 &    3 &    3 &    2 &   11\\
P & \multicolumn{5}{l}{$dE_{T}/d\eta/(.5N_{p})$ [GeV]}  & 2.09 & 2.06 & 2.06 & 2.05 & 2.03 & 2.00 & 1.95 & 1.96 & 1.93 & 1.87 & 1.84 & 1.78 & 1.73 & 1.70 & 1.59\\
P & \multicolumn{5}{l}{error [GeV]}                     & 0.12 & 0.12 & 0.12 & 0.12 & 0.12 & 0.12 & 0.11 & 0.11 & 0.11 & 0.11 & 0.12 & 0.14 & 0.15 & 0.17 & 0.18\\
R & \multicolumn{5}{l}{$dE_{T}/d\eta/(.5N_{p})$ [GeV]}  & 2.00 & 1.97 & 1.97 & 1.97 & 1.95 & 1.92 & 1.88 & 1.88 & 1.85 & 1.80 & 1.76 & 1.71 & 1.67 & 1.63 & 1.53\\
R & \multicolumn{5}{l}{error [GeV]}                     & 0.12 & 0.11 & 0.11 & 0.11 & 0.12 & 0.11 & 0.11 & 0.11 & 0.11 & 0.11 & 0.12 & 0.13 & 0.15 & 0.17 & 0.18\\
\hline
P & \Np                        &  381 &  355 &  310 &  268 &  231 &  199 &  171 &  145 &  124 &  105 &   87 &   72 &   58 &   46 &   36 &   27 &   20 &   13 &    9\\
P & error                      &   10 &    9 &    8 &    7 &    6 &    6 &    6 &    5 &    5 &    5 &    4 &    5 &    4 &    3 &    3 &    3 &    2 &    2 &   12\\
P & $dN_{ch}/d\eta/(.5N_{p})$  & 2.66 & 2.64 & 2.62 & 2.60 & 2.58 & 2.54 & 2.50 & 2.49 & 2.43 & 2.40 & 2.34 & 2.32 & 2.31 & 2.29 & 2.20 & 2.17 & 2.13 & 2.17 & 2.22\\
P & error                      & 0.09 & 0.09 & 0.10 & 0.10 & 0.10 & 0.10 & 0.11 & 0.11 & 0.11 & 0.12 & 0.14 & 0.17 & 0.19 & 0.21 & 0.23 & 0.25 & 0.26 & 0.29 & 0.31\\
R & $dN_{ch}/d\eta/(.5N_{p})$  & 2.13 & 2.11 & 2.10 & 2.08 & 2.07 & 2.03 & 2.00 & 2.00 & 1.94 & 1.92 & 1.88 & 1.85 & 1.85 & 1.83 & 1.76 & 1.74 & 1.71 & 1.74 & 1.78\\
R & error                      & 0.07 & 0.08 & 0.08 & 0.08 & 0.08 & 0.08 & 0.09 & 0.09 & 0.09 & 0.10 & 0.12 & 0.14 & 0.16 & 0.16 & 0.18 & 0.20 & 0.21 & 0.23 & 0.25\\
\hline
P & \multicolumn{4}{l}{\Np}                              &  383 &  359 &  315 &  276 &  242 &  211 &  185 &  160 &  140 &  123 &  106 &   91 &   78 &   66 &   56 &   49\\
P & \multicolumn{4}{l}{error}                            &   10 &    9 &    8 &    7 &    6 &    5 &    4 &    4 &    3 &    4 &    6 &    6 &    6 &    5 &    5 &   11\\
P & \multicolumn{4}{l}{\EN   [GeV]}                      & 0.78 & 0.78 & 0.79 & 0.80 & 0.80 & 0.81 & 0.79 & 0.79 & 0.81 & 0.81 & 0.81 & 0.80 & 0.79 & 0.78 & 0.77 & 0.77\\
P & \multicolumn{4}{l}{error [GeV]}                      & 0.06 & 0.06 & 0.06 & 0.06 & 0.06 & 0.06 & 0.06 & 0.06 & 0.06 & 0.06 & 0.06 & 0.06 & 0.06 & 0.06 & 0.06 & 0.06\\
R & \multicolumn{4}{l}{\EN   [GeV]}                      & 0.94 & 0.94 & 0.95 & 0.96 & 0.96 & 0.97 & 0.95 & 0.95 & 1.00 & 0.98 & 0.97 & 0.96 & 0.94 & 0.94 & 0.93 & 0.93\\
R & \multicolumn{4}{l}{error [GeV]}                      & 0.07 & 0.07 & 0.07 & 0.07 & 0.07 & 0.07 & 0.07 & 0.07 & 0.07 & 0.07 & 0.07 & 0.07 & 0.07 & 0.07 & 0.07 & 0.07\\
\end{tabular}
\end{ruledtabular}
\end{table*}

\subsection{NA45\label{sec:recalc_na45_cent}}
The NA45/CERES collaboration did not publish results for \dNch as a function of centrality at 
\sqns=\7. The data was taken from Fig.~6.5 in reference~\cite{ceres_1} and a 10\% error was assigned based 
on the analysis procedure. The number of participants was taken for the corresponding cross section 
bin reported by the NA50 results~\cite{na50_1}. At the lower energy, the results were originally published 
in~\cite{ceres_2} and then \Np was subsequently corrected, see e.g.:~\cite{ceres_3}. The 
results presented in Fig.~4 in reference~\cite{ceres_3} for charged hadrons $h^{-}$ and $(h^{+}-h^{-})$ 
were added together to get $dh/d\eta$ and then converted to \dNch in the C.M.S.
frame. The published and recalculated results are summarized in Table~\ref{tab:na45}.
\begin{table}
\caption{Published (P) NA45 results at \sqns=\7 taken from Fig.~6.5 in reference~\cite{ceres_1} and at \sqns=8.7~GeV from Fig.~4 
in reference~\cite{ceres_3} and recalculated (R) results plotted in Figs.~\ref{fig:comp_sps},~\ref{fig:sps_ags}.\label{tab:na45}}
\begin{ruledtabular}
\begin{tabular}{clccccccc}
\multicolumn{9}{c}{\sqns=\7}\\	      	      	    	   	   	   	     	       	 	     	 
P & bin                       &  0-2.3 & 2.3-5 &  5-8 & 8-12 & 12-18 & 18-23 & 23-35 \\
P & \dNch                     &    420 &   350 &  300 &  250 &   210 &   170 &   125 \\
R & \Np                       &  360   &   331 &  300 &  264 &   220 &   179 &   132 \\
R & error                     &  10.   &    10 &    9 &    8 &     7 &     7 &     6 \\
R & $dN_{ch}/d\eta/(.5N_{p})$ & 1.87   &  1.69 & 1.60 & 1.51 &  1.53 &  1.52 &  1.52 \\
R & error                     & 0.19   &  0.18 & 0.17 & 0.16 &  0.16 &  0.16 &  0.17 \\
\hline
\multicolumn{9}{c}{\sqns=8.7GeV}\\
P & \Np                       &    368 &   335 &  287 &  238 &   183 &  120 \\
P & $h^{-}$                   &    129 &   113 &   94 &   78 &    58 &   38 \\
P & error                     &     15 &    14 &   12 &   11 &     9 &    9 \\
P & $h^{+}-h^{-}$             &     52 &    46 &   39 &   30 &    22 &   15 \\
P & error                     &     12 &     8 &    8 &    7 &     6 &    6 \\
R & $dN_{ch}/d\eta/(.5N_{p})$ &   1.35 &  1.30 & 1.27 & 1.25 &  1.21 & 1.21 \\
R & error                     &   0.08 &  0.08 & 0.08 & 0.09 &  0.10 & 0.12 \\
\end{tabular}
\end{ruledtabular}
\end{table}

\subsection{NA50\label{sec:recalc_na50_cent}}
Results on \Np are taken from Tables 1 and 2 in reference~\cite{na50_1} and on 
multiplicity from Figs.~2~and~4 tabulated in captions in reference~\cite{na50_1}. 
The systematic errors are mentioned in the text. 
There is some discrepancy in the results of NA50 and NA45 as shown in 
Fig.~\ref{fig:sps_ags}.  In this respect the comparison made in Table~3 in 
reference~\cite{na50_1} is unclear. The results were converted to the C.M.S.
frame. Recalculated values are given in Table~\ref{tab:na50}.
\begin{table}
\caption{Recalculated NA50 results plotted in Figs.~\ref{fig:comp_sps}~and~\ref{fig:sps_ags}.\label{tab:na50}}
\begin{ruledtabular}
\begin{tabular}{lcccccc}
\multicolumn{7}{c}{\sqns=\7}\\	      	      	    	   	   	   	     	       	 	     	 
\Np                       & 354 &  294 &  246 &  205 &  173 &  129\\
error                     &  12 &   10 &    8 &    8 &    8 &   11\\
$dN_{ch}/d\eta/(.5N_{p})$ &1.98 & 1.98 & 1.94 & 1.95 & 1.95 & 2.10\\
error                     &0.16 & 0.16 & 0.16 & 0.16 & 0.16 & 0.17\\
\hline
\multicolumn{7}{c}{\sqns=8.7GeV}\\
\Np                       & 356 &  295 &  245 &  204 &  170 &  127\\
error                     &  12 &   10 &    8 &    8 &    8 &   11\\
$dN_{ch}/d\eta/(.5N_{p})$ &0.95 & 0.90 & 0.90 & 0.89 & 0.90 & 0.92\\
error                     &0.10 & 0.09 & 0.09 & 0.09 & 0.09 & 0.09\\
\end{tabular}
\end{ruledtabular}
\end{table}

\subsection{E802/E917\label{sec:recalc_e802_cent}}
The centrality dependence of $\pi^{+}$, $K^{+}$ yields and \mt were recalculated 
from Tables~V~and~VI in reference~\cite{e802_1}. Number of participants are taken from Table~II in the same publication.
The results are presented in Table~\ref{tab:e802}. 
\begin{table}
\caption{Centrality dependence of the identified particles measured by E802/E866/E917 collaborations. Number of 
participant pairs is published (P) in Table~II in  reference~\cite{e802_1}. $\pi^{+}$ and $K^{+}$ 
values are obtained by extrapolation (E) 
from E802 measurement very close to mid-rapidity. Data is taken from Tables~V~and~VI 
in reference~\cite{e802_1}. Proton data is a compilation of the results taken from Table~IV in reference~\cite{e802_1} 
and Fig.2 in reference~\cite{e917_4}. Recalculated values (R) plotted in 
Figs.~\ref{fig:comp_sps}~and~\ref{fig:sps_ags}.\label{tab:e802}}
\begin{ruledtabular}
\begin{tabular}{clcccccccc}
P & \Np pairs                                     & 181  &  168 &  152 &  134 &  113 & 89.5 & 62.5 & 26.9\\
E & $dN^{\pi^{+}}/dy$                             & 64.5 & 56.8 & 47.6 & 39.6 & 33.3 & 25.8 & 17.8 & 6.89\\
E & error                                         & 3.13 & 2.55 & 2.75 & 1.75 & 1.68 & 1.37 & 0.89 & 0.28\\
E & $\langle m_{T}^{\pi^{+}} \rangle$ [GeV]       & .398 & .392 & .387 & .385 & .375 & .365 & .362 & .361\\
E & error                                         & .013 & .013 & .014 & .012 & .041 & .011 & .010 & .011\\
E & $dN^{K^{+}}/dy$                               & 10.6 & 9.28 & 8.12 & 6.17 & 4.91 & 3.73 & 2.22 & 0.74\\ 
E & error                                         & 0.45 & 0.37 & 0.32 & 0.27 & 0.21 & 0.16 & 0.11 & 0.04\\
E & $\langle m_{T}^{K^{+}} \rangle$ [GeV]         & .809 & .787 & .774 & .785 & .770 & .740 & .743 & .685\\
E & error                                         & .034 & .032 & .028 & .028 & .030 & .027 & .025 & .021\\
E & $dN^{p}/dy$                                   & 62.8 & 57.0 & 49.4 & 43.0 & 33.7 & 25.2 & 16.5 &  6.2\\ 
E & error                                         &  1.7 &  1.5 &  1.4 &    1 &    1 &    1 &    1 &    1\\
E & $\langle m_{T}^{p} \rangle$ [GeV]             & 1.25 & 1.26 & 1.24 & 1.23 & 1.21 & 1.19 & 1.17 & 1.14\\
R & $dE_{T}/d\eta/(.5N_{p})$                      & .608 & .580 & .527 & .492 & .460 & .426 & .396 & .335\\
R & error [GeV]                                   & .146 & .138 & .125 & .116 & .111 & .103 & .098 & .131\\
R & $dN_{ch}/d\eta/(.5N_{p})$                     & .903 & .865 & .812 & .773 & .751 & .725 & .699 & .621\\
R & error                                         & .135 & .130 & .122 & .116 & .112 & .108 & .108 & .207\\
R & \EN [GeV]                                     & .673 & .670 & .649 & .636 & .612 & .588 & .567 & .540\\
R & error [GeV]                                   & .127 & .123 & .120 & .117 & .115 & .111 & .111 & .112\\
\end{tabular}
\end{ruledtabular}
\end{table}

$K^{-}/K^{+}$ ratio was assigned a value of 0.17 for all centralities based on Tables~II~and~III in reference~\cite{e802_2}. 
This is consistent with results reported in Fig.~6 in reference~\cite{e802_3} and Fig.~11 in reference~\cite{e802_5}. 
The proton production reported in Table~IV in reference~\cite{e802_1} was compared to measurements 
reported in Fig.~2 in reference~\cite{e917_4} and Fig.~10 in reference~\cite{e802_5} for different centrality 
bins. The results are consistent. 
$\bar{p}/p$ ratio was assigned a value of 0.0003 based on Fig.~11 in reference~\cite{e802_5}. A 25\% enhancement in $\pi^{-}/\pi^{+}$ ratio for low 
$m_{T}$ reported in~\cite{e802_6} for the most central bin is not clearly seen in Fig.~11 in  reference~\cite{e802_5} 
for all centralities. Such an enhancement would contribute an additional 8-9\% to the total particle and transverse 
energy production. This is less than the systematic error on the result and the recalculation error, and thus 
this effect is not considered. The resulting values shown in Table~\ref{tab:e802} are recalculated to mid-rapidity in the C.M.S. frame.

For the lower \sqn the information about particle yields and \mt was extracted for $\pi^{+}$ and $K^{+}$ from Table~II~and~I 
in reference~\cite{e802_7} respectively, for $K^{-}$ from Table~I in reference~\cite{e802_8} and for $p$ 
from Fig.~2 in reference~\cite{e917_4}. The same assumptions as above were made to recalculate values plotted 
in Figs.~\ref{fig:sqn}~and~\ref{fig:et_nc_sqn}. The numbers are given in Table~\ref{tab:e802_1}.
\begin{table}
\caption{Recalculated values from E802/E917 experiments plotted in Figs.~\ref{fig:sqn},~\ref{fig:et_nc_sqn}.\label{tab:e802_1}}
\begin{ruledtabular}
\begin{tabular}{lccc}
\sqn [GeV]                     &     4.84      &     4.27      &     3.81      \\
$dE_{T}/d\eta/(.5N_{p})$ [GeV] & .579$\pm$.087 & .498$\pm$.075 & .405$\pm$.061 \\
$dN_{ch}/d\eta/(.5N_{p})$      & .851$\pm$.128 & .787$\pm$.118 & .678$\pm$.102 \\
\EN [GeV]                      & .680$\pm$.068 & .634$\pm$.063 & .598$\pm$.060 \\
\end{tabular}
\end{ruledtabular}
\end{table}

\subsection{Averaging procedure at \sqns=8.7~GeV\label{sec:average_eight}}
 The averaging procedure is slightly different for this curve. First the 
average results of NA45 and NA50 are produced. Then at \Np=350, this result is 
combined with the NA49 measurement using the weighted error method. A scaling 
coefficient before and after NA49 averaging is calculated. The NA45/NA50 
combined result is scaled by this factor for all values of \Nps.

\subsection{FOPI\label{sec:recalc_fopi}}
The FOPI results for \Nch were calculated for 400A MeV based on the data plotted in 
Fig.~21 in reference~\cite{fopi_1}. The points were read at the angle corresponding to 
the mid-rapidity angle ($\theta = 55^{o}$) and then converted to $dN_{ch}/d\eta$ 
resulting in 39$\pm$4 at \sqns=2.053 GeV.
\begin{table}
\caption{Particle yields measured by FOPI experiment at mid-rapidity extracted from Fig.~21 in reference~\cite{fopi_1}.\label{fopi}}
\begin{ruledtabular}
\begin{tabular}{lccccc}
Z                 &  1  &  2  &  3  &   4 &  5-6 \\
$dM/d(cos\theta)$ & 43. & 12. &  2. & 0.5 & 0.25 \\
error             & 4.3 & 1.2 & 0.2 & .05 & .025 \\
\end{tabular}
\end{ruledtabular}
\end{table}
The corresponding number of participants for a 42~mb event sample is 359 based on Fig.~8 in reference~\cite{fopi_2}.
Data for 150A MeV were compiled based on the comparison between Fig.~13 and Fig.~14 in reference~\cite{fopi_1} and the 
used definition of rapidity $y$, resulting in \dNch=~40$\pm$5 at \sqns~=~1.937~GeV.

The estimate of the \Et production at 400A MeV is made based on a comparison of the total yields 
of the particles with $Z=1$ in~\cite{fopi_1} and yields of protons and deuterons 
published in~\cite{fopi_3}. That allowed to determine number of all pions at mid-rapidity to 
be 20.6 and number of all hadrons with Z=1 15.2. Assuming that the particle temperatures
are equal to $T$=40~MeV (exact numbers are published in~\cite{fopi_2,fopi_3}), one can estimate that 
the contribution to \Et from pions is $m_{\pi}+3/2T$ and from baryons is $3/2T$, 
according to the definition of \Et used in this paper. Resulting number of 5.0 GeV is a lower 
limit estimate because the contribution of heavier particles is not considered. 
A conservative error of 30\% is assigned to this number.

\subsection{PHOBOS measurement at \sqn=56 GeV \label{sec:recalc_phobos}}
The PHOBOS experiment published \dNch=408$\pm$12(stat)$\pm$30(syst) at \sqn=~56~GeV measured 
for \Np=330$\pm$4(stat)$^{+10}_{-15}$(syst) in~\cite{phobos1}. In the same paper \dNch per 
participant between 130 GeV and 56 GeV is measured to increase by 1.31$\pm$0.04(stat)$\pm$0.05(syst). 
That allows for use of the averaged value at \sqn=\3 consistent with the PHOBOS result published 
in~\cite{phobos2} to recalculate \dNch at \sqns~=~56~GeV with smaller systematic error. This value is 
plotted in Fig.~\ref{fig:sqn}.



\section{Output tables. \label{sec:tables}}

\squeezetable
\begin{table*}
\caption{Results of the measurements by PHENIX at \sqn=\2. Errors have the same dimension as preceding value.
Results are plotted in Figs.~\ref{fig:results_vs_npart},~\ref{fig:etra_mult_vs_npart}~and~\ref{fig:ebj}.
\label{tab:results_vs_npart_200}}
\begin{ruledtabular}
\begin{tabular}{lcccccccccccccc}
                                       bin [\%] &  0-5   &  5-10 & 10-15 & 15-20 & 20-25 & 25-30 & 30-35 & 35-40 & 40-45 & 45-50 & 50-55 & 55-60 & 60-65 & 65-70\\
\hline
                                            \Np &   353. &  300. &  254. &  215. &  181. &  151. &  125. &  103. &  83.3 &  66.7 &  52.5 &  40.2 &  30.2 &  22.0\\
                \multicolumn{1}{r}{syst. error} &    10. &   9.0 &   8.1 &   7.3 &   6.6 &   6.0 &   5.5 &   5.1 &   4.7 &   4.3 &   4.1 &   3.8 &   3.6 &   3.4\\
                            $A_{\perp}$[fm$^2$] &   140. &  125. &  112. &  100. &  90.8 &  82.2 &  73.9 &  66.8 &  60.0 &  54.3 &  49.3 &  45.1 &  40.9 &   37.5\\
                \multicolumn{1}{r}{syst. error} &    11. &   10. &   9.1 &   8.2 &   7.4 &   6.8 &   6.2 &   5.7 &   5.2 &   4.8 &   4.6 &   4.5 &   4.5 &   4.9\\
\hline
                           $dE_{T}/d\eta$ [GeV] &   606. &  493. &  402. &  328. &  266. &  216. &  173. &  137. &  107. &  81.8 &  60.4 &  43.9 &  31.1 &  21.1\\
                \multicolumn{1}{r}{stat. error} &    0.6 &   0.6 &   0.5 &   0.5 &   0.4 &   0.4 &   0.3 &   0.3 &   0.2 &   0.2 &   0.2 &   0.1 &   0.1 &   0.1\\
    \multicolumn{1}{r}{    bending syst. error} &    2.4 &   5.1 &   7.0 &   7.6 &   8.2 &   8.5 &   8.2 &   8.1 &   7.6 &   6.9 &   6.4 &   5.5 &   4.6 &   3.9\\
           \multicolumn{1}{r}{full syst. error} &    32. &   27. &   22. &   19. &   16. &   14. &   12. &   11. &   9.5 &   8.1 &   7.2 &   5.9 &   4.9 &   4.0\\
\hline
$\varepsilon_{Bj}\tau$ [GeV fm$^{-2}$ c$^{-1}$] &    5.4 &   4.9 &   4.5 &   4.1 &   3.7 &   3.3 &   2.9 &   2.6 &   2.2 &   1.9 &   1.5 &   1.2 &   1.0 &   0.7\\
           \multicolumn{1}{r}{full syst. error} &    0.6 &   0.5 &   0.5 &   0.5 &   0.4 &   0.4 &   0.4 &   0.3 &   0.3 &   0.3 &   0.2 &   0.2 &   0.2 &   0.2\\
\hline
                $dE_{T}/d\eta/(0.5N_{p})$ [GeV] &   3.43 &  3.28 &  3.16 &  3.05 &  2.94 &  2.86 &  2.76 &  2.66 &  2.57 &  2.45 &  2.30 &  2.18 &  2.06 &  1.92\\
    \multicolumn{1}{r}{    bending syst. error} &   0.01 &  0.03 &  0.05 &  0.07 &  0.09 &  0.12 &  0.14 &  0.17 &  0.19 &  0.22 &  0.27 &  0.31 &  0.36 &  0.43\\
           \multicolumn{1}{r}{full syst. error} &   0.21 &  0.20 &  0.20 &  0.20 &  0.21 &  0.22 &  0.23 &  0.25 &  0.27 &  0.29 &  0.33 &  0.36 &  0.41 &  0.47\\
\hline
                            $    dN_{ch}/d\eta$ &   687. &  560. &  457. &  372. &  302. &  246. &  197. &  156. &  124. &  95.3 &  70.9 &  52.2 &  37.5 &  25.6\\
                \multicolumn{1}{r}{stat. error} &    0.7 &   0.6 &   0.5 &   0.5 &   0.4 &   0.4 &   0.3 &   0.3 &   0.2 &   0.2 &   0.2 &   0.1 &   0.1 &   0.1\\
    \multicolumn{1}{r}{    bending syst. error} &    25. &   17. &   14. &   11. &   10. &   9.9 &   9.4 &   9.0 &   8.2 &   7.8 &   7.1 &   6.1 &   5.2 &   4.4\\
           \multicolumn{1}{r}{full syst. error} &    37. &   28. &   22. &   18. &   16. &   14. &   12. &   11. &   9.6 &   8.6 &   7.6 &   6.5 &   5.4 &   4.5\\
\hline
                     $dN_{ch}/d\eta/(0.5N_{p})$ &   3.89 &  3.73 &  3.59 &  3.45 &  3.34 &  3.25 &  3.15 &  3.05 &  2.96 &  2.86 &  2.70 &  2.60 &  2.48 &  2.33\\
    \multicolumn{1}{r}{    bending syst. error} &   0.14 &  0.12 &  0.11 &  0.10 &  0.12 &  0.14 &  0.16 &  0.19 &  0.21 &  0.25 &  0.30 &  0.35 &  0.41 &  0.49\\
           \multicolumn{1}{r}{full syst. error} &   0.23 &  0.22 &  0.21 &  0.21 &  0.21 &  0.22 &  0.24 &  0.26 &  0.28 &  0.32 &  0.36 &  0.41 &  0.46 &  0.55\\
\hline
                                      \EN [GeV] &  0.881 & 0.879 & 0.881 & 0.882 & 0.881 & 0.880 & 0.875 & 0.874 & 0.866 & 0.858 & 0.851 & 0.840 & 0.828 & 0.823\\
    \multicolumn{1}{r}{    bending syst. error} &  0.032 & 0.026 & 0.021 & 0.017 & 0.015 & 0.012 & 0.011 & 0.010 & 0.011 & 0.013 & 0.017 & 0.023 & 0.032 & 0.047\\
           \multicolumn{1}{r}{full syst. error} &  0.071 & 0.069 & 0.067 & 0.066 & 0.065 & 0.065 & 0.064 & 0.064 & 0.064 & 0.064 & 0.064 & 0.065 & 0.068 & 0.076\\
\end{tabular}
\end{ruledtabular}
\end{table*}

\squeezetable
\begin{table*}
\caption{Results of the measurements by PHENIX at \sqn=\3. Errors have the same dimension as preceding value.
Results are plotted in Figs.~\ref{fig:results_vs_npart},~\ref{fig:etra_mult_vs_npart}~and~\ref{fig:ebj}.
\label{tab:results_vs_npart_130}}
\begin{ruledtabular}
\begin{tabular}{lcccccccccccccc}
                                       bin [\%] &  0-5   &  5-10 & 10-15 & 15-20 & 20-25 & 25-30 & 30-35 & 35-40 & 40-45 & 45-50 & 50-55 & 55-60 & 60-65 & 65-70\\
\hline
                                            \Np &   348. &  294. &  250. &  211. &  179. &  150. &  125. &  103. &  83.2 &  66.3 &  52.1 &  40.1 &  30.1 &  21.9\\
                \multicolumn{1}{r}{syst. error} &   10.0 &   8.9 &   8.0 &   7.2 &   6.6 &   6.0 &   5.5 &   5.1 &   4.7 &   4.3 &   4.0 &   3.8 &   3.6 &   3.4\\
                            $A_{\perp}$[fm$^2$] &   138. &  123. &  110. &  99.5 &  89.4 &  80.6 &  72.8 &  65.8 &  59.5 &  54.3 &  49.0 &  44.8 &  40.9 &  37.4\\
                \multicolumn{1}{r}{syst. error} &    11. &   9.9 &   8.9 &   8.1 &   7.3 &   6.6 &   6.1 &   5.6 &   5.2 &   4.8 &   4.6 &   4.5 &   4.6 &   4.8\\
\hline
                           $dE_{T}/d\eta$ [GeV] &   523. &  425. &  349. &  287. &  237. &  191. &  154. &  122. &  96.0 &  73.3 &  55.5 &  41.0 &  30.2 &  21.4\\
                \multicolumn{1}{r}{stat. error} &    0.9 &   0.9 &   0.8 &   0.7 &   0.6 &   0.6 &   0.5 &   0.4 &   0.4 &   0.3 &   0.3 &   0.2 &   0.2 &   0.2\\
    \multicolumn{1}{r}{    bending syst. error} &    2.6 &   4.2 &   5.6 &   7.0 &   7.5 &   7.6 &   7.5 &   7.0 &   7.3 &   6.2 &   5.8 &   5.1 &   4.4 &   4.1\\
           \multicolumn{1}{r}{full syst. error} &    27. &   22. &   19. &   16. &   14. &   12. &   11. &   9.4 &   8.8 &   7.3 &   6.5 &   5.5 &   4.7 &   4.2\\
\hline
$\varepsilon_{Bj}\tau$ [GeV fm$^{-2}$ c$^{-1}$] &    4.7 &   4.3 &   3.9 &   3.6 &   3.3 &   3.0 &   2.6 &   2.3 &   2.0 &   1.7 &   1.4 &   1.1 &   0.9 &   0.7\\
           \multicolumn{1}{r}{full syst. error} &    0.5 &   0.5 &   0.4 &   0.4 &   0.4 &   0.3 &   0.3 &   0.3 &   0.3 &   0.2 &   0.2 &   0.2 &   0.2 &   0.2\\
\hline
                $dE_{T}/d\eta/(0.5N_{p})$ [GeV] &   3.01 &  2.89 &  2.80 &  2.72 &  2.65 &  2.56 &  2.47 &  2.37 &  2.31 &  2.21 &  2.13 &  2.05 &  2.01 &  1.95\\
    \multicolumn{1}{r}{    bending syst. error} &   0.01 &  0.03 &  0.05 &  0.07 &  0.09 &  0.10 &  0.12 &  0.14 &  0.19 &  0.20 &  0.24 &  0.29 &  0.34 &  0.45\\
           \multicolumn{1}{r}{full syst. error} &   0.18 &  0.18 &  0.18 &  0.18 &  0.19 &  0.20 &  0.21 &  0.22 &  0.25 &  0.26 &  0.30 &  0.34 &  0.39 &  0.49\\
\hline
                            $    dN_{ch}/d\eta$ &   602. &  488. &  403. &  329. &  270. &  219. &  176. &  139. &  109. &  84.1 &  64.3 &  48.4 &  35.2 &  25.3\\
                \multicolumn{1}{r}{stat. error} &    1.4 &   1.2 &   1.1 &   0.9 &   0.8 &   0.7 &   0.6 &   0.5 &   0.4 &   0.3 &   0.3 &   0.3 &   0.2 &   0.2\\
    \multicolumn{1}{r}{    bending syst. error} &    19. &   13. &   10. &   9.9 &   8.6 &   8.4 &   8.3 &   7.7 &   7.5 &   6.4 &   5.9 &   5.2 &   4.3 &   4.1\\
           \multicolumn{1}{r}{full syst. error} &    28. &   22. &   17. &   15. &   13. &   11. &   10. &   9.1 &   8.4 &   7.0 &   6.3 &   5.4 &   4.5 &   4.1\\
\hline
                     $dN_{ch}/d\eta/(0.5N_{p})$ &   3.46 &  3.32 &  3.23 &  3.12 &  3.03 &  2.93 &  2.82 &  2.70 &  2.63 &  2.54 &  2.47 &  2.41 &  2.34 &  2.31\\
    \multicolumn{1}{r}{    bending syst. error} &   0.11 &  0.09 &  0.08 &  0.09 &  0.10 &  0.12 &  0.14 &  0.16 &  0.19 &  0.21 &  0.25 &  0.30 &  0.35 &  0.47\\
           \multicolumn{1}{r}{full syst. error} &   0.19 &  0.18 &  0.17 &  0.18 &  0.18 &  0.19 &  0.21 &  0.22 &  0.25 &  0.27 &  0.31 &  0.35 &  0.41 &  0.52\\
\hline
                                      \EN [GeV] &  0.869 & 0.870 & 0.867 & 0.874 & 0.877 & 0.873 & 0.875 & 0.876 & 0.878 & 0.871 & 0.864 & 0.847 & 0.857 & 0.844\\
    \multicolumn{1}{r}{    bending syst. error} &  0.028 & 0.023 & 0.019 & 0.016 & 0.015 & 0.014 & 0.014 & 0.016 & 0.020 & 0.025 & 0.033 & 0.043 & 0.060 & 0.083\\
           \multicolumn{1}{r}{full syst. error} &  0.066 & 0.064 & 0.063 & 0.062 & 0.062 & 0.062 & 0.062 & 0.063 & 0.064 & 0.065 & 0.068 & 0.073 & 0.084 & 0.101\\
\end{tabular}
\end{ruledtabular}
\end{table*}

\begin{table*}
\caption{Results of the measurements by PHENIX at \sqn=\1. Errors have the same dimension as preceding value.
Results are plotted in Figs.~\ref{fig:results_vs_npart},~\ref{fig:etra_mult_vs_npart}~and~\ref{fig:ebj}.
\label{tab:results_vs_npart_019}}
\begin{ruledtabular}
\begin{tabular}{lcccccccccc}
                                       bin [\%] &  0-5   &  5-10 & 10-15 & 15-20 & 20-25 & 25-30 & 30-35 & 35-40 & 40-45 & 45-50\\
\hline
                                            \Np &   336. &  288. &  243. &  204. &  172. &  144. &  120. &  98.4 &  79.8 &  63.8\\
                \multicolumn{1}{r}{syst. error} &    9.7 &   8.8 &   7.9 &   7.1 &   6.4 &   5.9 &   5.4 &   5.0 &   4.6 &   4.3\\
                            $A_{\perp}$[fm$^2$] &   133. &  119. &  106. &  95.6 &  85.8 &  77.2 &  69.7 &  62.7 &  56.7 &  51.3\\
                \multicolumn{1}{r}{syst. error} &    11. &   9.6 &   8.6 &   7.8 &   7.0 &   6.4 &   5.8 &   5.3 &   4.9 &   4.6\\
\hline
                           $dE_{T}/d\eta$ [GeV] &   230. &  194. &  164. &  134. &  109. &  88.4 &  72.0 &  58.1 &  45.3 &  35.2\\
                \multicolumn{1}{r}{stat. error} &    0.8 &   0.8 &   0.8 &   0.7 &   0.5 &   0.5 &   0.4 &   0.4 &   0.3 &   0.3\\
    \multicolumn{1}{r}{    bending syst. error} &    1.7 &   2.6 &   2.9 &   4.0 &   3.8 &   3.8 &   3.9 &   3.9 &   3.9 &   3.4\\
           \multicolumn{1}{r}{full syst. error} &    14. &   12. &   11. &   9.3 &   7.8 &   6.7 &   6.0 &   5.3 &   4.8 &   4.0\\
\hline
$\varepsilon_{Bj}\tau$ [GeV fm$^{-2}$ c$^{-1}$] &    2.2 &   2.0 &   1.9 &   1.8 &   1.6 &   1.4 &   1.3 &   1.2 &   1.0 &   0.9\\
           \multicolumn{1}{r}{full syst. error} &    0.2 &   0.2 &   0.2 &   0.2 &   0.2 &   0.2 &   0.2 &   0.2 &   0.1 &   0.1\\
\hline
                $dE_{T}/d\eta/(0.5N_{p})$ [GeV] &   1.37 &  1.35 &  1.35 &  1.31 &  1.27 &  1.22 &  1.20 &  1.18 &  1.13 &  1.10\\
    \multicolumn{1}{r}{    bending syst. error} &   0.01 &  0.02 &  0.02 &  0.04 &  0.05 &  0.05 &  0.07 &  0.08 &  0.10 &  0.11\\
           \multicolumn{1}{r}{full syst. error} &   0.09 &  0.09 &  0.10 &  0.10 &  0.10 &  0.10 &  0.11 &  0.12 &  0.14 &  0.15\\
\hline
                            $    dN_{ch}/d\eta$ &   312. &  265. &  226. &  187. &  154. &  125. &  102. &  82.6 &  65.0 &  51.1\\
                \multicolumn{1}{r}{stat. error} &    0.9 &   0.9 &   0.9 &   0.8 &   0.7 &   0.6 &   0.5 &   0.4 &   0.4 &   0.3\\
    \multicolumn{1}{r}{    bending syst. error} &    5.3 &   4.5 &   4.4 &   5.4 &   4.9 &   5.3 &   5.2 &   5.1 &   5.1 &   4.2\\
           \multicolumn{1}{r}{full syst. error} &    21. &   18. &   15. &   13. &   11. &   9.7 &   8.5 &   7.4 &   6.6 &   5.4\\
\hline
                     $dN_{ch}/d\eta/(0.5N_{p})$ &   1.86 &  1.84 &  1.85 &  1.83 &  1.78 &  1.74 &  1.71 &  1.68 &  1.63 &  1.60\\
    \multicolumn{1}{r}{    bending syst. error} &   0.03 &  0.03 &  0.04 &  0.05 &  0.06 &  0.08 &  0.09 &  0.11 &  0.14 &  0.14\\
           \multicolumn{1}{r}{full syst. error} &   0.14 &  0.14 &  0.14 &  0.14 &  0.14 &  0.15 &  0.16 &  0.17 &  0.19 &  0.20\\
\hline
                                      \EN [GeV] &  0.738 & 0.733 & 0.728 & 0.720 & 0.711 & 0.705 & 0.704 & 0.704 & 0.697 & 0.690\\
    \multicolumn{1}{r}{    bending syst. error} &  0.027 & 0.023 & 0.020 & 0.017 & 0.015 & 0.014 & 0.014 & 0.016 & 0.019 & 0.024\\
           \multicolumn{1}{r}{full syst. error} &  0.078 & 0.076 & 0.075 & 0.073 & 0.072 & 0.071 & 0.071 & 0.071 & 0.072 & 0.072\\
\end{tabular}
\end{ruledtabular}
\end{table*} 

\begin{table*}
\caption{Ratios of measured quantities at \2/\3 and \2/\1. The number of \Np is the average between two energies. 
The data is plotted in Fig.~\ref{fig:ratios_vs_npart}.
 \label{tab:ratios_vs_npart}}
\begin{ruledtabular}
\begin{tabular}{lcccccccccccccc}
                                       bin [\%] &  0-5   &  5-10 & 10-15 & 15-20 & 20-25 & 25-30 & 30-35 & 35-40 & 40-45 & 45-50 & 50-55 & 55-60 & 60-65 & 65-70\\
\hline
\multicolumn{15}{c}{\2/\3}\\
\hline
                                            \Np &   350. &  297. &  252. &  213. &  180. &  150. &  125. &  103. &  83.2 &  66.5 &  52.3 &  40.2 &  30.2 &  22.0\\
                \multicolumn{1}{r}{syst. error} &    10. &   9.0 &   8.1 &   7.3 &   6.6 &   6.0 &   5.5 &   5.1 &   4.7 &   4.3 &   4.1 &   3.8 &   3.6 &   3.4\\
\hline
                      $dE_{T}/d\eta/(0.5N_{p})$ &   1.14 &  1.13 &  1.13 &  1.12 &  1.11 &  1.12 &  1.12 &  1.12 &  1.11 &  1.11 &  1.08 &  1.07 &  1.03 &  0.98\\
    \multicolumn{1}{r}{    bending syst. error} &   0.01 &  0.01 &  0.01 &  0.01 &  0.01 &  0.02 &  0.02 &  0.03 &  0.04 &  0.05 &  0.05 &  0.07 &  0.09 &  0.13\\
           \multicolumn{1}{r}{full syst. error} &   0.04 &  0.04 &  0.04 &  0.04 &  0.04 &  0.05 &  0.05 &  0.05 &  0.06 &  0.06 &  0.07 &  0.08 &  0.10 &  0.13\\
\hline
                     $dN_{ch}/d\eta/(0.5N_{p})$ &   1.12 &  1.12 &  1.11 &  1.11 &  1.10 &  1.11 &  1.12 &  1.13 &  1.13 &  1.13 &  1.10 &  1.08 &  1.06 &  1.01\\
    \multicolumn{1}{r}{    bending syst. error} &   0.01 &  0.01 &  0.01 &  0.01 &  0.01 &  0.01 &  0.02 &  0.02 &  0.03 &  0.03 &  0.03 &  0.03 &  0.04 &  0.07\\
           \multicolumn{1}{r}{full syst. error} &   0.04 &  0.04 &  0.04 &  0.04 &  0.04 &  0.04 &  0.04 &  0.04 &  0.04 &  0.04 &  0.04 &  0.05 &  0.05 &  0.08\\
\hline
\multicolumn{15}{c}{\2/\1}\\
\hline
                                            \Np &   344. &  294. &  249. &  210. &  177. &  148. &  122. &  101. &  81.6 &  65.2\\
                \multicolumn{1}{r}{syst. error} &    10. &   9.0 &   8.1 &   7.3 &   6.6 &   6.0 &   5.5 &   5.1 &   4.7 &   4.3\\
\hline
                      $dE_{T}/d\eta/(0.5N_{p})$ &   2.50 &  2.43 &  2.34 &  2.32 &  2.32 &  2.34 &  2.29 &  2.25 &  2.26 &  2.22\\
    \multicolumn{1}{r}{    bending syst. error} &   0.02 &  0.03 &  0.04 &  0.06 &  0.07 &  0.09 &  0.11 &  0.13 &  0.17 &  0.19\\
           \multicolumn{1}{r}{full syst. error} &   0.09 &  0.09 &  0.09 &  0.10 &  0.10 &  0.12 &  0.13 &  0.15 &  0.19 &  0.20\\
\hline
                     $dN_{ch}/d\eta/(0.5N_{p})$ &   2.09 &  2.03 &  1.94 &  1.89 &  1.87 &  1.87 &  1.84 &  1.81 &  1.82 &  1.79\\
    \multicolumn{1}{r}{    bending syst. error} &   0.08 &  0.06 &  0.06 &  0.06 &  0.06 &  0.07 &  0.08 &  0.10 &  0.12 &  0.13\\
           \multicolumn{1}{r}{full syst. error} &   0.15 &  0.14 &  0.13 &  0.13 &  0.13 &  0.14 &  0.14 &  0.15 &  0.17 &  0.17\\
\end{tabular}
\end{ruledtabular}
\end{table*} 

\begin{table*}
\caption{ Average values of \dNch/(0.5\Nps) at different \sqns. An 
additional 5\% error should be added to columns \7 through 4.8~GeV for 
the uncertainty related to recalculation to the Center of Mass system. 
The results are presented in Figs.~\ref{fig:comp_rhic}~and~\ref{fig:sps_ags}.
\label{tab:averages}}
\begin{ruledtabular}
\begin{tabular}{lccccccccccccccc}
\Np                      &  375 &  350 &  325 &  300 &  275 &  250 &  225 &  200 &  175 &  150 &  125 &  100 &   75 &   50 &   25 \\
\hline
                      \2 & 3.92 & 3.81 & 3.72 & 3.65 & 3.56 & 3.51 & 3.45 & 3.38 & 3.34 & 3.27 & 3.20 & 3.14 & 3.03 & 2.73 & 2.78 \\
\multicolumn{1}{r}{error}& 0.13 & 0.13 & 0.12 & 0.12 & 0.12 & 0.12 & 0.12 & 0.11 & 0.12 & 0.12 & 0.12 & 0.13 & 0.13 & 0.13 & 0.43 \\
                      \3 & 3.41 & 3.31 & 3.22 & 3.16 & 3.11 & 3.07 & 3.04 & 3.00 & 2.96 & 2.89 & 2.83 & 2.73 & 2.65 & 2.53 & 2.36 \\
\multicolumn{1}{r}{error}& 0.10 & 0.10 & 0.10 & 0.10 & 0.09 & 0.09 & 0.10 & 0.09 & 0.10 & 0.10 & 0.10 & 0.11 & 0.11 & 0.12 & 0.30 \\
                      \1 &      & 1.91 & 1.89 & 1.88 & 1.87 & 1.87 & 1.85 & 1.83 & 1.81 & 1.76 & 1.72 & 1.68 & 1.62 \\
\multicolumn{1}{r}{error}&      & 0.11 & 0.11 & 0.11 & 0.11 & 0.12 & 0.12 & 0.12 & 0.12 & 0.13 & 0.14 & 0.15 & 0.19 \\
                      \7 & 1.97 & 1.93 & 1.90 & 1.88 & 1.83 & 1.80 & 1.78 & 1.75 & 1.72 & 1.69 & 1.66 & 1.66 & 1.61 & 1.54 & 1.45 \\
\multicolumn{1}{r}{error}& 0.12 & 0.12 & 0.14 & 0.15 & 0.16 & 0.17 & 0.17 & 0.17 & 0.17 & 0.17 & 0.16 & 0.23 & 0.21 & 0.19 & 0.13 \\
                 8.7~GeV & 1.26 & 1.22 & 1.20 & 1.18 & 1.17 & 1.16 & 1.16 & 1.14 & 1.14 & 1.14 & 1.14 & 1.14 \\
\multicolumn{1}{r}{error}& 0.11 & 0.11 & 0.11 & 0.10 & 0.10 & 0.10 & 0.10 & 0.10 & 0.09 & 0.09 & 0.09 & 0.09 \\
                 4.8~GeV & 0.92 & 0.89 & 0.85 & 0.81 & 0.78 & 0.76 & 0.75 & 0.74 & 0.72 & 0.71 & 0.70 & 0.67 & 0.64 & 0.63 \\
\multicolumn{1}{r}{error}& 0.14 & 0.13 & 0.13 & 0.12 & 0.12 & 0.11 & 0.11 & 0.11 & 0.11 & 0.11 & 0.11 & 0.14 & 0.18 & 0.21\\

\end{tabular}
\end{ruledtabular}
\end{table*}

\end{document}